\documentclass[pre,aps]{revtex4}
\def\@llsymbol#1{\ifcase#1\or {}\or {'}\or {''}\or {'''}\or
   {''''}\or {'''''}\or  \else\@ctrerr\fi\relaz}
\newcounter{contador}
\newcommand{\letra}{
   \stepcounter{equation}
   \setcounter{contador}{\value{equation}}
   \setcounter{equation}{0}
   \renewcommand{\theequation}{\thecontador\alph{equation}}}
\newcommand{\antiletra}{
   \renewcommand{\theequation}{\arabic{equation}}
   \setcounter{equation}{\value{contador}}}

\usepackage{color}
\usepackage{indentfirst}
\usepackage{eucal}
\usepackage{amsmath}
\usepackage{amsfonts}
\usepackage{amssymb}
\usepackage{mathrsfs}
\begin{document}


%
\title{Integral relations for solutions of confluent Heun equations}
\author{L\'ea Jaccoud El-Jaick}
\email{leajj@cbpf.br}
\author{Bartolomeu D. B. Figueiredo}
\email{barto@cbpf.br}
\affiliation{Centro Brasileiro de Pesquisas F\'{\i}sicas (CBPF),\\
Rua Dr. Xavier Sigaud, 150, CEP 22290-180, Rio de Janeiro, RJ, Brasil}

\begin{abstract}
\noindent
{\bf Abstract: }Firstly, we construct kernels of integral relations among solutions of the
confluent Heun equation (CHE) and its limit, the reduced CHE (RCHE). In
both cases we generate additional kernels by systematically applying substitutions
of variables. Secondly, we establish integral
relations between known solutions of the CHE that are power series and
solutions that are series of special functions; and similarly for solutions of
the RCHE. Thirdly, by using one of the integral relations as an integral transformation
we obtain a new series solution of the spheroidal wave equation. From
this solution we construct new solutions of the general CHE, and show
that these are suitable for solving the radial part of the two-center problem in
quantum mechanics. Finally, by applying a limiting process to kernels for
the CHEs we obtain kernels for {two} double-confluent Heun equations.
{ As a result, we deal with kernels of four equations of the Heun family, each
equation presenting a distinct structure of singularities. In addition, we find that the known kernels
for the Mathieu equation are special instances of kernels of the RCHE.}


\end{abstract}

\maketitle

%
%




\tableofcontents


%


\section*{1. Introductory remarks}

Recently we have found that  
the transformations of variables which preserve the form of the general Heun 
equation correspond to transformations which preserve the form
of the equation for the kernels of integral relations among solutions of the Heun equation \cite{lea}.
In fact, by using the known transformations of the Heun equation \cite{heun,maier} we have found 
prescriptions for transforming kernels and, in this manner, 
we have generated several new kernels for the equation.

The above correspondence can be extended to   
the confluent equations of the Heun family, that is, to the (single) confluent, 
double-confluent, biconfluent and triconfluent Heun equations
\cite{ronveaux,nist}, as well as to the reduced forms of such equations
\cite{kazakov-1,lay}. In the present study
we consider only the confluent Heun equation (CHE) and equations
connected to the CHE by limiting processes. 
Specifically:
 \begin{itemize}
 \itemsep-3pt
\item we deal with the construction and  
 transformations of integral kernels for CHE and its limit called
 reduced confluent Heun equation (RCHE);
\item
 from some of these kernels we  
  establish integral relations 
 between known solutions for the CHE; 
 \item 
 using one of the relations as an integral transformation  we obtain new 
 solutions in series of confluent hypergeometric functions for the CHE;
 \item
 we show that the previous solutions are suitable to solve the radial
 part of the Schr\"{o}dinger equation for an electron in the field of
 two Coulomb centres \cite{wilson0} (two-centre problem);
   \item
   finally, from kernels of the { CHE and RCHE  we find   
    kernels for the double-confluent Heun equation (DHE) and for 
    the reduced DHE (RDHE).}
 \end{itemize}



We write the CHE as \cite{leaver}
\begin{eqnarray}
\label{gswe}
\displaystyle z(z-z_{0})\frac{d^{2}U}{dz^{2}}+(B_{1}+B_{2}z)\frac{dU}{dz}
+\displaystyle\left[B_{3}-2\omega\eta(z-z_{0})+\omega^{2}z(z-z_{0})\right]U=0,
\end{eqnarray}
where $z_{0}$, $B_{i}$, $\eta$ and $\omega$ are constants.
This equation is called {\it generalized spheroidal wave equation}  by  Leaver  \cite{leaver}
but sometimes such expression refers to a particular case of the CHE \cite{nist,wilson}.  
Excepting the special case represented by the Mathieu equation, 
the CHE is the most studied of the confluent Heun equations and embraces the 
(ordinary) spheroidal equation as a particular case \cite{nist}. However, further studies  
are necessary due to the recent emergence of several classes of quantum two-state systems 
ruled by the CHE \cite{ishkhanyan}. 
On the other side, the reduced confluent 
Heun equation (RCHE) is written as
\begin{equation}\label{incegswe}
z(z-z_0)\frac{d^{2}U}{dz^{2}}+(B_{1}+B_{2}z)
\frac{dU}{dz}+
\left[B_{3}+q(z-z_{0})\right]U=0,
\end{equation}
where $z_0$, $B_{i}$ and $q$ ($q\neq0$) are constants. The RCHE describes 
the angular part of the Schr\"odinger equation 
for an electron in the field of a point electric dipole \cite{levy,volcano}. 
It appears as well in the study of two-level systems \cite{two-level}, 
polymer dynamics \cite{fene} and theory of gravitation \cite{branas}. 
The form (\ref{incegswe}) for the RCHE results from the CHE (\ref{gswe}) by means of the limits 
\begin{eqnarray}\label{ince}
\omega\rightarrow 0, \quad
\eta\rightarrow
\infty \quad \mbox{such that }\quad \ 2\eta \omega =-q,
\qquad [\text{Whittaker-Ince limit}].
\end{eqnarray}
In both equations, $z=0$ and $z=z_{0}$ are regular singular
points with exponents ($0,1+B_{1}/z_{0}$)
and ($0,1-B_{2}-B_{1}/z_{0}$), respectively, that is, {from ascending 
power series solutions we find
\begin{eqnarray}\label{frobenius}
\begin{array}{l}
 z\to 0:\quad U(z)\sim1\quad \text{or}\quad
U(z)\sim z^{1+\frac{B_{1}}{z_{0}}};\qquad
z\to z_0:\quad
U(z)\sim 1\quad \text{or}\quad
U(z)
\sim (z-z_{0})^{1-B_{2}-\frac{B_{1}}{z_{0}}}.
\end{array}
\end{eqnarray}
In contrast, at the irregular singular point $z=\infty$, the behaviour of the solutions 
is different for each equation since 
\begin{eqnarray}\label{thome1}\begin{array}{l}
z\to \infty:\quad 
U(z)\sim e^{\pm i\omega z}\ z^{\mp i\eta-(B_{2}/2)}\ 
\text { for the CHE (\ref{gswe}) and}
\quad
U(z)\sim
e^{\pm 2i\sqrt{qz}}\ z^{(1/4)-(B_{2}/2)} \text{ for the 
RCHE (\ref{incegswe})},\end{array}
\end{eqnarray}
as follow from the normal and the subnormal Thom\'e solutions \cite{olver} 
for the CHE and RCHE, respectively. }

According to the concepts of Ref. \cite{lay},
the s-rank of the singularity at $z=\infty$ is 2 for the CHE, and 
$3/2$ for the RCHE. However, more important is the fact 
that the solutions exhibit the above behavior predicted by the normal or subnormal
Thom\'e solutions, and the fact that the Whittaker-Ince limit (\ref{ince}) may generate solutions
to the RCHE. In effect, most of the known solutions for the RCHE \cite{lea-2,eu,coulomb}
has been obtained from solutions of the CHE by means of the limit (\ref{ince}).  
Despite this, the main part of the present study is restricted to integral relations 
concerning the CHE. Relations for RCHE are relegated to an 
appendix. In appendices we also present kernels for double-confluent Heun equations
which are obtained by taking $z_0=0$ in Eqs. (\ref{gswe})  and (\ref{incegswe}). 

Integral relations are important because, in principle, they make possible the  
transformation of known solutions into solutions 
with different properties. However, apart from the Mathieu equation, only in 
rare cases this task has been
accomplished successfully. One case is constituted by the expansions of the Lam\'e
functions in series of associated Legendre functions \cite{lame}, obtained by Erd\'elyi 
from Fourier-Jacobi series for the Lam\'e equation; however,
as far we are aware, his solutions have not been extended
for the general Heun equation (of which Lam\'e equation is a particular case). 
Another example is a Leaver expansion in series of irregular confluent 
hypergeometric functions for the CHE \cite{leaver}, 
obtained from a power series; the integral transformation was 
originally constructed for a particular case of CHE but the expansion has been generalized for any CHE.

To establish integral relations for solutions it is necessary 
to get appropriate integral kernels. To this end, in section 2 we proceed as in 
case of the general Heun equation \cite{lea}. In other words, firstly 
we insert into the integral connecting two solutions a weight function $w(z,t)$  
which allows to write the CHE and the equation for its kernels in terms of 
differential operators functionally
identical (respecting $z$ and $t$). In this manner, 
by examining each variable substitution which leaves invariant the form of the 
CHE (one variable, $z$) we find prescriptions for the  variables transformations which 
preserve the form of the equation for the kernels (two variables, $z$ and $t$). By using 
these substitutions, we may systematically convert a given (initial) kernel 
into new kernels. {As initial kernels we use the 
ones obtained as limits of kernels of the general 
Heun equation \cite{lea}, adapting them to the form (\ref{gswe}) for the CHE.}


In section 3 we find integral relations  
which transform  the Jaff\'e power-series  solutions \cite{jaffe} into 
expansions in series of irregular confluent
hypergeometric functions, including the 
aforementioned solution given by Leaver. In the second
place, we find that the power-series solutions
of Baber and Hass\'e \cite{baber} are transformed into expansions in
series of regular confluent hypergeometric functions. 
These are integral transformations among known solutions of the CHE. 
In both examples, power-series solutions are converted into 
series of confluent hypergeometric functions. {However, there 
are the non-integral transformations (involving only substitutions of variables)
which do not modify the type of series: these transform, for example,
a power-series solution into another power-series solution, and
an expansion in series of hypergeometric functions into another 
expansion in series hypergeometric functions. Integral relations 
among these two types of modified series demand the use of kernels transformed in accordance with 
the prescriptions mentioned in the previous paragraph.}


Analogously, in section 4 we apply an 
integral transformation to an asymptotic (Thom\'e) solution
of the spheroidal equation and obtain a new solution
in series of irregular confluent hypergeometric functions.  That 
solution is extended to any CHE (not just the spheroidal equation); then, by substitutions
of variables, we obtain a group of solutions for the CHE with domains 
of convergence different of the ones of the asymptotic solutions. Therefore,
by combining integral and non-integral transformations we get new solutions for the CHE; as a test, 
we show that some of these solutions
afford bounded and convergent solutions to the radial
part of the two-center problem. 
 
In section 5, we present concluding 
remarks and mention open issues. In appendix A we write some formulas concerning special
functions, while in appendix
B we discuss the convergence of asymptotic 
solutions for the CHE. {In appendices C, D and E we obtain, respectively,  kernels for the
reduced CHE (RCHE), for the double-confluent Heun equation (DHE) 
 and for the  reduced DHE (RDHE).}
\section*{2. Kernels for the confluent Heun equation}
In this section we regard kernels for the CHE (\ref{gswe}). In particular,
\begin{itemize}
  \itemsep-3pt
  \item
  in section 2.1 we get the correspondences among substitutions of variables which preserve the form of the
  CHE and the substitutions which preserve the equation of the kernels of the CHE;
   \item   
     in section 2.2 we construct a group of kernels with an
   arbitrary constant of separation $\lambda$, given by 
   products of two confluent hypergeometric functions and elementary functions;
    \item 
    in section 2.3 we find another group of kernels with an
    arbitrary constant of separation $\lambda$, given by products of confluent hypergeometric functions 
   and Gauss hypergeometric functions (and elementary functions);
 \item
{in sections 2.4, 2.5 and 2.6, by taking suitable values for $\lambda$ we get 
  kernels given by products of elementary and  special functions};
  thus, in sections 2.4 and 2.5 we find products of elementary and confluent
  hypergeometric functions and, in section 2.6, products of elementary and
  Gauss hypergeometric functions.
  \end{itemize}

 
Later on, in section 4, we will need kernels for the ordinary spheroidal wave 
equation \cite{nist}
\begin{equation}\label{esferoidal}
\begin{array}{l}
\frac{d}{dx}\left[\left( 1-x^2\right) \frac{dX(x)}{dx} \right]+
\left[ \gamma^2(1-x^2)+\bar{\lambda}-
\frac{\mu^2}{1-\
x^2}\right] X(x)=0.
\end{array}
\end{equation}
Such kernels are obtained from the ones of the CHE
through the substitutions
\begin{eqnarray}\label{esferoidal-2}
x=1-2z,\qquad X(x)=z^{{\mu}/{2}
}\ (z-1)^{{\mu}/{2}}U(z),
\end{eqnarray}
which give
\letra
\begin{eqnarray}\label{esferoidal-2a}
\begin{array}{l}
z(z-1)\frac{d^{2}U}{dz^{2}}+
\left[
-\left(\mu+1\right)+2
\left(\mu+1\right)z
\right]
\frac{dU}{dz}+
\left[
\mu\left(
\mu+1\right)-\bar{\lambda}+4\gamma^{2}z(z-1)
\right]U=0,\end{array}
\end{eqnarray}
that is, the CHE (\ref{gswe}) with
\begin{eqnarray}\label{esferoidal-3}
z_0=1,\qquad B_2=-2B_1=2(\mu+1),
\qquad B_3=\mu(\mu+1)-\bar{\lambda},\qquad
 \eta=0,
\qquad \omega^2=4\gamma^2.
\end{eqnarray}
Thus, the spheroidal equation (\ref{esferoidal})
will be treated as a CHE with $z_0=1$, $\eta=0$ and $B_2=-2B_1$,
namely, 
\antiletra
\begin{eqnarray}\label{CHE-esferoidal}
\displaystyle z(z-1)\frac{d^{2}U}{dz^{2}}+(B_{1}-2B_{1}z)\frac{dU}{dz}
+\displaystyle\left[B_{3}+\omega^{2}z(z-1)\right]U=0.
\end{eqnarray}

\subsection*{2.1. Transformations of the CHE 
and its kernels}

Defining the operator $L_{z}$ by
\begin{equation}
\label{Lz1}
\begin{array}{l}
L_{z}=z[z-z_0]\frac{\partial^{2}}{\partial z^{2}}+
\left[B_{1}+B_{2}z\right]
\frac{\partial}{\partial z}
+\left[\omega^{2}z(z-z_{0})-
2\omega\eta z\right]
\end{array}
\end{equation}
and interpreting this as an ordinary differential operator,
the CHE (\ref{gswe}) reads
\begin{eqnarray}
\label{gswe2}
[L_{z}+B_{3}+2\eta \omega z_{0}]U(z)=0.
\end{eqnarray}
The adjoint operator $\bar{L}_{z}$ corresponding to
${L}_{z}$ is \cite{ince2}
\begin{eqnarray}\label{Lz2}
\bar{L}_{z}=
z(z-z_{0})\frac{\partial^{2}}{\partial z^{2}}+\left[-2z_{0}-B_{1}
+(4-B_{2})z\right]\frac{\partial}{\partial z}
+
\left[\omega^{2}z(z-z_{0})-2\omega\eta z+2-B_{2}\right].
\end{eqnarray}

On the other side, if $U(z)$ is a known {solution of
the CHE}, we seek new solutions
$\mathcal{U}(z)$ having the form
\begin{eqnarray}
\label{integral}
&\mathcal{U}(z)=\int_{t_{1}}^{t_{2}}K(z,t)U(t)dt=
\int_{t_{1}}^{t_{2}}w(z,t)G(z,t)U(t)dt
=\int_{t_{1}}^{t_{2}}t^{-1-\frac{B_1}{z_0}}
(t-z_0)^{B_2+\frac{B_1}{z_0}-1}G(z,t)U(t)dt,
\\
& w(z,t)=t^{-1-\frac{B_1}{z_0}}
(t-z_0)^{B_2+\frac{B_1}{z_0}-1},\nonumber
\end{eqnarray}
where the kernel $K(z,t)$ or $ G(z,t)$
is determined from
a partial differential equation. The
general theory is usually established for the function
$K(z,t)$ \cite{ince2}, but to study the transformations
of kernels we will deal with $G(z,t)$. 
If the integration endpoints $t_1$ and $t_2$ 
are independent of $z$, 
by applying $L_{z}$ to integral (\ref{integral})
we find
\begin{eqnarray}
\label{integral2}
L_{z}\mathcal{U}(z)=\int_{t_{1}}^{t_{2}}
\left[ L_{z} K(z,t)\right] U(t)dt
=\int_{t_{1}}^{t_{2}}U(t)
\left[ L_{z}-\bar{L}_{t}\right] K(z,t)dt
+\int_{t_{1}}^{t_{2}} U(t)\bar{L}_{t}K(z,t)dt,
\end{eqnarray}
$\bar{L}_{t}$ being obtained from
$\bar{L}_{z}$ by replacing $z$ with $t$.
Now we demand that
\begin{eqnarray}\label{nucleus-gswe}
\left[L_{z}-\bar{ L}_{t}\right] K(z,t)=0 \quad \Leftrightarrow\quad
%
%
\left[L_{z}-{ L}_{t}\right] G(z,t)=0.
\end{eqnarray}
Thence, by using the Lagrange identity
\[U(t)\bar{L}_{t}K(z,t)-K(z,t) L_{t}U(t)=
\frac{\partial}{\partial t}P(z,t),\]
%
where the bilinear concomitant  $P(z,t)$  is given by
\begin{eqnarray}
\label{concomitant}
P(z,t)&=&
t(t-z_{0})\left[U(t)
\frac{\partial K(z,t)}{\partial t}
-K(z,t)\frac{dU(t)}{d t}\right]  -
\left[\left(B_{2}-2\right)t +B_{1}+z_{0}\right]U(t)K(z,t)
\nonumber\\
&=&\begin{array}{l}t^{-\frac{B_1}{z_0}}(t-z_0)^{B_2+\frac{B_1}{z_0}}
\left[U(t)\frac{\partial G(z,t)}{\partial t}-G(z,t)\frac{d U(t)}{d t}\right],\end{array}
\end{eqnarray}
Eq. (\ref{integral2}) reduces to
\begin{eqnarray*}
L_{z}\mathcal{U}(z)=
\int_{t_{1}}^{t_{2}}\left[K(z,t)L_{t}U(t)+
\frac{\partial{P(z,t)}}{\partial{t}}\right]dt
=
-(B_3+2\eta\omega z_0)
\int_{t_{1}}^{t_{2}}K(z,t)U(t)dt+
\int_{t_{1}}^{t_{2}}\frac{\partial{P(z,t)}}{\partial{t}}dt,
\end{eqnarray*}
where in the last step we have used equation (\ref{gswe2}).
Using equation  (\ref{integral}) as well, this yields
\begin{eqnarray}
\label{condition1}
[L_{z}+B_{3}+2\eta\omega z_0]\mathcal{U}(z)=
P(z,t_{2})-P(z,t_{1}).
\end{eqnarray}
Therefore, $\mathcal{U}(z)$ is also a solution of the CHE if:  (i)
the kernel satisfies Eq. (\ref{nucleus-gswe}),
(ii) the integral (\ref{integral}) exists and
(iii) the right-hand side of Eq. (\ref{condition1}) vanishes.

Now let us examine the transformations of the solutions
$U(z)$ and kernels $G(z,t)$.
If $U(z)=U(B_{1},B_{2},B_{3}; z_{0},\omega,\eta;z)$
denotes one solution of the CHE, the following transformations
\cite{lea-2,decarreau1,decarreau2} -- $T_{1},\ T_{2},\ T_{3}$ and
$T_{4}$ --  leave invariant the form of the CHE:
%
\begin{eqnarray}\label{transformacao1}
\begin{array}{ll}
T_{1}U(z)=z^{1+\frac{B_1}{z_0}}
U(C_{1},C_{2},C_{3};z_{0},\omega,\eta;z),\qquad&
%
T_{2}U(z)=(z-z_{0})^{1-B_{2}-\frac{B_1}{z_0}}\ U(B_{1},D_{2},D_{3};
z_{0},\omega,\eta;z),\vspace{3mm}\\
%
T_{3}U(z)=U(B_{1},B_{2},B_{3}; z_{0},-\omega,-\eta;z),&
%
T_{4}U(z)
=U(-B_{1}-B_{2}z_{0},B_{2},
B_{3}+2\eta\omega z_{0};z_{0},-\omega,
\eta;z_{0}-z),
\end{array}
\end{eqnarray}
where
%
\begin{eqnarray}\label{constantes-C-D}
&\begin{array}{l}
C_{1}=-B_{1}-2z_{0}, \qquad
C_{2}=2+B_{2}+\frac{2B_{1}}{z_{0}},
\qquad C_{3}=B_{3}+
\left[1+\frac{B_{1}}{z_{0}}\right]
\left[B_{2}+\frac{B_{1}}{z_{0}}\right], \end{array} \nonumber\\
&\begin{array}{l}
D_{2}=2-B_{2}-\frac{2B_{1}}{z_{0}},\qquad
D_{3}=B_{3}+
\frac{B_{1}}{z_{0}}\left(\frac{B_{1}}{z_{0}}
+B_{2}-1\right).\end{array}
\end{eqnarray}
By composition of these  transformations,
from an initial solution  we may generate a group
containing up to  16 solutions. 
To get the corresponding transformations
for the kernels, we notice that
the operators $L_z$ and $L_t$ (which appear in the CHE
(\ref{gswe2}) and in $[L_z-L_t]G(z,t)=0$) have the same
functional form. Hence, 
if $G(z,t)=G(B_{1},B_{2}; z_{0},\omega,\eta;z,t)$
is a solution of the Eq. (\ref{nucleus-gswe}), we find
that the  transformations $R_{1}$, $R_{2}$, $R_{3}$ and
$R_{4}$, given by
%
\begin{eqnarray}\label{transformacaofor-K1}
\begin{array}{ll}
R_{1}G(z,t)=\left( {z}{t}\right)^{1+\frac{B_1}{z_0}}
G(C_{1},C_{2};z_{0},\omega,\eta;z,t),\qquad&
R_{2}G(z,t)=\left[ ({z-z_0})({t-z_0})\right]  ^{1-B_{2}-\frac{B_1}{z_0}}
G(B_{1},D_{2};
z_{0},\omega,\eta;z,t),\vspace{3mm}\\
%
R_{3}G(z,t)=G(B_{1},B_{2}; z_{0},-\omega,-\eta;z,t),&
R_{4}G(z,t)=
G(-B_{1}-B_{2}z_{0},B_{2};z_{0},-\omega,
\eta;z_{0}-z,z_0-t) 
\end{array}
\end{eqnarray}
do not change the form of the kernel equation (\ref{nucleus-gswe}). 
These transformations  
may generate a group containing up to 16 kernels
when applied to an initial kernel. 

For another version of the CHE we have obtained { 
initial kernels as limits of kernels 
for the general Heun equation \cite{lea}. 
For the version (\ref{gswe}), in the following we reobtain  these kernels
by solving Eq. (\ref{nucleus-gswe}) and use the 
transformations (\ref{transformacaofor-K1})
to generate groups of kernels closed under such transformations}.

\subsection*{2.2. First group of kernels: products of two confluent hypergeometric functions}

{Kernels  with products  
of two confluent hypergeometric functions 
have already appeared in a paper  
\cite{masuda} which considers a particular problem obeying a CHE. Here we are
concerned with the general case. }

 In the first place we show that {the kernel equation  
$[L_z-L_t]G(z,t)=0$, given in (\ref{nucleus-gswe})}, is satisfied by 16 of such products,
denoted by $G_{\ 1}^{(i,j)}$ and defined as 
($i,j=1,2,3,4$)
%
\begin{eqnarray}\label{quarto-grupo}
G_{\ 1}^{(i,j)}(z,t)=
e^{-i\omega(z+t)}
\varphi^{i}(\xi)\times\bar{\varphi}^{j}(\zeta),
\end{eqnarray}
where $\varphi^{i}(\xi)$ and $\bar{\varphi}^{j}(\zeta)$ are
the confluent hypergeometric functions
(\ref{confluent1}), 
having the following arguments and parameters:
\letra
\begin{eqnarray}\label{quarto-grupo-b}%
\begin{array}{l} 
\varphi^{i}(\xi):\quad \xi=-\frac{2i\omega }{z_0}(z-z_0)(t-z_0),\qquad
{a}=\frac{B_2}{2}-i\eta-\lambda,\qquad
{c}=B_2+\frac{B_1}{z_0};
\end{array}
\end{eqnarray}
\begin{eqnarray}
\begin{array}{l}
\bar{\varphi}^{j}(\zeta):\quad \zeta=\frac{2i\omega }{z_0}zt, \quad
a= \lambda,\quad
c=-\frac{B_1}{z_0},
    \end{array}
\end{eqnarray}
where $\lambda$ is an arbitrary constant of separation. In 
the second place, by the transformation $R_3$
we may get another set of kernels, $G_{\ 2}^{(i,j)}$, given by
\antiletra
\begin{eqnarray}
G_{\ 2}^{(i,j)}(z,t)=R_3G_{\ 1}^{(i,j)}(z,t)=
G_{\ 1}^{(i,j)}(z,t)\big|_{(\eta,\omega)\mapsto (-\eta,-\omega)}, 
\end{eqnarray}
The transformations $R_1$, $R_2$ and 
$R_4$ are superfluous in this case.

To obtain the kernels (\ref{quarto-grupo}),   
first we write
\begin{eqnarray}\label{nulceus-4}
G(z,t)=e^{-i\omega(z+t)}f(z,t),
\end{eqnarray}
in Eq. (\ref{nucleus-gswe}). This leads to
%
%
%
\begin{eqnarray}\label{N}
&&
z(z-z_{0})\frac{\partial^{2}f}{\partial z^{2}}+\Big[B_{1}+
(B_{2}+2i\omega z_{0})z-
2i\omega z^2\Big]\frac{\partial f}{\partial z}
\nonumber\\
&&
-t(t-z_{0})\frac{\partial^{2}f}{\partial t^{2}}-\Big[B_{1}+
(B_{2}+2i\omega z_{0})t-
2i\omega t^2\Big]\frac{\partial f}{\partial t}
-2i\omega \left(\frac{B_{2}}{2}-i\eta\right)(z-t)f=0.\quad
\end{eqnarray}
Then, by the substitutions
\begin{equation} \label{substituicoes1}
\begin{array}{l}
\xi=-\frac{2i\omega (z-z_0)(t-z_0)}{z_0},\qquad
 \zeta=\frac{2i\omega zt}{z_0},\quad f=X(\xi)Y(\zeta)
 \end{array}
 \end{equation}
we find the confluent hypergeometric equations
\begin{equation}\label{nulceus-4-separadas}
\begin{array}{l}
\xi\frac{d^2X}{d\xi^2}+
\left[ B_2+\frac{B_1}{z_0}-\xi\right] \frac{dX}{d\xi}-
\left[\frac{B_2}{2}-i\eta -\lambda \right] X=0,\qquad
%
%
\zeta\frac{d^2Y}{d\zeta^2}+
\left[ -\frac{B_1}{z_0}-\zeta\right] \frac{dY}{d\zeta}
-\lambda Y=0,
\end{array}
\end{equation}
where $\lambda$ is the constant of separation. The
solutions for the above equations are: $X(\xi)=\varphi^{i}(\xi)$ 
with $a=(B_2/2)-i\eta-\lambda$ and $c=B_2+(B_1/z_0)$;
and $Y(\zeta)=\bar{\varphi}^{j}(\zeta)$  with $a=\lambda$ 
and $c=-B_1/z_0$. Inserting these solutions into
(\ref{nulceus-4}) and  (\ref{substituicoes1})
we find the kernels (\ref{quarto-grupo}). 
Thence, the kernels given by regular confluent hypergeometric functions are
%
\begin{equation}
\begin{array}{l}
G_{\ 1}^{(1,1)}(z,t)=
e^{-i\omega(z+t)}
\Phi\left[\frac{B_2}{2}-i\eta-\lambda,B_2+\frac{B_1}{z_0};
-\frac{2i\omega }{z_0}(z-z_0)(t-z_0)\right] 
\Phi\left[\lambda,-\frac{B_1}{z_0};
\frac{2i\omega }{z_0}zt\right],
\end{array}
\end{equation}
\begin{eqnarray}
G_{\ 1}^{(1,2)}(z,t)=
e^{-i\omega(z+t)+\frac{2i\omega}{z_0}zt}
[zt]^{1+\frac{B_1}{z_0}} 
%
\Phi\left[\frac{B_2}{2}-i\eta-\lambda,B_2+\frac{B_1}{z_0};
-\frac{2i\omega }{z_0}(z-z_0)(t-z_0)\right]
\Phi\left[1-\lambda,2+\frac{B_1}{z_0};
{-}\frac{2i\omega }{z_0}zt\right],\quad
\end{eqnarray}
\begin{eqnarray}
G_{\ 1}^{(2,1)}(z,t)&=&
e^{- i\omega(z+t)-\frac{2i\omega}{z_0}(z-z_0)(t-z_0)}
\left[ (z-z_0)(t-z_0)\right]^{1-B_2-\frac{B_1}{z_0}}
\nonumber\\
&\times&
\Phi\left[1+i\eta+\lambda-\frac{B_2}{2},2-B_2-\frac{B_1}{z_0};
\frac{2i\omega }{z_0}(z-z_0)(t-z_0)\right]
\Phi\left[\lambda,-\frac{B_1}{z_0};
\frac{2i\omega }{z_0}zt\right],
\end{eqnarray}
\begin{eqnarray}
G_{\ 1}^{(2,2)}(z,t)&=&
e^{i\omega(z+t)}
[zt]^{1+\frac{B_1}{z_0}}
\left[ (z-z_0)(t-z_0)\right]^{1-B_2-\frac{B_1}{z_0}}
\nonumber\\
&\times&
\begin{array}{l}
\Phi\left[1+i\eta+\lambda-\frac{B_2}{2},2-B_2-\frac{B_1}{z_0};
\frac{2i\omega }{z_0}(z-z_0)(t-z_0)\right]
\Phi\left[1-\lambda,2+\frac{B_1}{z_0};{-}
\frac{2i\omega }{z_0}zt\right].\quad\end{array}
\end{eqnarray}
The remaining kernels are obtained by replacing
one or both functions $\Phi$ by $\Psi$. 
In this manner we obtain the 16 kernels.
The transformations $R_1$, $R_2$ and $R_4$
are superfluous because they simply rearrange these kernels. For instance,
\begin{eqnarray*}
R_1G_{\ 1}^{(1,1)}&=&\begin{array}{l}
e^{-i\omega(z+t)+\frac{2i\omega}{z_0}zt}[zt]^{1+\frac{B_1}{z_0}}
\Phi\left[2+\frac{B_1}{z_0}-\lambda_1,2+\frac{B_1}{z_0};
-\frac{2i\omega }{z_0}zt\right]\end{array}\\
&\times&\begin{array}{l}
\Phi\left[\frac{B_1}{z_0}+\frac{B_2}{2}+1-i\eta-\lambda_1,B_2+\frac{B_1}{z_0};
-\frac{2i\omega }{z_0}(z-z_0)(t-z_0)\right],
\end{array}
\end{eqnarray*}
where we have transformed $\lambda$ into $\lambda_1$. 
By putting $\lambda_1=\lambda+1+(B_1/z_0)$,
we see that $R_1G_{\ 1}^{(1,1)}=G_{\ 1}^{(1,2)}$. 
%
%

%

\subsection*{2.3. Second group: product of hypergeometric and confluent hypergeometric
functions}

Now we find a group of kernels $G^{(i,j)}$ given by
products of the {confluent hypergeometric functions 
$\varphi^{i}$  written in (\ref{confluent1}) 
with the Gauss hypergeometric
functions} $F^j$ written in (\ref{hiper-1}), (\ref{hiper-2})
and (\ref{hiper-3}). These kernels take the form
\begin{eqnarray}\label{quinto}
G^{(i,j)}(z,t)=
e^{-i\omega(z+t)}\left(z+t-z_0\right)^{-\lambda}
\varphi^{i}(\xi)\times  F^{j}(\zeta),\qquad
[i=1,\cdots,4;\qquad
j=1,\cdots,6]
\end{eqnarray}
where $\lambda$ is a constant of separation, whereas the 
arguments and parameters for the hypergeometric functions are
\letra
\begin{eqnarray}\label{quinto-grupo-b}
\varphi^{i}(\xi):\; \xi=2i\omega(z+t-z_0), \qquad
{a}=\frac{B_2}{2}-i\eta-\lambda,\qquad 
{c}=B_2-2\lambda;
\end{eqnarray}
 \begin{eqnarray}\label{quinto-grupo-b2}
F^{j}(\zeta):\; \zeta=\frac{zt}{z_0(z+t-z_0)},\qquad 
a= \lambda,&b=B_2-1-\lambda, \qquad
c=-\frac{B_1}{z_0}.
\end{eqnarray}
%
%
%
We can show that the transformations $R_i$
do not generate new kernels. 

The above kernels are constructed 
by inserting 
\antiletra
\begin{eqnarray}
G(z,t)=e^{-i\omega(z+t)}f(z,t)=e^{-i\omega(z+t)}g(\xi,\zeta)
\end{eqnarray}
into $[L_z-L_t]G(z,t)=0$, where $\xi$ and $\zeta$ are defined in 
Eqs. (\ref{quinto-grupo-b}) and (\ref{quinto-grupo-b2}) . Thus we find
\begin{eqnarray*}
\begin{array}{l}
\xi\left[ \xi\frac{\partial^2g}{\partial\xi^2}+
\left(B_2-\xi\right) \frac{\partial g}{\partial\xi}-
\left(\frac{B_2}{2}-i\eta \right) g\right] +
\zeta(1-\zeta)\frac{\partial^2g}{\partial\zeta^2}+
\left( -\frac{B_1}{z_0}-B_2\zeta\right)\frac{\partial g}{\partial\zeta}=0.
\end{array}
\end{eqnarray*}
The separation of variables $g(\xi,\zeta)=X(\xi)Y(\zeta)$ leads to
\begin{eqnarray}\label{X-Y}
\begin{array}{l}
\xi\frac{d^2X}{d\xi^2}+
\left[ B_2-\xi\right] \frac{dX}{d\xi}-
\left[\frac{B_2}{2}-i\eta -\frac{\bar{\lambda}}{\xi} \right] X=0,
\qquad 
\zeta(1-\zeta)\frac{d^2Y}{d\zeta^2}-
\left[ \frac{B_1}{z_0}+B_2\zeta\right] \frac{dY}{d\zeta}
-\bar{\lambda} Y=0,
\end{array}
\end{eqnarray}
where $\bar{\lambda}$ is a constant of separation. Putting $\bar{\lambda}=\lambda(B_2-1-\lambda)$,
we find that $Y(\zeta)$ is given by hypergeometric
functions $Y(\zeta)=F^{j}(\zeta)$ as in Eqs. (\ref{quinto})
and (\ref{quinto-grupo-b2}), while $X(\xi)$
obeys the equation
\begin{eqnarray*}
\begin{array}{l}
\xi\frac{d^2X}{d\xi^2}+
\left[ B_2-\xi\right] \frac{dX}{d\xi}-
\left[\frac{B_2}{2}-i\eta -\frac{\lambda(B_2-1-\lambda)}{\xi} \right] X=0.
\end{array}
\end{eqnarray*}
The substitution $X(\xi)=\xi^{-\lambda}\bar{X}(x)$
gives the confluent hypergeometric equation
\begin{eqnarray}\label{Ver-eta-zero}
\begin{array}{l}
\xi\frac{d^2\bar{X}}{d\xi^2}+
\big[ B_2-2\lambda-\xi\big] \frac{d\bar{X}}{d\xi}-
\left[\frac{B_2}{2}-i\eta -\lambda \right] \bar{X}=0,
\end{array}
\end{eqnarray}
whose solutions are $\bar{X}(\xi)=\varphi^{i}(\xi)$.
In this manner, by inserting the previous
solutions for $\bar{X}(\xi)$ and
$Y(\zeta)$ into
\begin{eqnarray}\label{stack}
G(z,t)=e^{-i\omega(z+t)}(z+t-z_0)^{-\lambda}\ \bar{X}(\xi)
\ Y(\zeta)
\end{eqnarray}
we obtain kernels having the form (\ref{quinto}). 

%
%
The kernels $G^{(1,j)}$ and  $G^{(2,j)}$ in terms of regular confluent
hypergeometric functions are
\begin{eqnarray}\label{quinto-grupo}
\begin{array}{l}
G^{(1,j)}(z,t)
=
e^{-i\omega(z+t)}\left[z+t-z_0\right]^{-\lambda}F^{j}(\zeta)\;
%
\Phi\left[\frac{B_2}{2}-i\eta-\lambda,B_2-2\lambda;
2i\omega (z+t-z_0)\right], 
\end{array}
\end{eqnarray}
\begin{eqnarray}\label{quinto-grupo-bb}
G^{(2,j)}(z,t)=
e^{i\omega(z+t)}\left[z+t-z_0\right]^{1-B_2+\lambda}
F^{j}(\zeta)\;
\Phi\left[1+i\eta+\lambda-\frac{B_2}{2},2+
2\lambda-B_2;
-2i\omega (z+t-z_0)\right],
\end{eqnarray}
whereas $G^{(3,j)}$ and  $G^{(4,j)}$
in terms of irregular functions are obtained
by substituting $\Psi(a,c;u)$ for $\Phi(a,c;u)$,
that is,
\begin{eqnarray}
G^{(3,j)}(z,t)= G^{(1,j)}(z,t)\big|_{\Phi\mapsto\Psi},
\qquad
G^{(4,j)}(z,t)= G^{(2,j)}(z,t)\big|_{\Phi\mapsto\Psi}.
\end{eqnarray}
The functions $F^{j}(\zeta)$ are given by
%
\begin{eqnarray}\label{1-hipergeometricas}
\begin{array}{l}
F^{1}(\zeta)=F\left[\lambda,B_2-1-\lambda;-\frac{B_1}{z_0};
\frac{zt}{z_0(z+t-z_0)} \right], 
\end{array}
\end{eqnarray}
\begin{eqnarray}
\begin{array}{l}
F^{2}(\zeta)=
\left[ \frac{zt}{z_0(z+t-z_0)}\right]^{1+\frac{B_1}{z_0}}
%
F\left[\lambda+1+\frac{B_1}{z_0},B_2+\frac{B_1}{z_0}-\lambda;
2+\frac{B_1}{z_0};
\frac{zt}{z_0(z+t-z_0)} \right], \qquad
\end{array}
\end{eqnarray}
\begin{eqnarray}
\begin{array}{l}
F^{3}(\zeta)=F\left[\lambda,B_2-1-\lambda;
B_2+\frac{B_1}{z_0};
\frac{(z-z_0)(t-z_0)}{z_0(z_0-z-t)} \right], 
\end{array}
\end{eqnarray}
\begin{eqnarray}
F^{4}(\zeta)=
\left[ \frac{(z-z_0)(t-z_0)}{z_0(z+t-z_0)}\right]
^{1-B_2-\frac{B_1}{z_0}}
F\left[-\lambda-\frac{B_1}{z_0},\lambda+1-B_2-\frac{B_1}{z_0};
2-B_2-\frac{B_1}{z_0};
\frac{(z-z_0)(t-z_0)}{z_0(z_0-z-t)} \right], 
\end{eqnarray}
\begin{eqnarray}
\begin{array}{l}
F^{5}(\zeta)=
\left[ \frac{z_0(z+t-z_0)}{zt}\right]^{\lambda}
F\left[\lambda,\lambda+1+\frac{B_1}{z_0};2+2\lambda-B_2;
\frac{z_0(z+t-z_0)}{zt} \right],
\end{array}
\end{eqnarray}
\begin{eqnarray}\label{6-hipergeometricas}
\begin{array}{l}
F^{6}(\zeta)=
\left[ \frac{z_0(z+t-z_0)}{zt}\right]^{B_2-1-\lambda}
F\left[B_2+\frac{B_1}{z_0}-\lambda,B_2-1-\lambda;B_2-2\lambda;
\frac{z_0(z+t-z_0)}{zt} \right].
\end{array}
\end{eqnarray}
By using the explicit form for the kernels and
the fact that the separation constant is arbitrary,
it is possible to show that the transformations
$R_i$ simply rearrange
the previous kernels. For instance, we get
\begin{eqnarray*}\begin{array}{l}
R_3G^{(1,j)}(z,t)=
e^{i\omega(z+t)}\left[z+t-z_0\right]^{-{\lambda_3}}
\Phi\left[\frac{B_2}{2}+i\eta-{\lambda_3},
B_2-2{\lambda_3};
-2i\omega (z+t-z_0)\right]
H^{j}(\zeta),\end{array}
\end{eqnarray*}
where $H^{j}(\zeta)$ is obtained
by substituting ${\lambda_3}$ for
${\lambda}$ in ${F}^{j}(\zeta)$. Thence,
putting ${\lambda_3}=B_2-\lambda-1$
and taking into account that $F(a,b;c;u)=
F(b,a;c;u)$, we find that
$H^{5}(\zeta)={F}^{6}(\zeta)$,
$H^{6}(\zeta)={F}^{5}(\zeta)$
and $H^{j}(\zeta)={F}^{j}(\zeta)$
if $j=1,2,3,4$. For this reason,
$R_3G^{(1,j)}$ is equivalent to
$G^{(2,j)}$.
%
%

%

%
%
\subsection*{2.4. Third group: confluent hypergeometric functions}
An initial set 
has the form
%
\begin{equation}\label{lambe}
G_{\ 1}^{(i)}(z,t)=
e^{-i\omega(z+t)}
\varphi^{i}(\xi),\qquad \left[i=1,2,3,4\right]
\end{equation}
where the $\varphi^{i}(\xi)$ denote the four solutions 
(\ref{confluent1}) 
for the confluent hypergeometric equation
with the following argument and parameters:
\begin{equation}
\begin{array}{l}
\xi=-\frac{2i\omega}{z_0}(z-z_0)(t-z_0),\qquad a=\frac{B_2}{2}-i\eta,
\qquad 
c=B_2+\frac{B_1}{z_0}.\end{array}
\end{equation}

The set (\ref{lambe}) is obtained  by putting $\lambda=0$
and  $Y$ constant in (\ref{nulceus-4-separadas}). 
Besides this, from 
(\ref{lambe}) we form four sets by
using the rules $R_2$ and $R_4$, namely,
%
\begin{eqnarray}
G_{\ 1}^{(i)}(z,t),\qquad
G_{\ 2}^{(i)}(z,t)=R_1G_{\ 1}^{(i)}(z,t),\qquad
%
G_{\ 3}^{(i)}(z,t)=R_4G_{\ 2}^{(i)}(z,t),\qquad
G_{\ 4}^{(i)}(z,t)=R_2G_{\ 3}^{(i)}(z,t).
\label{sequencia}
\end{eqnarray}
The four pairs in terms of regular confluent
hypergeometric functions $\Phi(a,c;u)$ read
%
%
\begin{eqnarray}\label{G1}
\begin{array}{l}
G_{\ 1}^{(1)}(z,t)=e^{-i\omega (z+t)}
\Phi\left[ \frac{B_{2}}{2}-i\eta, B_2+\frac{B_{1}}{z_{0}};
-\frac{2i\omega }{z_{0}}(z-z_0) (t-z_0)\right],\vspace{3mm}\\
{G}_{\ 1}^{(2)}(z,t)=e^{i\omega (z+t)- \frac{2i\omega zt}{z_0}}
[(z-z_0)(t-z_0)]^{1-B_2-\frac{B_1}{z_0}} 
%
\Phi\left[1+i\eta -\frac{B_{2}}{2}, 2-B_2-\frac{B_1}{z_0};
\frac{2i\omega }{z_0}(z-z_0)(t-z_0) \right];\hspace{1.55cm}
\end{array}
\end{eqnarray}
%
%
\begin{eqnarray}\label{G2}
\begin{array}{l}
G_{\ 2}^{(1)}(z,t)=e^{-i\omega (z+t)}
[zt]^{1+\frac{B_1}{z_0}}
\Phi \left[1-i\eta+\frac{B_1}{z_0}
+\frac{B_2}{2},B_2+\frac{B_1}{z_0};
-\frac{2i\omega }{z_{0}}(z-z_0) (t-z_0)\right],\vspace{3mm}\\
G_{\ 2}^{(2)} (z,t)=
e^{i\omega (z+t)- \frac{2i\omega zt}{z_0}}
[zt]^{1+\frac{B_1}{z_0}}
\left[(z-z_0)(t-z_0) \right]^{1-B_2-\frac{B_1}{z_0}}
%
\Phi\left[ i\eta-\frac{B_1}{z_0} -\frac{B_{2}}{2},
2-B_2-\frac{B_1}{z_0};
\frac{2i\omega }{z_0} (z-z_0)(t-z_0)\right];
\end{array}
\end{eqnarray}
%
%
%
\begin{eqnarray}\label{G3}
\begin{array}{l}
G_{\ 3}^{(1)}(z,t)=
e^{-i\omega (z+t)}[(z-z_0)(t-z_0)]^{1-B_2-\frac{B_1}{z_0}}
\Phi\left( 1-i\eta-\frac{B_1}{z_0}-\frac{B_{2}}{2}, -\frac{B_{1}}{z_{0}};
\frac{2i\omega zt}{z_{0}} \right),\ \ \vspace{3mm}\\
{G}_{\ 3}^{(2)} (z,t)=
e^{-i\omega (z+t)+ \frac{2i\omega zt}{z_0}}
(zt)^{1+\frac{B_1}{z_0}}
[(z-z_0)(t-z_0)]^{1-B_2-\frac{B_1}{z_0}} 
%
\Phi\left(i\eta+\frac{B_1}{z_0} +\frac{B_{2}}{2}, 2+\frac{B_{1}}{z_{0}};
-\frac{2i\omega zt}{z_{0}} \right);
\end{array}
\end{eqnarray}
%
%
\begin{eqnarray}\label{G4}
\begin{array}{l}
G_{\ 4}^{(1)}(z,t) =e^{-i\omega (z+t)}
\Phi\left( \frac{B_{2}}{2}-i\eta, -\frac{B_{1}}{z_{0}};
\frac{2i\omega zt}{z_{0}} \right),\quad\vspace{3mm}\\
{G}_{\ 4}^{(2)}(z,t)=e^{-i\omega (z+t)+ \frac{2i\omega zt}{z_0}}
(zt)^{1+\frac{B_1}{z_0}}
\Phi
\left(1+i\eta -\frac{B_{2}}{2}, 2+\frac{B_{1}}{z_{0}};
-\frac{2i\omega zt}{z_{0}} \right).
\end{array}
\end{eqnarray}

We get the pairs in terms of irregular confluent
hypergeometric functions by replacing $\Phi(a,c;u)$ by $\Psi(a,c;u)$:
\begin{eqnarray}\label{pares-irregulares}
G_{\ i}^{(3)}(z,t) =G_{\ i}^{(1)}(z,t)\Big|_{\Phi\to\Psi},\qquad
G_{\ i}^{(4)}(z,t) =G_{\ i}^{(2)}(z,t)\Big|_{\Phi\to\Psi},\qquad
[i=1,2,3,4].
\end{eqnarray}
Thus, by using also $R_3$, we find that this group is constituted by
32 kernels. Notice that $R_1$, $R_2$ and
$R_4$ generate only four sets of kernels
instead of 16 sets because in some cases these
transformations rearrange the kernels of a given set
in a different order: we can test this by computing,
for example, $R_2G_{\ 1}^{(i)}$ or $R_2G_{\ 2}^{(i)}$.
Notice that kernels whose arguments of the hypergeometric functions
are $\pm(2i\omega zt/z_0)$ have been known since long 
\cite{lambe}.

{For the spheroidal equation 
($\eta=0,\ z_0=1,\ B_2=-2B_1$), sixteen of the
previous kernels reduce to four kernels
in terms of elementary functions, namely, 
\begin{eqnarray}\label{kernel-esferoidal}
G_{\ 1}^{(\pm)}(z,t)=e^{\pm i\omega(z+t)\mp 2i\omega zt},
\qquad
G_{\ 2}^{(\pm)}(z,t)=e^{\pm i\omega(z+t)\mp 2i\omega zt}
[zt(z-1)(t-1)]^{1+B_1}.
\end{eqnarray}
For instance,
\begin{eqnarray*}
\begin{array}{ll}
G_{\ 1}^{(1)}(z,t)\propto G_{\ 1}^{(4)}(z,t)\propto  G_{\ 1}^{(+)}(z,t),
\qquad&
G_{\ 4}^{(1)}(z,t)\propto G_{\ 4}^{(4)}(z,t)\propto  G_{\ 1}^{(-)}(z,t),\vspace{3mm}\\
G_{\ 2}^{(2)}(z,t)\propto G_{\ 2}^{(4)}(z,t)\propto  G_{\ 2}^{(+)}(z,t),
&
G_{\ 3}^{(2)}(z,t)\propto G_{\ 3}^{(4)}(z,t)\propto  G_{\ 2}^{(-)}(z,t).
\end{array}
\end{eqnarray*}
The kernel $G_{\ 2}^{(-)}(z,t)$ will be used in section 4.1.}

%
\subsection*{2.5. Fourth group: confluent hypergeometric functions again}

To obtain new kernels given by confluent hypergeometric functions
we take
\begin{eqnarray*}
\begin{array}{l}
G_{\ 1}^{(i)}(z,t)=G^{(i,1)}(z,t)\big|_{\lambda=0},
%
\end{array}
\end{eqnarray*}
where the $G^{(i,1)}$ denote the kernels (\ref{quinto})
with $j=1$. Since the above choice for $\lambda$ eliminates the 
Gauss hypergeometric function [$F(0,b;c;\zeta)=1$] we find
\letra
\begin{eqnarray}\label{inicial}
\displaystyle
G_{\ 1}^{(i)}(z,t)=
e^{-i\omega(z+t)}
\varphi^{i}(\xi),\qquad [i=1,2,3,4]
\end{eqnarray}
where $\varphi^{i}(\xi)$ denote the solutions 
(\ref{confluent1}) for the confluent hypergeometric equation with
\begin{eqnarray}
\xi=2i\omega(z+t-z_0),\qquad a=\frac{B_2}{2}-i\eta,\qquad
c=B_2\qquad[\text{see Eq. (\ref{quinto-grupo-b})}].
\end{eqnarray}
Other choices for $\lambda$ 
also lead to kernels in terms of confluent hypergeometric
functions. However, such kernels are obtained from the initial 
set (\ref{inicial}) by using the transformations $R_i$.
In this manner we find four sets, namely, 
\antiletra
%
%
\begin{equation}
\begin{array}{l}
{G}_{\ 1}^{(i)}(z,t),\quad
{G}_{\ 2}^{(i)}(z,t)=R_1{G}_{\ 1}^{(i)}(z,t),
\quad
%
{G}_{\ 3}^{(i)}(z,t)=R_2{G}_{\ 2}^{(i)}(z,t),\quad
{G}_{\ 4}^{(i)}(z,t)=R_1{G}_{\ 3}^{(i)}(z,t),
\end{array}
\label{sequencia2}
\end{equation}
since $R_4$ does not generate new kernels. 
The kernels given by regular
confluent hypergeometric functions are
%
%
\begin{eqnarray}\label{GG1}
\begin{array}{l}
\begin{array}{l}
{G}_{\ 1}^{(1)}(z,t)
=e^{-i\omega (z+t)}
\Phi\left[ \frac{B_{2}}{2}-i\eta, B_2;
2i\omega(z+t-z_0) \right],\end{array}\vspace{2mm}\\
\begin{array}{l}
G_{\ 1}^{(2)}(z,t) =
e^{i\omega (z+t)}[z+t-z_0]^{1-B_2}
\Phi\left[1+i\eta -\frac{B_{2}}{2}, 2-B_2;
-2i\omega(z+t-z_0) \right];\hspace{.7cm}
\end{array}\end{array}
\end{eqnarray}
%
%
%
%
\begin{eqnarray}\label{GG2}
 \begin{array}{l}
 \begin{array}{l}
G_{\ 2}^{(1)}(z,t) =
e^{-i\omega (z+t)}[zt]^{1+\frac{B_1}{z_0}}
\Phi\left[1-i\eta+\frac{B_1}{z_0}+ \frac{B_{2}}{2},
2+ B_2+\frac{2B_1}{z_0};
2i\omega(z+t-z_0) \right],\end{array} 
\vspace{3mm}\\
\begin{array}{l}
G_{\ 2}^{(2)}(z,t) =
e^{i\omega (z+t)}[zt]^{1+\frac{B_1}{z_0}}
[z+t-z_0]^{-1-B_2-\frac{2B_1}{z_0}} 
%
\Phi\left[i\eta -\frac{B_1}{z_0}-\frac{B_{2}}{2},-B_2-\frac{2B_1}{z_0} ;
-2i\omega(z+t-z_0) \right];\end{array}
\end{array}
\end{eqnarray}
%
%
\begin{eqnarray}\label{GG3}
\begin{array}{l}
\begin{array}{l}
G_{\ 3}^{(1)}(z,t) =
e^{-i\omega (z+t)} [zt]^{1+\frac{B_1}{z_0}}
[(z-z_0)(t-z_0)]^{1-B_2-\frac{B_1}{z_0}}\end{array} 
\begin{array}{l}
\Phi\left[2-i\eta- \frac{B_{2}}{2},
4- B_2;
2i\omega(z+t-z_0) \right],\end{array}\vspace{3mm}\\
\begin{array}{l}G_{\ 3}^{(2)}(z,t) =
e^{i\omega (z+t)}[zt]^{1+\frac{B_1}{z_0}}
[(z-z_0)(t-z_0)]^{1-B_2-\frac{B_1}{z_0}}
[z+t-z_0]^{B_2-3} 
\Phi\left[i\eta -1+\frac{B_{2}}{2},B_2-2 ;
-2i\omega(z+t-z_0) \right];
\end{array}\end{array}
\end{eqnarray}
%
%
%
\begin{eqnarray}\label{GG4}
\begin{array}{l}
\begin{array}{l}G_{\ 4}^{(1)}(z,t) =
e^{-i\omega (z+t)}[(z-z_0)(t-z_0)]^{1-B_2-\frac{B_1}{z_0}} 
%
\Phi\left[1-i\eta-\frac{B_1}{z_0}- \frac{B_{2}}{2},
2- B_2-\frac{2B_1}{z_0};
2i\omega(z+t-z_0) \right],\end{array}\vspace{3mm}\\
\begin{array}{l}G_{\ 4}^{(2)}(z,t) =
e^{i\omega (z+t)}[(z-z_0)(t-z_0)]^{1-B_2-\frac{B_1}{z_0}}
[z+t-z_0]^{B_2+\frac{2B_1}{z_0}-1} 
%
\Phi\left[i\eta+\frac{B_1}{z_0}+
\frac{B_{2}}{2},B_2+\frac{2B_1}{z_0} ;
-2i\omega(z+t-z_0) \right].
\end{array}\end{array}
\end{eqnarray}
Replacing $\Phi(a,c;u)$ by $\Psi(a,c;u)$ and using
the transformation $R_3$, once more we
get a group with 32 kernels. Some particular 
cases of these kernels 
are already known \cite{ronveaux}. Furthermore,
if $\eta=0$ this group can be expressed in terms of
Bessel functions by means of (\ref{definition-J}).
%

%
%
%
%
%
%
%
%
%
%
\subsection*{2.6. Fifth group: hypergeometric functions}
To get kernels given by hypergeometric functions
we take $G_{\ 1}^{(i)}(z,t)=G^{(1,i)}(z,t)|_{\lambda=({B_2}/{2})-i\eta}$, 
where $G^{(1,i)}$  are the kernels given in  
(\ref{quinto-grupo}). In fact, for this choice for $\lambda$  
we obtain $\Phi(0,c;\xi)=1$ and, thence,
%
%
\begin{eqnarray}\label{5-a}
\begin{array}{l}
G_{\ 1}^{(i)}(z,t)=
e^{-i\omega(z+t)}[z+t-z_0]^{i\eta-\frac{B_2}{2}}
F^{i}(\zeta),\qquad
%
\zeta=zt/[z_0(z+t-z_0)],
\end{array}
\end{eqnarray}
where the hypergeometric functions $F^{i}(\zeta)$ are obtained by
putting $\lambda=(B_2/2)-i\eta$ in Eqs. (\ref{1-hipergeometricas}-\ref{6-hipergeometricas}). 
Explicitly 
\begin{eqnarray}\label{G3-1}
G_{\ 1}^{(1)}(z,t)=
e^{-i\omega(z+t)}\left[z+t-z_0\right]^{i\eta-\frac{B_2}{2}}
\begin{array}{l}
F\left[\frac{B_2}{2}-i\eta,
\frac{B_2}{2}+i\eta-1;-\frac{B_1}{z_0};
\frac{zt}{z_0(z+t-z_0)}\right],
\end{array}
\end{eqnarray}
\begin{eqnarray}\label{G3-2}
{G}_{\ 1}^{(2)}(z,t)=
e^{-i\omega(z+t)}
\left[z+t-z_0\right]^{i\eta-1-\frac{B_1}{z_0}-\frac{B_2}{2}}
\left[zt\right] ^{1+\frac{B_1}{z_0}} 
%
\begin{array}{l}
F\left[1-i\eta+\frac{B_1}{z_0}+\frac{B_2}{2},
i\eta+\frac{B_1}{z_0}+\frac{B_2}{2};2+\frac{B_1}{z_0};
\frac{zt}{z_0(z+t-z_0)}\right],
\end{array}
\end{eqnarray}
\begin{eqnarray}\label{G3-3}
\begin{array}{l}G_{\ 1}^{(3)}(x,t)=
e^{-i\omega(z+t)}
\left[z+t-z_0\right]^{i\eta-\frac{B_2}{2}}
F\left[\frac{B_2}{2}-i\eta,
\frac{B_2}{2}+i\eta-1;B_2+\frac{B_1}{z_0};
\frac{(z-z_0)(t-z_0)}{z_0(z_0-z-t)}\right],\end{array}
\end{eqnarray}
\begin{eqnarray}\label{G3-4}
G_{\ 1}^{(4)}(x,t)&=&
e^{-i\omega(z+t)}
\left[z+t-z_0\right]^{i\eta-1+\frac{B_1}{z_0}+\frac{B_2}{2}}
\left[(z-z_0)(t-z_0)\right]^{1-B_2-\frac{B_1}{z_0}} 
\nonumber\\
&\times&
\begin{array}{l}
F\left[i\eta-\frac{B_1}{z_0}-\frac{B_2}{2},
1-i\eta-\frac{B_1}{z_0}-\frac{B_2}{2};2-B_2-\frac{B_1}{z_0};
\frac{(z-z_0)(t-z_0)}{z_0(z_0-z-t)}\right],\hspace{2.0cm}
\end{array}
\end{eqnarray}
\begin{eqnarray}\label{G3-5}
\begin{array}{l}G_{\ 1}^{(5)}(x,t)=
e^{-i\omega(z+t)}
\left[zt\right] ^{i\eta-\frac{B_2}{2}}
F\left[\frac{B_2}{2}-i\eta,
1-i\eta+\frac{B_1}{z_0}+\frac{B_2}{2};2-2i\eta;
\frac{z_0(z+t-z_0)}
{zt}\right],
\end{array}
\end{eqnarray}
\begin{eqnarray}\label{G3-6}
\begin{array}{l}G_{\ 1}^{(6)}(x,t)=
e^{-i\omega(z+t)}
\left[z+t-z_0\right]^{2i\eta-1}[zt]^{1-i\eta-\frac{B_2}{2}}
\end{array}
%
\begin{array}{l}
F\left[i\eta+\frac{B_1}{z_0}+\frac{B_2}{2},i\eta-1+\frac{B_2}{2}
;2i\eta;
\frac{z_0(z+t-z_0)}
{zt}\right].
\end{array}
\end{eqnarray}
%

The transformations $R_1$, $R_2$ and $R_4$ 
 at most rearrange the preceding kernels. For example,
\begin{eqnarray*}
R_1G_{\ 1}^{(1)}(z,t)=G_{\ 1}^{(2)}(z,t),\qquad 
R_2G_{\ 1}^{(1)}(z,t)=G_{\ 1}^{(1)}(z,t),\qquad
R_4G_{\ 1}^{(1)}(z,t)=G_{\ 1}^{(3)}(z,t).
\end{eqnarray*}
However, we find six additional kernels $G_{\ 2}^{(j)}(z,t)$ by
using the transformations $R_3$ as
\begin{eqnarray}
G_{\ 2}^{(i)}(z,t)=R_3G_{\ 1}^{(i)}(z,t)\ 
\end{eqnarray}
So, $G_{\ 2}^{(i)}$ is obtained by replacing 
$(\eta,\omega)$ by $(-\eta,-\omega)$ in $G_{\ 1}^{(i)}$.
%


%

%

%
%

\section*{3. Integral relations between known solutions}

In this section we use some kernels
to obtain integral relations among solutions of the CHE. 
We find that: 
%
 %
\begin{itemize}
\itemsep-3pt
\item the Jaff\'e solutions in power series are tranformed into Leaver's
expansions in series of irregular confluent hypergeometric series;
\item  the Baber-Hass\'e solutions 
in power series are transformed into
solutions given by series of regular confluent hypergeometric
functions.
\end{itemize}
%
Relations for solutions generated by transformations
of the CHEs may be obtained by transforming also the 
kernel, since each transformation
of a solution corresponds to a transformation of a kernel.

\subsection*{3.1. Jaff\'e's solutions in power series and Leaver's solutions }
By $U_{ 1}^{J}(z)$  and  $U_{ 1}^{L}(z)$ we denote respectively the Jaff\' e  \cite{jaffe} and
the Leaver \cite{leaver} solutions for the CHE, namely,
\letra
\begin{eqnarray}\label{jaffe-1}
&&U_{ 1}^{J}(z)=e^{i\omega z}z^{-i\eta-\frac{B_{2}}{2}}\sum_{n=0}^{\infty}a_{n}^{1}
\begin{array}{l}\left(\frac{z-z_{0}}{z}
\right)^{n}\end{array},\hspace{5cm}
\end{eqnarray}
\begin{eqnarray}\label{leaver-1}
&&U_{ 1}^{L}(z)=e^{i\omega z}\sum_{n=0}^{\infty}a_{n}^{1}\begin{array}{l}
\Gamma\left(n+B_{2}+\frac{B_{1}}{z_{0}}\right)
%
\Psi\left(n+i\eta+\frac{B_{2}}{2},-\frac{B_{1}}{z_{0}};-2i\omega z\right),\end{array}\quad
\end{eqnarray}
where the recurrence relations for the $a_{n}^{1}$
are ($a_{-1}^{1}=0$)
\begin{eqnarray*}
&\begin{array}{l}
(n+1)\left[n+B_{2}+\frac{B_{1}}{z_{0}}\right]a_{n+1}^{1}+
\Big[ -2n\Big(n+B_{2}+\frac{B_{1}}{z_{0}}+i\eta-i\omega z_{0}\Big)+B_{3}+
\left(B_{2}+\frac{B_{1}}{z_{0}}\right)
\left(i\omega z_{0}-i\eta -\frac{B_{2}}{2}\right)\Big]a_{n}^{1}
\end{array}
\nonumber\\
& \begin{array}{l}
+
%
\left[n-1+i\eta+\frac{B_{2}}{2}\right]\left[n+\frac{B_{2}}{2}+
\frac{B_{1}}{z_{0}}+i\eta \right]a_{n-1}^{1}=0.
\end{array}\quad
\end{eqnarray*}

{The convergence of solutions (\ref{jaffe-1}) and (\ref{leaver-1}) is discussed 
in the Leaver paper \cite{leaver},
where it is used the minimal solution for the coefficients 
$a_{n}^{1}$. In fact, three-term recurrence relations as the above ones
admit two independent solutions, say, $f_n$ and $g_n$. If $\lim_{n\to\infty}(f_n/g_n)=0$, 
$f_n$ is called  minimal solution \cite{gautschi,gautschi-2}. 
In addition, it is necessary to suppose that there is an arbitrary
parameter in the CHE. The series converges only for special values of
that parameter, determined from a transcendental (characteristic) equation
which results from the recurrence relations \cite{leaver}.}

By supposing that $U_{ 1}^{J}(z)$ {converges} for
 $|z|\geq|z_0|$ and by using Eq. (\ref{integral}), we find the relation 
\antiletra
\begin{eqnarray}\label{int-above}
\displaystyle 
U_{ 1}^{L}(z)=C_1 \int_{z_0}^{\infty}
t^{-1-\frac{B_1}{z_0}}[t-z_0]^{B_2+\frac{B_1}{z_0}-1}G_{\ 4}^{(4)}(z,t)U_{ 1}^{J}(t)dt,
\qquad
\text{Re}\left[n+B_2+\frac{B_1}{z_0} \right]>0,\qquad
\text{Re}[i\omega z]<0, 
\end{eqnarray}
where $C_1$ is a constant and $G(z,t)=G_{\ 4}^{(4)}(z,t)$ is the kernel indicated 
in (\ref{pares-irregulares}).  In fact,
by setting $y=t/z_0$, we find that the right-hand side of (\ref{int-above})  is 
equivalent to
\begin{eqnarray*}
\displaystyle 
e^{-i\omega z}
z^{1+\frac{B_1}{z_0}}
%
\sum_{n=0}^{\infty}a_{n}^{1}
\int_{z_0}^{\infty}dy
\bigg[ e^ {2i\omega z y}\left( y-1\right) ^{n+B_{2}+\frac{B_{1}}{z_{0}}-1}
 y^{-n-i\eta-\frac{B_2}{2}} 
%
\begin{array}{l}
\Psi\left(1+i\eta- \frac{B_{2}}{2}, 2+\frac{B_{1}}{z_{0}};
-2i\omega z y \right)\bigg]\end{array}.
\end{eqnarray*}
Then, by using the integral \cite{gradshteyn}
\begin{eqnarray}
\int_{1}^{\infty}
e^{-ay}(y-1)^{\mu-1}y^{\alpha+k-\mu-\frac{1}{2}}
\Psi\left(\frac{1}{2}+\alpha-k,2\alpha+1;ay\right)dy
=\Gamma(\mu)e^{-a}
\Psi\left(\frac{1}{2}+\alpha+\mu-k,
2\alpha+1;a\right),
\\
%
\qquad[\text{Re}\ {\mu}>0, \ \text{Re}\ {a}>0],\nonumber
\end{eqnarray}
we obtain the relation (\ref{int-above}).

For the bilinear concomitant (\ref{concomitant}) 
we find
\begin{eqnarray*}
P_1(z,t)&=&z^{1+\frac{B_1}{z_0}}e^{i\omega z(\frac{2t}{z_0}-1)}\
t^{-i\eta-\frac{B_{2}}{2}}\ (t-z_{0})^{B_{2}+\frac{B_{1}}{z_{0}}}
%
\sum_{n=0}^{\infty}a_{n}\left(\frac{t-z_{0}}{t}
\right)^{n}\nonumber\\
&\times &\begin{array}{l}
\bigg\{\bigg[2i\omega\left(\frac{z}{z_0}-1\right)t-\frac{nz_0}{t-z_0}+
 \frac{1}{t}\left(i\eta+1+\frac{B_1}{z_0}+\frac{B_2}{2} \right) \bigg]
\Psi+t\frac{\partial\Psi}{\partial t}\bigg\},\end{array}
\end{eqnarray*}
where
\begin{eqnarray*}\begin{array}{l}
\Psi=
\Psi\left(1+i\eta-\frac{B_2}{2},2+\frac{B_1}{z_0};
-\frac{2i\omega z t}{z_0} \right),\qquad
\frac{\partial\Psi}{\partial t}=\frac{2i\omega z}{z_0}
\left( 1+i\eta-\frac{B_2}{2}\right) \Psi\left(2+i\eta-\frac{B_2}{2},3+\frac{B_1}{z_0};
-\frac{2i\omega z t}{z_0} \right).\end{array}
\end{eqnarray*}
Since $\Psi(a,b;y)= y^{-a}$ when $|y|\rightarrow \infty$
and $\text{Re}(i\omega z)<0$, the exponential factor assures that
$P_1(z,t)$ vanishes when $t/z_0\to \infty$. 
On the other hand, the condition  $\text{Re}[B_2+\frac{B_1}{z_0}]>0$ 
assures that $P_1(z,t)$ vanishes also for $t=z_{0}$ since 
$(t-z_0)^{B_2+B_1/z_0} \to 0$.

In this manner, we have extended the results of 
Leaver \cite{leaver} who has considered only 
relations between solutions with 
$i\eta=\pm(B_2/2-1)$. Notice also that the conditions given in (\ref{int-above})
are necessary only to assure the integral relation between the solutions.
In fact the Leaver solutions can be derived 
directly from the differential
equation without imposing those conditions \cite{leaver}.

{
For the present case the transformation $T_1$ is ineffective and, so,
from $\big(U_1^{{J}},U_1^{{L}}\big)$ we can obtain
only 8 pairs of solutions by composition of the transformations 
(\ref{transformacao1}); to each pair 
corresponds a kernel generated by the 
transformations (\ref{transformacaofor-K1}). For example, taking  
$U_{ 2}^{J}(z)=T_2U_{ 1}^{J}(z)$ and $U_{ 2}^{L}(z)=T_2U_{ 1}^{L}(z)$, we find
\letra
\begin{eqnarray}\label{jaffe-2}
&&U_{ 2}^{J}(z)=
e^{i\omega z}(z-z_0)^{1-B_2-\frac{B_2}{2}}
z^{-i\eta-1+\frac{B_1}{z_0}+\frac{B_{2}}{2}}\sum_{n=0}^{\infty}a_{n}^{2}
\begin{array}{l}\left(\frac{z-z_{0}}{z}
\right)^{n}\end{array},\\
&&U_{ 2}^{L}(z)=e^{i\omega z}(z-z_0)^{1-B_2-\frac{B_2}{2}}
\sum_{n=0}^{\infty}\begin{array}{l}a_{n}^{2}
\Gamma\left(n+2-B_{2}-\frac{B_{1}}{z_{0}}\right)
\Psi\left(n+i\eta+1-\frac{B_{2}}{2}-\frac{B_{1}}{z_{0}},-\frac{B_{1}}{z_{0}};
-2i\omega z\right),\end{array}
\end{eqnarray}
where the recurrence relations for $a_{n}^{2}$
are ($a_{-1}^{2}=0$)
\antiletra
\begin{eqnarray*}
&\begin{array}{l}
(n+1)\left[n+2-B_{2}-\frac{B_{1}}{z_{0}}\right]a_{n+1}^{2}+
\Big[ -2n\Big(n+2+i\eta-i\omega z_{0}-B_{2}-\frac{B_{1}}{z_{0}}\Big)+B_{3}+
\left(2-B_{2}+\frac{B_{1}}{z_{0}}\right)
\left(i\omega z_{0}-i\eta\right) +\frac{B_{1}}{z_0}
\end{array}\\
& \begin{array}{l}
+
\left(1-\frac{B_2}{2}\right)\left(1+\frac{B_1}{z_0}-\frac{B_2}{2}\right)\Big]a_{n}^{2}+
\left[n+1+i\eta-\frac{B_{2}}{2}\right]\left[n+i\eta-\frac{B_{2}}{2}-
\frac{B_{1}}{z_{0}}\right]a_{n-1}^{2}=0.
\end{array}
\end{eqnarray*}
Using Eq. (\ref{integral}), we find that
\antiletra
\begin{eqnarray}
\displaystyle 
U_{ 2}^{L}(z)=C_2 \int_{z_0}^{\infty}dt\;
t^{-1-\frac{B_1}{z_0}}[t-z_0]^{B_2+\frac{B_1}{z_0}-1}U_{ 2}^{J}(t)R_2G_{\ 4}^{(4)}(z,t),\qquad
\text{Re}\left[n+2-B_2-\frac{B_1}{z_0} \right]>0,\quad
\text{Re}[i\omega z]<0,\nonumber
\end{eqnarray}
where $C_2$ is a constant, $G_{\ 4}^{(4)}(z,t)$ is the kernel indicated 
in (\ref{pares-irregulares}), and the transformation $R_2$ is 
given in (\ref{transformacaofor-K1}); then,
\begin{equation*}
 \begin{array}{l}
R_{2}G_{\ 4}^{(4)}=e^{-i\omega (z+t)+\frac{2i\omega zt}{z_0}}
\left[ ({z-z_0})({t-z_0})\right]^{1-B_{2}-\frac{B_1}{z_0}}(zt)^{1+\frac{B_1}{z_0}}
\Psi\left(i\eta+\frac{B_1}{z_0}+\frac{B_2}{2},2+\frac{B_1}{z_0};-\frac{2i\omega z t}{z_0}\right).\end{array}
\end{equation*}
}
{We have supposed that the Jaff\'e solutions converge for
$|z|\geq|z_0|$, but we must be careful about the point $z=\infty$, since \cite{leaver}
\begin{eqnarray}\label{jaffe-razao}
\begin{array}{l}\displaystyle
 \lim_{z\to\infty}
U_{ 1}^{J}(z)=e^{i\omega z}z^{-i\eta-\frac{B_{2}}{2}}\sum_{n=0}^{\infty}a_{n}^{1}\quad
\text{with}\quad 
\lim_{n\rightarrow \infty}\frac{a_{n+1}^{1}}{a_{n}^{1}}=
1-\frac{\sqrt{-2i\omega z_0}}{\sqrt{n}}+\frac{i(\eta-\omega z_0)-(3/4)}{n},
\end{array}
\end{eqnarray}
where the ratio ${a_{n+1}^{1}}/{a_{n}^{1}}$ holds for the minimal solution
of the recurrence relations. Thus, the D'Alambert test is inconclusive
as to the convergence of $\sum a_{n}^{1}$. For the radial part of the two-center
problem we could use the Raabe test for convergence, as in Eq. (\ref{jaffe-razao-radial}).}

\subsection*{3.2. 
Solutions in power series and solutions in series of confluent hypergeometric functions}
We find another pair of solutions for the
CHE which are again connected by the integral 
(\ref{integral}). By one side we have the Baber-Hass\'e 
expansion \cite{lea-2,baber}
\letra
\begin{eqnarray}\label{barber}
U_{1}^{\text{baber}}(z)=e^{i\omega z}\displaystyle \sum_{n=0}^{\infty}a_{n}^{1}
(z-z_{0})^{n},
\end{eqnarray}
where the coefficients satisfy the relations ($a_{-1}^{1}=0$)
\begin{eqnarray}\label{barber-recorrencia}
\begin{array}{l}
 z_{0}\left(n+B_{2}+\frac{B_{1}}{z_{0}}\right)
\left(n+1\right)a_{n+1}^{1}+\beta_n^{1}a_n^{1}+
2i\omega\left(n+i\eta+\frac{B_{2}}{2}-1\right)
a_{n-1}^{1}=0,\end{array}
\end{eqnarray}
with $\beta_{n}^{1}=n(n+B_{2}-1+2i\omega z_{0})+B_{3}+i\omega
z_{0}\left[B_{2}+{B_{1}}/{z_{0}}\right]$. 
%
%
The minimal solutions for $a^1_n$ yield solutions convergent 
for any finite value of $z$. 
On the other side, if $(B_2/2)-i\eta$ is not zero or negative
integer we have the solution \cite{lea-2}
\antiletra\letra
\begin{eqnarray}\label{segunda}
U_{1}(z)=
e^{-i\omega z}\displaystyle\sum_{n=0}^{\infty}
b_{n}^{1}
\ {\Phi}\left(\frac{B_{2}}{2}-i\eta,n+B_{2};2i\omega z\right),\ \
\end{eqnarray}
where the recurrence relations for $b_n^{1}$ are obtained from the previous 
ones by taking 
\begin{eqnarray*}\begin{array}{l}
 b_{n}^{1}=\frac{C(-z_0)^n\Gamma(n+B_2+B_1/z_0)}
{\Gamma(n+B_2)} a_{n}^{1},\qquad C=\text{constant}.\end{array}
\end{eqnarray*}
This yields
\begin{eqnarray}\label{truncadaR2}
\begin{array}{l}
-(n+B_2)(n+1)b_{n+1}^{1}+\beta_{n}^{1}
b_{n}^{1}
-
2i\omega z_{0}\frac{\left(n+B_{2}+\frac{B_{1}}{z_{0}}
-1\right)\left(n+i\eta+\frac{B_{2}}{2}-1\right)}{n+B_2-1}
b_{n-1}^{1}=0.
\end{array}
\end{eqnarray}
Now, if we insert $U_1^{\text{baber}}(t)$ and the kernel $G_{ \ 4}^{(1)}(z,t)$
given in (\ref{G4}) into Eq. (\ref{integral}),
we find the solution $U_{1}(z)$, that is,
\antiletra
\begin{eqnarray}
\displaystyle 
U_{1}=K \int_{t_1}^{t_2}
t^{-1-\frac{B_1}{z_0}}[t-z_0]^{B_2+\frac{B_1}{z_0}{-}1}G_{\ 4}^{(1)}(z,t)U_1^{\text{baber}}(t)dt,
\qquad 
\text{Re}\left[n+B_2+\frac{B_1}{z_0} \right]>0,\quad
\text{Re}[-\frac{B_1}{z_0}]>0, 
\end{eqnarray}
where $K$ is a constant. In effect, by taking $t_1=0$ and $t_2=z_0$, the above integral is
proportional to 
\begin{eqnarray*}
e^{-i\omega z}\sum_{n=0}^{\infty}(-z_0)^na_{n}^{(1)}\int_{0}^{1}
\begin{array}{l}
d\left( \frac{t}{z_0}\right)
\bigg[\left(\frac{t}{z_0} \right)^{-1-\frac{B_1}{z_0}}
\left(1-\frac{t}{z_0} \right)^{n+B_2-1+\frac{B_1}{z_0}}
\Phi\left( \frac{B_2}{2}-i\eta,-\frac{B_1}{z_0};2i\omega z\frac{t}{z_0}\right)
\bigg]\end{array}
\end{eqnarray*}
Then, by using the relation \cite{gradshteyn}
\begin{eqnarray}\label{int-1}
&\int_{0}^{1}\left[x^{\lambda-1}(1-x)^{2\mu-\lambda}
\Phi\left(\frac{1}{2}+\mu-\nu,\lambda; yx\right)\right]dx
=
\frac{\Gamma(\lambda)\Gamma(1+2\mu-\lambda)}
{\Gamma(1+2\mu)}
\Phi\left( \frac{1}{2}+\mu-\nu,1+2\mu;y\right),\\
%
%
&\begin{array}{l}
\left[ \text{Re}(\lambda)>0,\ \ \text{Re}(1+2\mu-\lambda)>0\right] ,
\end{array}\nonumber
\end{eqnarray}
%
%
we find the solution 
$U_1(z)$ given in (\ref{segunda}) provided that $\text{Re}\left[n+B_2+({B_1}/{z_0}) \right]>0$ and 
$\text{Re}[-B_1/z_0]>0$.
%
%
%
%
On the other side,
%
%
%
from $d\Phi(a,b;\xi)/d\xi=(a/b)\Phi(a+1,b+1;\xi)$
for $\xi=2i\omega zt/z_0$, $a=B_2/2-i\eta$ and $b=-B_1/z_0$,
we find that the bilinear concomitant (\ref{concomitant}) is given by
%
%
%
%
\begin{eqnarray*}
P_{1}(z,t)&=&-e^{-i\omega z}t^{-\frac{B_1}{z_0}}
(t-z_0)^{B_2+\frac{B_1}{z_0}} \\
&\times&\bigg\{\Phi(a,b;\xi)\sum_{n=1}^{\infty}na_{n}^{1}
(t-z_{0})^{n-1}
+2i\omega
\Big[\Phi(a,b;\xi)
-\frac{az}{bz_0}\Phi(a+1,b+1;\xi)\Big] \sum_{n=0}^{\infty}a_{n}^{1}
(t-z_{0})^{n}\bigg\}.
\end{eqnarray*}
Therefore, $P_1(z,t=0)=P_1(z,t=z_0)=0$ due to the conditions
$\text{Re}(-B_1/z_0)>0$ and $\text{Re}(B_2+B_1/z_0)>0$.
%

Observe that from the pair $\big(U_1^{\text{baber}},U_1\big)$ we can obtain
16 pairs of solutions by using the four transformations 
(\ref{transformacao1}) and composition of them: to each pair 
corresponds a kernel which is obtained by using the 
transformations (\ref{transformacaofor-K1}).

\section*{4. New solutions for the confluent equation}

In  section 4.1, by an integral transformation we find a new solution 
in series of irregular confluent hypergeometric functions for the ordinary spheroidal equation. 
Then, in section 4.2 we extend that solution
to the general case (no restriction on the parameters of the CHE).
In this manner, we obtain an initial solution, $\mathcal{U}_1(z)$, which
allows to generate a group of solutions $\mathcal{U}_i(z)$ for the CHE by 
by means of transformations (\ref{transformacao1}). Finally, in section 4.3,
we show that the new solutions are suitable
for the radial part of the two-center problem of
the quantum mechanics.

 Initially we make some comments on the recurrence relations and the ratio test for convergence.
 As in the preceding section, the three-term recurrence relations 
 for the series coefficients $b_{n}^i$  of $\mathcal{U}_i(z)$ have the form 
 %
\begin{eqnarray}\label{r1a}
\begin{array}{l}
{\alpha}_{0}^i \  {b}_{1}^i+
{\beta}_{0}^i\  {b}_{0}^i=0,\qquad
{\alpha}_{n}^i\  {b}_{n+1}^i+{\beta}_{n}^i\ 
{b}_{n}^i+
{\gamma}_{n}^i\  {b}_{n-1}^i=0\quad (n\geq1)
\end{array}
\end{eqnarray}
where $\alpha_{n}^i$, $\beta_{n}^i$ and $\gamma_{n}^i$
depend on the parameters of the
differential equation and on the summation index $n$. 
By omitting the superscripts, 
these relations take the form
\begin{eqnarray}\label{matriz-tridiagonal}
\left[
\begin{array}{lcccr|lllr}
\beta_{0} & \alpha_{0} & 0 & &� &� &\\
\gamma_{1}&\beta_{1} & \alpha_{1} & & �&
\\
0�&\gamma_{2} & \beta_{2}&\alpha_{2}& &� \vspace{1mm}\\
& & \ddots �& \ddots &  \ddots &�& \vspace{1mm}\\
&� & �&\gamma_{\text{\tiny N}} & \;\beta_{\text{\tiny N}}&
\alpha_{\text{\tiny N}}\\
\hline
 & & �& & \gamma_{\text{\tiny N}+1}&
\beta_{\text{\tiny N}+1}& \alpha_{\text{\tiny N}+1} \\
 & &� & �& & \ddots &\ddots
& \ddots 
%
\end{array}
\right]
\left[\begin{array}{l}
{b}_{0} �\\
{b}_{1} \\
{b}_{2} \\
\vdots\\
%
%
{b}_{\text{\tiny N}}\\
\hline
{b}_{\text{\tiny N}+1}\\
%
%
\vdots
\end{array}
\right]=
\left[\begin{array}{c}
0 \\
0 \\
0 \\
\vdots\\
0\\
\hline
0\\
%
%
\vdots
\end{array}
\right],
\end{eqnarray}
where we have split the matrix into blocks. 
This system of homogeneous linear equations 
has nontrivial solutions for $b_{n}$ only if the
determinant of the above tridiagonal
matrix vanishes: this demands some arbitrary parameter
in the matrix elements and, as a consequence, in the differential equation.
The condition on the determinant can also be expressed by an 
(characteristic) equation given by the continued
fraction \cite{leaver}
\begin{eqnarray}\label{characteristic1}
\beta_{0}=\frac{\alpha_{0}\gamma_{1}}{\beta_{1}-}\
\frac{\alpha_{1}\gamma_{2}}
{\beta_{2}-}\ \frac{\alpha_{2}\gamma_{3}}{\beta_{3}-}\cdots.
\end{eqnarray}
{The solution of the characteristic equation and the computation of
the series coefficients are important aspects concerning 
applications of  the CHE \cite{hodge,liu}. 
The problem is simplified  
if ${\gamma}_{n=\text{\tiny N}+1}^i=0$ 
for some $\text{\small N}\geq 0$; then, the series
terminates at $n=\text{\small N}$ leading to  
a finite-series solution with $0\leq n\leq N$ 
(see page 146 of \cite{arscott}) which is called polynomial
or quasi-polynomial solution. 
In this case, only the left upper block of the matrix
is relevant.}

On the other side, the convergence of a series like $\sum_{n=0}^{\infty}f_n(z)$
is obtained by computing the limit {of 
%
%
\letra
\begin{equation}\label{ratio-1}
L(z)=\left|\frac{f_{n+1}(z)}{f_{n}(z)}\right|\quad \text{when}\quad n\to\infty.
\end{equation}
By the D'Alembert ratio test the series converges 
in the region where $L(z)<1$ and diverges where $L_1(z)>1$. If
$L(z)=1$, the D'Alembert test is inconclusive; however, by the Raabe test \cite{watson,knopp}, if 
\begin{eqnarray}\label{Raabe-1}
L(z)=1+\frac{A}{n}+O\left(\frac{1}{n^2}\right)
\end{eqnarray}}
(where $A$ is a constant) the series converges if 
$A<-1$ and diverges if $A>-1$; the  
test is inconclusive if  $A=-1$. 
%
%
%

\subsection*{4.1. An integral transformation for the spheroidal equation} 

{For the spheroidal equation in the form (\ref{CHE-esferoidal}) we will find a 
solution $\mathcal{U}_{1}(z)$ given by
\antiletra\letra
\begin{eqnarray}\label{sph-2a}
\begin{array}{l}
\mathcal{U}_{1}(z)=e^{i\omega z}z^{1+B_1}(z-1)^{1+B_1}
\displaystyle\sum_{n=0}^{\infty}{b}_n^{1}\Psi(2+B_1,2+B_1-n; - 2i\omega z)\qquad
 [B_1\neq -2,-3,\cdots]
\end{array}
\end{eqnarray}
where} the coefficients ${b}_{n}^{1}$ satisfy 
the relations (${b}_{-1}^{1}=0$)
\begin{eqnarray}\label{sph-2b}
& -2i\omega(n+1){b}_{n+1}^{1}+
\big[ n\left(n+1+ 2i\omega \right)+ 
i\omega (2+B_1)-B_1\left(1+B_1\right)+B_3\big]{b}_{n}^{1} 
%
-n\left(n+B_1+1\right){b}_{n-1}^{1}=0.
\end{eqnarray}
$\mathcal{U}_{1}(z)$ is not valid if $B_1= -2,-3,\cdots$,  because 
in these cases the function $\Psi(a,c;y)$ becomes
a polynomial of fixed degree and, accordingly, (\ref{sph-2a}) 
is not a series expansion. This follows from the relation \cite{erdelii}
\antiletra
\begin{eqnarray} \label{laguerre-1}
\Psi(-l,\alpha+1;y)=(-1)^l\;l!\ 
L_l^{\alpha}(y),\qquad [l=0,1,\cdots]
\end{eqnarray}
where the $L_l^{(\alpha)}(y)$ denote Laguerre polynomials of degree $l$.  
Besides this, the above expansion in general does not hold
at $z=0$ because in most cases $\Psi(a,c;y)$ 
goes to infinity at $z=0$ \cite{nist}. The convergence of
$\Psi$ for $z\neq 0$ will be discussed later on.
%

We get the expansion (\ref{sph-2a}) by applying 
an integral transformation to 
the asymptotic expansion $\mathcal{W}_2(z)$ 
given in Eq. (\ref{second-wilson}). First, 
for the spheroidal equation, 
by writing $W(z)=\mathcal{W}_2(z)$ and $a_n^2={b}_n^1$, we find
\begin{eqnarray} \label{assintotica-esf}
\begin{array}{l}
W(z) =e^{ i\omega z}\ (z-1)^{1+B_{1}}
\displaystyle \sum_{n=0}^{\infty}{b}_{n}^{1}
\ z^{-n-1},\qquad [ \text{Eq. (\ref{second-wilson}) for the spheroidal equation}]
\end{array}\end{eqnarray}
where the coefficients ${b}_{n}^{1}$ satisfy (\ref{sph-2b}). 
In the second place, the solution $\mathcal{U}_1$ is 
obtained by inserting $U(t)=W(t)$ 
and $G(z,t)=G_2^{-}(z,t)$ -- see 
Eq. (\ref{kernel-esferoidal}) --   
into the right-hand side of Eq. (\ref{integral}), and by 
integrating from $t=1$ to $t=\infty$, that is,
{
\begin{eqnarray*}
\mathcal{U}_{1}(z)\stackrel{\text{(\ref{integral}})}{=}
\int_{1}^{\infty}t^{-1-B_1}
(t-1)^{-1-B_1}G_2^{-}(z,t)\ W(t)dt
\stackrel{\text{(\ref{kernel-esferoidal})}}{=}
e^{-i\omega z}[z(z-1)]^{1+B_1}\int_{1}^{\infty}e^{-i\omega t+2i\omega zt}\ W(t)dt
\end{eqnarray*}
which gives
\begin{eqnarray} 
\displaystyle
\mathcal{U}_{1}(z)
\stackrel{\text{(\ref{assintotica-esf})}}{=}e^{-i\omega z}[z(z-1)]^{1+B_1}\sum_{n=0}^{\infty}{b}_n^1
\int_{1}^{\infty}
e^{2i\omega zt}(t-1)^{1+B_1}t^{-n-1}dt.
\end{eqnarray}
Thence,} we obtain (\ref{sph-2a}) by using \cite{nist}
\begin{eqnarray*}
\label{A1}
\int_{1}^{\infty}e^{-yt}(t-1)^{a-1}t^{c-a-1}dt=
\Gamma(a)e^{-y}\Psi(a,c;y), \qquad [\text{Re}\,{a}>0, \ \ \text{Re}\,{y}>0].
\end{eqnarray*}
The integrability conditions on the right-hand side require
that
\begin{eqnarray}\label{vanish}
\text{Re}[2+B_1]>0 \text{ and } \text{Re}[i\omega z]<0.
\end{eqnarray}
On the other side, the bilinear concomitant (\ref{concomitant}) reads 
\begin{eqnarray}
P(z,t)&=&
\begin{array}{l}t^{-{B_1}}(t-1)^{-B_1}
\left[W(t)\frac{\partial G_2^{-}(z,t)}{\partial t}-G_2^{-}(z,t)
\frac{d W(t)}{d t}\right],\end{array}\nonumber\\
&=&
\begin{array}{l}
e^{i\omega z(2t-1)}[z(z-1)]^{1+B_1}
(t-1)^{2+B_1}
\end{array} 
\begin{array}{l}\displaystyle
\left\lbrace \big[2i\omega(z-1)t+B_1\big]\sum_{n=0}^{\infty}
{b}_n^1 t^{-n-1}+\sum_{n=0}^{\infty}
n{b}_n^1 t^{-n-1}
\right\rbrace
\end{array}
\end{eqnarray}
Since the series converge at $t=\infty$, the conditions (\ref{vanish})
assure that $P(z,t=\infty)=0$. 
However, the concomitant is undetermined  at $t=1$ because [for $\text{Re}(2+B_1)>0$]
$P(z,t)$ is given by the product of the vanishing factor $(t-1)^{2+B_1}$ by a divergent 
series. Despite this, we can check 
directly \cite{arxiv} that $\mathcal{U}_{1}(z)$ 
is indeed a solution of the spheroidal equation (\ref{CHE-esferoidal})
{regardless of the conditions (\ref{vanish}).}

{
Now we use the ratio test to get the convergence of 
$\mathcal{U}_1$.} Thus, when $n\to\infty$,
we find that the minimal solution of (\ref{sph-2b}) satisfies \cite{arxiv}
\begin{eqnarray}\label{minimal-bn}
\begin{array}{l}
  \frac{{b}_{n+1}^{1}}{{b}_{n}^{1}}\sim
1+\frac{B_1}{n}\quad
\Rightarrow\quad
\frac{{b}_{n-1}^{1}}{{b}_{n}^{1}}\sim
1-\frac{B_1}{n}.
\end{array}
\end{eqnarray}
To get the ratio between successive $\Psi$, we use the relation \cite{nist}
\[(a+1-c)\Psi(a,c-1;y)+(c-1+y)
\Psi(a,c;y)-y\Psi(a,c+1;y)=0.\]
Hence, by taking 
\[a= 2+B_1,\quad c=2+B_1-n, \quad y=-2i\omega z, \quad 
\Psi_n(y)= \Psi( 2+B_1, 2+B_1-n;-2i\omega z)\]
we obtain
\begin{eqnarray*} \label{relation-2}
\begin{array}{l}
\left(n+1\right) \frac{\Psi_{n+1}}{\Psi_{n}}
-\left(n-1-B_1+2i\omega z\right)
+2i\omega z
\frac{\Psi_{n-1}}{\Psi_{n}}=0.
\end{array}
\end{eqnarray*}
If $z$ is bounded (that is, if $2i\omega z/n\to 0$), 
then when $n\to\infty$ this equation is satisfied by 
\begin{eqnarray}\label{letra-1}
\begin{array}{lll}
\frac{\Psi_{n+1}}{\Psi_{n}}\sim 1-
\frac{1}{n}\left(B_1+2\right)\quad& \Leftrightarrow
& \quad
\frac{\Psi_{n-1}}{\Psi_{n}}\sim
 1+\frac{1}{n}\left(B_1+2\right) \text{ or }
\end{array}
\end{eqnarray}
\begin{eqnarray*}\label{letra-2}
\begin{array}{lll}\frac{ \Psi_{n+1}}{\Psi_{n}}\sim
 \frac{2i\omega z}{n}\left(1+\frac{B_1}{n}\right)\qquad& \Leftrightarrow &\qquad
 \frac{\Psi_{n-1}}{\Psi_{n}}\sim
  \frac{n}{2i\omega z}\left[1-\frac{1}{n}\left(1+B_1\right)\right].
\end{array}
\end{eqnarray*}
Only the first ratio 
is consistent with the fact that, if $|c|\to\infty$ while $a$
and $y$ remain fixed and bounded,  then \cite{erdelii}
\begin{eqnarray*}\label{criterio}
&\begin{array}{l}
\Psi(a,c;y)=
c^{-a}\left[
(-1)^{-a}+\frac{\sqrt{2\pi}}{\Gamma(a)}
\ \left(\frac{c}{ey} \right) ^{c+a-\frac{3}{2}}y^{a-\frac{1}{2}}\ e^{y+a-\frac{3}{2}}\right]
\left[1+O\left(\frac{1}{|c|}\right)\right],
\end{array}\\
&\left[c\to \infty; \ \ a\neq 0,-1,-2,\cdots;\ \ |\arg(\pm c)|<\pi\right]. \nonumber \ \
\end{eqnarray*}
Thus, using (\ref{minimal-bn}) and (\ref{letra-1}), we find {that
\letra
\begin{eqnarray}\label{limite-1}
\begin{array}{l}
\text{when }n\to
\infty,\quad 
\displaystyle \frac{{b}_{n+1}^1\Psi_{n+1}}
{{b}_n^1\Psi_{n}}= 1-
\frac{2}{n}+O\left(\frac{1}{n^2}\right)\quad \text{in} \quad \mathcal{U}_1.
\end{array}
\end{eqnarray}
Therefore, by the Raabe test the series
may converge for any finite value of $z$ (the ratios
(\ref{letra-1}) are valid if $z$ is finite); however, 
we must exclude the point $z=0$ because
in general the function $\Psi(a,c;y)$ goes to infinity at $y=0$. On the other side, 
{since $\Psi(a,c;y)$$\sim$$y^{-a}$ when $y\to\infty$, we find that for $z\to\infty$
\begin{eqnarray}\label{limite-2}
\mathcal{U}_{1}(z)\sim e^{ i\omega z}\ z^{B_1}
\sum_{n=0}^{\infty}{b}_n^1,\qquad 
\left|
\frac{{b}_{n+1}^{1}}{{b}_{n}^{1}}\right|\stackrel{(\ref{minimal-bn})}{\;=\;}
1+\frac{\text{Re}\,B_1}{n}+O\left(\frac{1}{n^2}\right)\quad \text{when}\quad {n\to\infty}.
\end{eqnarray}}
Thus, according to the 
 Raabe test, the series $\sum {b}_n^1$
converges  only if $\text{Re}(B_1)<-1$, and this condition assures that
$\mathcal{U}_{1}(z)$ converges at $z=\infty$. }



%
\subsection*{4.2. Solutions for the confluent Heun equation}
Now the solution $\mathcal{U}_{ 1}(z)$ for the spheroidal equation,
given in (\ref{sph-2a}), is extended 
for any CHE. In fact, we can construct a group of solutions
 $\mathcal{U}_{ i}(z)$ whose series coefficient $b_n^i$ satisfy 
 the relations (\ref{r1a}). To this end, in the right-hand side of  (\ref{sph-2a})
we perform the substitutions
%
\begin{eqnarray*}
z^{1+B_1} (z-1)^{1+B_1}\;\mapsto\; z^{1+\frac{B_1}{z_0}} (z-z_0)^{1-B_2-\frac{B_1}{z_0}},
\qquad
%
\Psi(2+B_1,2+B_1-n; - 2i\omega z)\;\mapsto\; \Psi(\alpha,\beta-n; - 2i\omega z),
\end{eqnarray*}
{where  we have used the exponents} $1+{B_1}/{z_0}$ and $1-B_2-{B_1}/{z_0}$
because these are indicial exponents at $z=0$ and  $z=z_0$, respectively. 
By using the properties
of $\Psi(a,c;y)$ we find that $\alpha=2+i\eta-{B_2}/{2}$ and $\beta=2+B_1/z_0$
 \cite{arxiv}. Thus, $\mathcal{U}_1$ is given
%
\antiletra\letra
\begin{eqnarray}\label{che-primeiro-set}
\begin{array}{l}
\begin{array}{l}
\mathcal{U}_1(z)=e^{i\omega z}z^{1+\frac{B_1}{z_0}}
[z-z_0]^{1-B_2-\frac{B_1}{z_0}}\end{array}
\displaystyle\sum_{n=0}^{\infty}
{b}_{n}^{1}
\begin{array}{l}
\Psi\left(2+i\eta-\frac{B_2}{2},2+\frac{B_1}{z_0}-n; - 2i\omega z\right)\end{array},
\end{array}\quad
\begin{array}{l}
 [i\eta-B_2/2\neq -2,-3,\cdots] 
\end{array}
\end{eqnarray}
where the coefficients ${b}_n^1$ satisfy  
the recurrence relations (\ref{r1a}) 
with \cite{arxiv}
\begin{eqnarray}\label{alfa}
&\begin{array}{l}
{\alpha}_n^1=- 2i\omega z_0(n+1), \qquad
{\beta}_n^1=n\left[n+1-B_2-\frac{2B_1}{z_0}+
2i\omega z_0\right]+
\left[i\omega z_0-1-\frac{B_1}{z_0}\right] \left[2-B_2-
\frac{B_1}{z_0}\right]+2-B_2+B_3,\end{array}
\nonumber\\
&\begin{array}{l}
{\gamma}_n^1=-\left[n+i\eta-\frac{B_1}{z_0}-\frac{B_2}{2}\right]
\left[n+1-B_2-\frac{B_1}{z_0}\right].\end{array}
\end{eqnarray}

%
By the transformations (\ref{transformacao1}), $ \mathcal{U}_1$ 
produces a group constituted by 16 solutions, $\mathcal{U}_i$. 
Eight of these can be constructed as
\antiletra
\begin{eqnarray}\label{TRANS}
\begin{array}{llll}
\mathcal{{U}}_1(z),\qquad&
\mathcal{{U}}_2(z)=T_1\mathcal{{U}}_1(z),\quad&
\mathcal{{U}}_3(z)=T_2\mathcal{{U}}_2(z),\quad&
\mathbb{{U}}_4(z)=T_1\mathcal{{U}}_3(z);\vspace{2mm}\\
\mathcal{{U}}_5(z)=T_4\mathcal{{U}}_1(z),\quad&
\mathcal{{U}}_6(z)=T_4\mathcal{{U}}_2(z),&
\mathcal{{U}}_7(z)=T_4\mathcal{{U}}_3(z),&
\mathcal{{U}}_8(z)=T_4\mathcal{{U}}_4(z),
\end{array}
\end{eqnarray}
while the others result by the transformation
$T_3$ which changes ($\eta,\omega$)
by ($-\eta,-\omega$) in the above solutions. 
Thus, 
\letra
\begin{eqnarray}\label{che-segundo-set}
\begin{array}{l}
\mathcal{U}_2(z)=e^{i\omega z}[z-z_0]^{1-B_2-\frac{B_1}{z_0}}\end{array}
\displaystyle\sum_{n=0}^{\infty}
\begin{array}{l}{b}_n^{2}
\Psi\left(1+i\eta-\frac{B_2}{2}-\frac{B_1}{z_0},-\frac{B_1}{z_0}-n; - 2i\omega z\right)
\end{array},\quad
\begin{array}{l}
 [i\eta-\frac{B_2}{2}-\frac{B_1}{z_0}\neq -1,-2,\cdots] 
\end{array}
\end{eqnarray}
where, in the recurrence relations (\ref{r1a}) for ${b}_n^{2}$,
\begin{eqnarray}\label{alfa-2}
&\begin{array}{l}
{\alpha}_n^2=-2i\omega z_0(n+1), \qquad
{\beta}_n^2=n\left[n+3-B_2+2i\omega z_0)\right]+
 i \omega z_0\left[2-B_2-\frac{B_1}{z_0}\right]+2-B_2+B_3
],\end{array}\nonumber\\
&\begin{array}{l}
{\gamma}_n^2=-\left[n+i\eta+1-\frac{B_2}{2}\right]
\left[n+1-B_2-\frac{B_1}{z_0}\right. 
\end{array}
\end{eqnarray}
The third solution reads
\antiletra\letra
\begin{equation}\label{che-third-set}
\begin{array}{l}
\mathcal{U}_3(z)=e^{i\omega z}\end{array}
\displaystyle\sum_{n=0}^{\infty}
{b}_n^{3}
\begin{array}{l}
\Psi\left(i\eta+\frac{B_2}{2},-\frac{B_1}{z_0}-n; - 2i\omega z\right),
\qquad\left[i\eta+\frac{B_2}{2}\neq 0,-1,\cdots,\right]
\end{array}
\end{equation}
with
\begin{eqnarray}\label{alfa-3}
&\begin{array}{l}
{\alpha}_n^3= -2i\omega z_0(n+1), \qquad
{\beta}_n^3=n\left[n+1+B_2+
\frac{2B_1}{z_0}+2i\omega z_0\right]+
\left[B_2+\frac{B_1}{z_0}\right]\left[1+\frac{B_1}{z_0}+i\omega z_0\right]+B_3, 
\end{array}
\nonumber\\
&\begin{array}{l}
{\gamma}_n^3=-\left[n+i\eta+\frac{B_2}{2}+\frac{B_1}{z_0}\right]
\left[n-1+B_2+\frac{B_1}{z_0}\right].
\end{array}
\end{eqnarray}
At last, we write 
\antiletra\letra
\begin{eqnarray}\label{che-fourth-set}
\begin{array}{l}
\mathcal{U}_4(z)=e^{i\omega z}z^{1+\frac{B_1}{z_0}}\end{array}
\displaystyle\sum_{n=0}^{\infty}
\begin{array}{l}{b}_n^{4}
\Psi\left(1+i\eta+\frac{B_2}{2}+\frac{B_1}{z_0} ,2+\frac{B_1}{z_0}-n; -2i\omega z\right),\end{array}
\quad
\begin{array}{l}
 [i\eta+B_2/2+B_1/z_0\neq -1,-2,\cdots] 
\end{array}
\end{eqnarray}
with
\begin{eqnarray}\label{alfa-4}
&\begin{array}{l}
{\alpha}_n^4= -2i\omega z_0(n+1), \qquad
{\beta}_n^4=n\left[n-1+B_2+2i\omega z_0\right]+i\omega z_0
\left[B_2+\frac{B_1}{z_0}\right]+B_3,\end{array}
\nonumber\\
&\begin{array}{l}
{\gamma}_n^4=-\left[n+i\eta-1+\frac{B_2}{2}\right]\left[n-1+B_2+\frac{B_1}{z_0}\right].
\end{array}
\end{eqnarray}
%
%
%
%

The relation (\ref{limite-1}) is valid also for the 
present case, whereas for large values of $z $ 
Eq. (\ref{limite-2}) is replaced by \cite{arxiv}
\antiletra
\begin{eqnarray}
\mathcal{U}_1(z)\sim e^{ i\omega z}\ z^{-i\eta-\frac{B_2}{2}}
\sum_{n=0}^{\infty}{b}_n^1, \qquad 
\left|\frac{{b}_{n+1}^{1}}{{b}_{n}^{1}}\right|=
1+\frac{1}{n}\text{Re}\left(i\eta-\frac{B_2}{2}\right)+O\left(\frac{1}{n^2}\right) \text{ if } n\to\infty.
\end{eqnarray}
Then, $\mathcal{U}_1$ converges at $z=\infty$ 
if $\text{Re}(i\eta-B_2/2)<-1$. By using the transformations as in (\ref{TRANS}), 
we find that all the $\mathcal{U}_i$ 
converge for finite values of $z$, excepting possibly the points $z= 0$ 
(if $i=1,2,3,4$) and $z=z_0$ (if $i=5,6,7,8$). According to the Raabe test, these $\mathcal{U}_i$  
converge also at $z=\infty$ if 
\begin{eqnarray}\label{regiao-1}
\begin{array}{ll}
\text{Re}\left[i\eta-\frac{B_2}{2}+1\right]< 0:\;\mathcal{U}_1,\,\mathcal{U}_5; \qquad &
\text{Re}\left[i\eta-\frac{B_1}{z_0}-\frac{B_2}{2}\right]< 0:\;
\mathcal{U}_2,\,\mathcal{U}_8;\\
\text{Re}\left[i\eta+\frac{B_2}{2}-1\right]< 0:\;\mathcal{U}_3,\,\mathcal{U}_7; \qquad &
\text{Re}\left[i\eta+\frac{B_1}{z_0}+\frac{B_2}{2}\right]< 0:\; 
 \mathcal{U}_4,\,\mathcal{U}_6.
\end{array}
\end{eqnarray}

\subsection*{4.3. The radial part of the two-center problem}

Now we consider the equations of the two-center problem of quantum mechanics, as the
one describing the electron of the ionized hydrogen molecule. Using Leaver`s 
conventions \cite{leaver}, the wave
function $\psi$ of the time-independent Schr\"{o}dinger equation
for an electron in the field of two Coulombian centers has the form
\begin{eqnarray}
\label{tc2}
\psi=e^{im\varphi}\;\bar{R}(\lambda)\;\bar{S}(\mu) ,
\qquad \lambda=\frac{r_{1}+r_{2}}{2a},\qquad
\mu=\frac{r_{1}-r_{2}}{2a},\qquad m=0,\;\pm1,\;\pm2,\;\cdots,
\end{eqnarray}
where $r_{1}$ and $r_{2}$ are the distances from the electron
to the two nuclei, and $2a$ the intercenter distance.
By the definitions
\begin{eqnarray}
\begin{array}{l}\label{RS}
\mathrm{S}(z)=\bar{S}(\lambda)=z^{\frac{m}{2}}(2-z)^{\frac{m}{2}}U^{-}(z),\qquad
 z=\mu+1, \qquad [0\leq z\leq 2],
\vspace{3mm}\\
\mathrm{R}(z)=\bar{R}(\mu)=z^{\frac{m}{2}}(z-2)^{\frac{m}{2}}U^{+}(z),\qquad
z=\lambda+1,\qquad [z\geq 2],
\end{array}
\end{eqnarray}
Leaver obtained CHEs in the form (\ref{gswe}) for $U^{\pm}$, with
the parameters \big($\eta^{\pm}$ for $\eta$ , $B_3^{\pm}$ for $B_3$\big)
\begin{eqnarray}\label{2centros}
&z_{0}=2, \qquad \omega^2=2a^2E,\qquad \omega 
\eta^{\pm}=-a(N_{1}\pm N_{2}), \qquad B_{1}=-2(m+1),
\qquad B_{2}=2(m+1),\nonumber
\vspace{3mm}\\
&B_{3}^{\pm}=\omega^2+
2a(N_{1}\pm N_{2})+m(m+1)-A_{lm}.
\end{eqnarray}
where $A_{lm}$ is a separation constant, and $N_{1}$ and $N_{2}$ are the charges 
on the two nuclei. Thus, there are two CHEs, one for the ``angular'' coordinate 
$\mu$ and one for the ``radial'' coordinate $\lambda$. Each CHE is associated with
a characteristic equation (\ref{characteristic1}) which determines the possible values of
the constants $A_{lm}$ and $E$.

Now we consider $\mathrm{R}(z)$, the radial solution  
given in (\ref{RS}). For bound states ($E<0$) 
we take 
\begin{eqnarray}\label{iomega}
 i\omega=-a\sqrt{2|E|}\quad \Rightarrow \quad i\eta=i\eta^+=-(N_1+N_2)\big/\sqrt{2|E|}
\end{eqnarray}
in order to assure that the factor $\exp{(i\omega z)}$ remains finite when $z\to\infty$. 
Then, if $|E|$ is finite,
\begin{eqnarray*}
\begin{array}{l}
\text{Re}\left[i\eta-\frac{B_1}{z_0}-\frac{B_2}{2}\right]=
\text{Re}\left[i\eta+\frac{B_1}{z_0}+\frac{B_2}{2}\right]
=-\frac{N_1+N_2}{\sqrt{2|E|}}< 0,
\end{array}
\end{eqnarray*}
and, consequently, four of the solutions listed in (\ref{regiao-1}) converge at $z=\infty$. 
To get wavefunctions bounded also at $z=2$, we select 
$\mathcal{U}_2$ if $m\leq 0$ and $\mathcal{U}_4$ if $m\geq 0$. Thus, we find
\letra
\begin{eqnarray}\label{R}
\begin{array}{l}
\mathrm{R}(z)=e^{-a\sqrt{2|E|}\;z}\;z^{-\frac{|m|}{2}}\;(z-2)^{\frac{|m|}{2}}
\displaystyle\sum_{n=0}^{\infty}
\begin{array}{l}
{b}_{n}^{2}
\Psi\left(1-\frac{N_1+N_2}{\sqrt{2|E|}},1-|m|-n;a\sqrt{8|E|}\;z \right)\end{array}
\end{array}
\end{eqnarray}
where the coefficients $b_n^2$ satisfy the relations (\ref{r1a}) with
\begin{eqnarray}\label{alfa-2-center}
&\begin{array}{l}{\alpha}_n^2=\sqrt{8|E|}\;a(n+1), \qquad
{\beta}_n^2=n\left[n+1+2|m|-2a\sqrt{8|E|}\right]+
\Big[|m|+1\Big]\left[|m|-a\sqrt{8|E|}\right]+
\end{array}\nonumber\\
&\begin{array}{l}2a\Big[N_1+N_2-a|E|\Big]-A_{lm},\qquad
{\gamma}_n^2=-\Big[n+|m|\Big]
\left[n+|m|-\frac{N_1+N_2}{\sqrt{2|E|}}\right].\end{array}
\end{eqnarray}
The expansion (\ref{R}) holds only if
\antiletra
\begin{eqnarray}\label{thecondition}
{(N_1+N_2)}/{\sqrt{2|E|}}\neq l+1,\qquad [l=0,2,\cdots]
\end{eqnarray}
a condition which assures that $\Psi(a,b;y)$ is not a polynomial 
of degree $l$ in $y$. 

The condition (\ref{thecondition})
is also required by the Jaff\'e expansions. In effect, by using
the solutions $U_1^J$ (if $m\geq 0$) and $U_2^J$ (if $m\leq 0$) given in 
Eqs (\ref{jaffe-1}) and (\ref{jaffe-2}), respectively, we find
\letra
\begin{eqnarray}\label{jaffe-2centros}
\begin{array}{l}
\mathrm{R}^{J}(z)=e^{-a\sqrt{2|E|}\;z}\;z^{-1-\frac{|m|}{2}+\frac{N_1+N_2}{\sqrt{2|E|}}}\;(z-2)^{\frac{|m|}{2}}
\displaystyle\sum_{n=0}^{\infty} 
{a}_{n}^{1}\left(\frac{z-2}{z}\right)^n,
\end{array}
\end{eqnarray}
where the recurrence relations for $a_n^1$ have the form (\ref{r1a}) with
\begin{eqnarray}
&\begin{array}{l}
\alpha_n^1=(n+1)(n+|m|+1),\qquad
{\beta}_n^1=-2n\left[n+1+|m|+a\sqrt{8|E|}-\frac{N_1+N_2}{\sqrt{2|E|}}\right]+
\Big[|m|+1\Big]\left[\frac{N_1+N_2}{\sqrt{2|E|}}-a\sqrt{8|E|}-1\right]+
\end{array}\nonumber\\
&\begin{array}{l}
2a\Big[N_1+N_2-a|E|\Big]-A_{lm},\qquad
%
{\gamma}_n^1=-\Big[n+|m|-\frac{N_1+N_2}{\sqrt{2|E|}}\Big]
\left[n-\frac{N_1+N_2}{\sqrt{2|E|}}\right].\end{array}
\end{eqnarray}
Thus, $\gamma_{l+1}=0$ if ${(N_1+N_2)}/{\sqrt{2|E|}}= l+1$ 
and, then, $\mathrm{R}^J(z)$
becomes a finite-series solution with $0\leq n\leq l$, as stated
after Eq. (\ref{characteristic1}). In this case,
the constant $A_{lm}$ would be determined from the characteristic 
equation associated with the recurrence relations for $a_n^1$.
However, if $E$ and $A_{lm}$ are both determined from the radial
solution, we cannot satisfy the characteristic equation corresponding to 
the angular solutions (these are usually 
given by series where the summation begins at $n=0$ and, so, present recurrence relations having
the form (\ref{matriz-tridiagonal})). Therefore, also for the Jaff\'e solutions it is necessary 
that ${(N_1+N_2)}/{\sqrt{2|E|}}\neq l+1$. The same is 
true respecting Hylleraas' expansions in series
of Laguerre polynomials \cite{leaver,hylleraas}.

The convergence of solution (\ref{R}) follows immediately from the Raabe test. 
As to the Jaff\'e solution (\ref{jaffe-2centros}), we have to examine 
its behavior at $z=\infty$. 
By using (\ref{2centros}) together with (\ref{iomega}), the expressions (\ref{jaffe-razao}) 
imply that
\antiletra
\begin{eqnarray}\label{jaffe-razao-radial}
 \lim_{z\to\infty}\mathrm{R}^{J}(z)=e^{-a\sqrt{2|E|}\;z}\;z^{\frac{N_1+N_2}{\sqrt{2|E|}}-1}
\displaystyle\sum_{n=0}^{\infty} 
{a}_{n}^{1},\; \text{ with }\; 
%
\lim_{n\rightarrow \infty}\frac{a_{n+1}^{1}}{a_{n}^{1}}
\begin{array}{l}
=1-\frac{1}{n}\left[\frac{3}{4}+2\sqrt{an\sqrt{2|E|}}-2a\sqrt{2|E|}+\frac{N_1+N_2}{\sqrt{2|E|}}\right].
\end{array}
\end{eqnarray}
Then, by a convenient choice of $n$, the constant $A$ which appears in (\ref{Raabe-1}) 
becomes less than $-1$ and so, by the Raabe test, the solution converges at $z=\infty$.

\section*{5. Concluding remarks}
%

By inserting a suitable weight function 
$w(z,t)$ into the integral relation (\ref{integral})
we have found the kernel equation (\ref{nucleus-gswe})
where the differential operators $L_z$ and $L_t$ {depend on $z$ and $t$ in same manner as  
the operator of the CHE (\ref{gswe2}); this  fact allows} to get transformations 
of the kernels by examining the known transformations of the solutions for the CHE. 
As mentioned, this is an extension of a similar correspondence found in 2011 
for the general Heun equation (HE) \cite{lea}.

Actually, in 1942 Erd\'{e}lyi used the appropriate weight function 
for the HE but he could not infer how to transform the kernels 
because the transformations of the HE were fully established only
in 2007 \cite{maier}. On the other side, transformations 
of confluent Heun equations are known since 1978
\cite{decarreau1,decarreau2} but have not been applied 
to transform kernels -- see, for example, references \cite{ronveaux,kazakov-1,lay,masuda}. 
In the present study we have considered transformations of
kernels of the CHE {and limiting cases}. The initial kernels (to be transformed) come from 
kernels of the HE by a process of confluence
\cite{lea}; however, for the sake of completeness, in section 2  
we have reobtained them by solving the kernel equation.

By separation of variables we have found    
two groups of kernels presenting an
arbitrary constant of separation. One group, with  
products of two confluent hypergeometric functions,
includes some particular kernels already known in the literature \cite{masuda}; 
the other group, with products of confluent hypergeometric functions 
and Gauss hypergeometric functions, is new as far as we {know}. By 
ascribing particular values to the constant of separation 
we have obtained three groups given by product of elementary functions
with one special function: this is  
represented by confluent hypergeometric functions (two groups) and 
by Gauss hypergeometric functions (one group).

In section 3 we have found some integral transformations
among known solutions of the confluent Heun equations.  
We have used two singularities as 
endpoints of integration 
and supposed that the solutions to be transformed 
are convergent at both endpoints 
(this assures that the bilinear concomitants vanish there).
If the solutions are modified by the rules 
(\ref{transformacao1}), the kernels 
must be modified by the rules (\ref{transformacaofor-K1}).
This emphasizes that the correspondence between the transformations
of the Heun equations and  
of the respective kernels are important parts
of the transformation theory.
%

The applications of section 3 simply interconnect known solutions 
without affording new solutions. In contrast, in section 4, 
by means of an integral
transformation we have obtained a new 
solution for the spheroidal wave equation, which in turn leads to a group of new solutions for the CHE.
We have seen that these solutions
may be used to compute the radial part of the wavefunctions for 
bound states of hydrogen 
moleculelike ions and, by this reason, can play the role 
of the expansions in series of Laguerre polynomials proposed by Hylleraas 
in 1931 \cite{hylleraas} and the Jaff\'e power-series {solutions \cite{jaffe} which have been 
used from 1934 up to now \cite{kepler}.}

{It is possible to realize further properties  of the solutions by considering other problems, as 
the Lorentzian model of a quantum two-state system given by Ishkhanyan and Gregoryan \cite{ishkhanyan}.
This is ruled by a CHE with $z=(1+it)/2$ and $z_0=1$, where $t$ denotes the time. 
According to the authors, for certain values of a parameter $R$, the problem 
admits finite-series solutions which are bounded for any admissible value of $t$
and assure that the system returns to the initial state
after the interaction. By using the solutions of section 4.2, 
we have verified that the previous statement is correct; it seems that no other
known solution of the CHE permits to prove the statement.  In addition, since in this case $|z|\geq 1/2$,
it would be interesting to check if there are {\it infinite-series} solutions 
suitable for some range of the parameter $R$ (the Hylleraas and Jaff\'e solutions do not
converge for $|z|<1$).}

%
We have omitted details concerning the derivation of the new solutions 
of the CHE. {In addition, the solutions}  must be improved as follows: (i)
by considering also expansions in series of regular confluent hypergeometric 
functions, we will get solution valid in the neighborhood of $z=0$ \cite{arxiv},
(ii) by using the Whittaker-Ince limit as in Ref. \cite{lea-2}, we can expect solutions in series of
Bessel functions for the RCHE (\ref{incegswe}), (iii) by inserting a ``characteristic'' 
parameter $\nu$ and letting that the series summation
runs from minus to plus infinite \cite{arxiv} (two-sided series),  we can obtain 
solutions for a CHE without free parameters \cite{lea-2}.

As mentioned, besides the RCHE, there are two other equations which are associated 
with the CHE by formal limits. These are the double-confluent Heun equation (DHE) and
the reduced DHE (RDHE) which appear when we allow that 
$z_0\to 0$ in the CHE and RCHE, respectively, that is,
\begin{eqnarray}\label{dche-1}
\begin{array}{lll}
\text{DHE}:&\qquad z^{2}\frac{d^{2}U}{dz^{2}}+
\left(B_{1}+B_{2}z\right)\frac{dU}{dz}+
\left(B_{3}-2\eta \omega z+\omega^{2}z^{2}\right)U=0,&\qquad \left[B_{1}\neq 0, \ \omega\neq 0\right]\vspace{3mm}\\
\text{RDHE}:&\qquad z^2\frac{d^{2}U}{dz^{2}}+(B_{1}+B_{2}z)
\frac{dU}{dz}+
\left(B_{3}+qz\right)U=0,&\qquad [q\neq0, \ B_{1}\neq 0]
\end{array}
\end{eqnarray}
where now $z=0$ and $z=\infty$ are irregular
singularities. At $z=\infty$ the behaviour is again
given by Eq. (\ref{thome1}), that is,
\begin{eqnarray*}
\lim_{z\rightarrow  \infty}U(z)\sim e^{\pm i\omega z}\ z^{\mp i\eta-(B_{2}/2)}\ 
\text { for the DHE (\ref{gswe})},
\qquad
\lim_{z\rightarrow  \infty}U(z)\sim
e^{\pm 2i\sqrt{qz}}\ z^{(1/4)-(B_{2}/2)} \text{ for the 
RDHE (\ref{incegswe})},
\end{eqnarray*}
while at $z=0$ the normal Thom\'e solutions affords \cite{lea-2}
\begin{eqnarray*}
\displaystyle\lim_{z\rightarrow  0}U(z)\sim 1\qquad \mbox{or}\qquad
%
 \displaystyle\lim_{z\rightarrow  0}U(z)\sim e^{B_{1}/ z}z^{2-B_{2}} \qquad \text{for DHE and RDCE}.
\end{eqnarray*}
%

{Starting with kernels of the CHE}, in appendices C, D and E we 
have found that the Whitakker-Ince limit (\ref{ince}) and the 
Leaver limit ($z_{0}\rightarrow 0$) lead to new kernels for the RCHE, DHE and RDHE, 
in accordance with a previous conjecture \cite{lea}. 
However, by integrating the kernel equations we have also found kernels which are not connected with 
known kernels of the CHE: 
for the RCHE we have a group of
kernels expressed by products of Bessel and hypergeometric functions, while for the DHE and RDHE
we have kernels given in terms of elementary functions. 
{Therefore, the limiting procedures do not exhaust the
possibilities for generating kernels.}

{
In the appendix C we have {found} that the usual kernels of the Mathieu 
equation turn out to be particular cases of kernels of the RCHE.
{Furthermore}, in appendix D, we have noticed that for DHE and RDHE in general 
{ it is convenient} to use integral relations with variable limits of integration; this fact leads to an  
additional term in the bilinear concomitant - see Eq. (\ref{condition2}).}

 

%
%
%
\section*{Appendix A. Hypergeometric functions}
\protect\label{A}
\setcounter{equation}{0}
\renewcommand{\theequation}{A.\arabic{equation}}
The regular and
irregular confluent hypergeometric functions are denoted by $\Phi({a,c};u)$
and $\Psi({a,c};u)$, respectively. They satisfy the
confluent hypergeometric equation \cite{erdelii}
\begin{eqnarray}\label{confluent0}
u\frac{d^2\varphi(u)}{du^2}+\left( {c}-u\right)
\frac{d\varphi(u)}{du} -{a}\varphi(u)=0
\end{eqnarray}
which admits the solutions
\begin{equation}\label{confluent1}
\varphi^{1}(u)=\Phi({a,c};u),\quad
 \varphi^{2}(u)=e^{u}u^{1-c}\Phi(1-a,2-c;-u),\quad
%
\varphi^{3}(u)=\Psi(a,c;u),\quad
 \varphi^{4}(u)=e^{u}u^{1-c}\Psi(1-a,2-c;-u).
\end{equation}
All of them are defined and distinct only if
$\mathrm{c}$ is not an integer.
Alternative forms for these solutions
follow from the relations
\begin{eqnarray}\label{kummer}
\Phi({a,c};u)=e^{u}\Phi({c-a,c};-u),\qquad
\Psi({a,c};u)=u^{{1-c}}
\Psi(1+{a-c,2-c};u).
\end{eqnarray}
On the other side, solutions for the (Gauss) hypergeometric equation  \cite{erdelii},
\begin{eqnarray}
\label{hypergeometric}
u(1-u)\frac{d^{2}F}{du^{2}}+\big[{c}-({a}+{b}+1)u
\big]\frac{dF}{du}-{a}{b}F=0,
\end{eqnarray}
are given by hypergeometric functions
$F({a,b;c};u)=F({b,a;c};u)$.
In fact, in the vicinity of the singular points $0$,
$1$ and $\infty$, the formal solutions for the hypergeometric
equation (\ref{hypergeometric}) are, respectively,
\begin{eqnarray}\label{hiper-1}
\begin{array}{l}
F^{1}(u)=F\left( {a,b;c;}
u\right) ,
\qquad
F^{2}(u)=u^{1-{c}}
F\left( {a+1-c,b+1-c;
2-c};u\right);
\end{array}
\end{eqnarray}
\begin{eqnarray}\label{hiper-2}
\begin{array}{l}
F^{3}(u)=F\left( {a,b;
a+b+1-c};1-u \right),\end{array}\qquad 
%
\begin{array}{l}
F^{4}(u)=
 \left( 1-u\right)^{{c-a-b}}
F\left({c- a,c-b;
1+c-a-b};1-u\right);\end{array}
\end{eqnarray}
\begin{equation}\label{hiper-3}
\begin{array}{l}
F^{5}(u)=u^{-{a}}
F\left( {a,a+1-c;
a+1-b};\frac{1}{u}\right),\qquad
F^{6}(u)=u^{-\mathrm{b}}
F\left({b+1-c,b;
b+1-a};\frac{1}{u}\right).
\end{array}
\end{equation}
Each of these may be written in four forms
by using the relations
\begin{eqnarray}\label{euler-1}
\begin{array}{l}
F({a,b;c;u})=(1-u)^{{c-a-b}}
F({c-a,c-b;c;u}),\qquad
%
F({a,b;c;u})=(1-u)^{{-a}}
F\left[ {a,c-b;c;{u}/({u-1}})\right] .
\end{array}
\end{eqnarray}

On the other side, the usual form for the Bessel equation is \cite{nist}
\begin{eqnarray}\label{bessel}
y^2\frac{d^2Z(y)}{dy^2}+
y\frac{dZ(y)}{dy}+
\left[y^2-\alpha^2 \right] Z(y)=0.
\end{eqnarray}
The solutions for this
equation are denoted by $Z_{\ \alpha}^{(j)}(y)$  
according as \cite{nist,arscott}
\begin{eqnarray}\label{Z2}
Z_{\ \alpha}^{(1)}(y)=J_{\alpha}(y),\qquad
Z_{\ \alpha}^{(2)}(y)=Y_{\alpha}(y),
\qquad
Z_{\ \alpha}^{(3)}(y)=H_{\ \alpha}^{(1)}(y),\qquad
Z_{\ \alpha}^{(4)}(y)=H_{\ \alpha}^{(2)}(y)
\end{eqnarray}
where $J_{\alpha}(y)$ and $Y_{\alpha}(y)$ 
 are the Bessel functions of the first and second
 kind, respectively; $H_{\  \alpha}^{(1)}(y)$
and $H_{\  \alpha}^{(2)}(y)$ are the first and the second Hankel functions. 
There are formulas connecting these functions \cite{nist}.
For example, 
\begin{eqnarray}
Y_{\alpha}=\frac{1}{2i}\left[H_{\ \alpha}^{(1)}-
H_{\ \alpha}^{(2)}\right]=\frac{\cos(\alpha\pi)
J_{\alpha}-J_{-\alpha}}{\sin(\alpha\pi)}.
\end{eqnarray}
Bessel and confluent
hypergeometric functions are connected by \cite{erdelii}
\begin{eqnarray}\label{definition-J}
\begin{array}{l}\Phi\left(\alpha+\frac{1}{2},
2\alpha+1;- 2iy\right)\end{array}&=&\begin{array}{l}\Gamma(\alpha+1)\
e^{-iy}\left(\frac{y}{2}\right)^{-\alpha}J_{\alpha}(y),\quad
%
\Psi\left(\alpha+\frac{1}{2},
2\alpha+1;- 2iy\right)=\frac{i\sqrt{\pi}}{2}
e^{-i(y-\alpha\pi)}\left(2y\right)^{-\alpha}H_{\alpha}^{(1)}(y),\end{array}\nonumber\\
%
%
\begin{array}{l}
\Psi\left(\alpha+\frac{1}{2},
2\alpha+1;2iy\right)\end{array}
&=&\begin{array}{l}
-\frac{i\sqrt{\pi}}{2}
e^{i(y-\alpha\pi)}\left(2y\right)^{-\alpha}H_{\alpha}^{(2)}(y).
\end{array}
\end{eqnarray}
In addition, we have the relations \cite{erdelii}
\begin{eqnarray}\label{J}\begin{array}{l}
\displaystyle
\lim_{a\rightarrow  \infty}
\Phi\left(a,c;-\frac{y}{a}\right)=
\Gamma(c)\ y^{(1-c)/2}J_{c-1}\big(2\sqrt{y}\big),\vspace{2mm}\\
\displaystyle
\lim_{a\rightarrow  \infty}\left[\Gamma(a+1-c)\ \Psi\left(a,c;
-\frac{y}{a}\right)\right]
=\begin{cases}\displaystyle
-i\pi e^{i\pi c}y^{(1-c)/2}H_{c-1}^{(1)}\big(2\sqrt{y}\big),
\ \ &\text{Im}\ y>0,
\vspace{2mm}\\
\displaystyle
i\pi e^{-i\pi c}y^{(1-c)/2}H_{c-1}^{(2)}\big(2\sqrt{y}\big),
\ \ &\text{Im}\ y<0.
\end{cases}
\end{array}
\end{eqnarray}
%

%

\section*{Appendix B. Wilson's asymptotic expansions for 
the CHE}
\protect\label{B}
\setcounter{equation}{0}
\renewcommand{\theequation}{B.\arabic{equation}}
{Such solutions were considered in 1928 by Wilson \cite{wilson}. 
Actually they are given by 
8 asymptotic Thom\'e expansions \cite{olver} which we denote by
$\mathcal{W}_i(z)$ ($i=1,2,3,4$) and  
$T_3\mathcal{W}_i(z)$. For the CHE in the form (\ref{gswe})
we find 
%
%
\begin{eqnarray}\label{first-wilson} \begin{array}{l}
\mathcal{W}_{1}(z)=e^{ i\omega z}\ z^{- i\eta-
\frac{B_{2}}{2}}\displaystyle \sum_{n=0}^{\infty}a_{n}^{1}z^{-n}
\end{array}\end{eqnarray}
where the coefficients $a_{n}^{1}$ satisfy 
the three-term recurrence relations
($a_{-1}^{1}=0$)
\begin{eqnarray}\label{w2}
&&\begin{array}{l}
2i\omega(n+1)a_{n+1}^{1}-
\Big[ n\left(n+1+ 2i\eta+ 2i\omega z_{0}\right)+
i\omega z_{0}\Big(B_{2}+\frac{B_{1}}{z_{0}}\Big)+B_{3}+
\Big(\frac{B_{2}}{2}+i\eta\Big)\left(1+ i\eta -
\frac{B_{2}}{2}\right)\Big]a_{n}^{1}
\end{array}\nonumber\\
&&\begin{array}{l}
+ z_{0}\left(n+ i\eta+\frac{B_{1}}{z_{0}}+
\frac{B_{2}}{2}
\right)\left(n+ i\eta+\frac{B_{2}}{2}-1\right)a_{n-1}^{1}=0.
\qquad
\end{array}
\end{eqnarray}
For the other solutions we take
\begin{eqnarray}\label{W-CONV}
\mathcal{W}_2(z)=T_2\mathcal{W}_1(z),\qquad
\mathcal{W}_3(z)=T_4\mathcal{W}_1(z),\qquad
\mathcal{W}_4(z)=T_4\mathcal{W}_2(z)=T_1\mathcal{W}_3(z)
\end{eqnarray}
and $\mathcal{W}_{i+4}=T_3\mathcal{W}_i$ ($i=1,\cdots 4$). 
%
%
%
Thus, from the first solution we get
\begin{eqnarray}
\label{second-wilson}
\begin{array}{l}
\mathcal{W}_{2}(z) =e^{i\omega z}\ (z-z_{0})^{1-B_{2}-\frac{B_{1}}{z_{0}}}\ z^{-i\eta-1+\frac{B_{1}}{z_{0}}+\frac{B_{2}}{2}}
\displaystyle \sum_{n=0}^{\infty}a_{n}^{2}
z^{-n},\qquad \text{where}
\end{array}
\end{eqnarray}
%
%
\begin{eqnarray}
&&\begin{array}{l} 2i\omega(n+1)a_{n+1}^{2}-
\Big[ n\left(n+1+ 2i\eta+ 2i\omega z_{0}\right)+
i\omega z_{0}\Big(2-B_{2}-\frac{B_{1}}{z_{0}}\Big) +B_{3}+
\Big(\frac{B_{2}}{2}+ i\eta\Big)
\Big(1+ i\eta -\frac{B_{2}}{2}\Big)\Big]a_{n}^{2}
\end{array}
\nonumber\\
&&\begin{array}{l} 
%
+ z_{0}\Big(n+i\eta-\frac{B_{1}}{z_{0}}-
\frac{B_{2}}{2}
\Big)\Big(n+ i\eta+1-\frac{B_{2}}{2}\Big)a_{n-1}^{2}=0.\qquad
\end{array}
\end{eqnarray}
This $\mathcal{W}_{2}(z)$ is the only solution relevant for section 5.}
For this reason we omit the other solutions.

{By the D'Alembert test the solutions $\mathcal{W}_{ 1}(z)$
and $\mathcal{W}_{ 2}(z)$ converge for 
$|z|>|z_0|$, whereas $\mathcal{W}_{ 3}(z)$ and
$\mathcal{W}_{ 4}(z)$ converge for
$|z-z_0|>|z_0|$. However, by the Raabe test they converge also at
$|z|=|z_0|$ and $|z-z_0|=|z_0|$ provided that
\begin{equation}\label{convergencia1-0}
|z|\geq|z_0|  \text{ if }
\begin{cases}\text{Re}\left[  B_2+\frac{B_1}{z_0}\right] <1 \mbox{ in }
\mathcal{W}_{1}(z),\vspace{2mm} \\
\text{Re}\left[  B_2+\frac{B_1}{z_0}\right] >1\mbox{ in }
\mathcal{W}_{2}(z);
\end{cases}\ \ 
|z-z_0|\geq|z_0| \text{ if }
\begin{cases}\text{Re}\left[ \frac{B_1}{z_0}\right] >{-}1\mbox{ in }
\mathcal{W}_{3}(z),
\vspace{2mm} \\
\text{Re}\left[ \frac{B_1}{z_0}\right] <{-}1\mbox{ in }
\mathcal{W}_{4}(z),
\end{cases}
\end{equation}
where the restrictions on parameters of the equation
are necessary only to assure convergence at $|z|=|z_0|$ or $|z-z_0|=|z_0|$.}

%
The above regions of convergence suppose
the minimal solutions for the series coefficients \cite{olver}.
In the following we consider only the series
which appears in $W_1(z)$, the convergence for the other
solutions being obtained by using the transformations
as indicated above. 
Thus, when $n \to \infty$ in $W_1(z)$ we  have
\begin{eqnarray*}
&&\begin{array}{l}
2i\omega \frac{a_{n+1}^{1}}{a_{n}^{1}}- 
\left(n+1+ 2i\eta+ 2i\omega z_{0}\right)
%
+ z_{0}\left(n+ 2i\eta+B_2+\frac{B_{1}}{z_{0}}-1
\right)\frac{a_{n-1}^{1}}{a_{n}^{1}}=0
\qquad
\end{array}
\end{eqnarray*}
whose minimal solution for $a_{n+1}^{1}/{a_{n}^{1}}$  { satisfies
\begin{eqnarray*}
&&
\begin{array}{l}\displaystyle\frac{a_{n+1}^{1}}{a_{n}^{1}}\sim
z_0\left[1+\frac{1}{n}\left(B_2+\frac{B_1}{z_0}-2\right)\right]
\end{array}\Rightarrow
\begin{array}{l}\displaystyle
\frac{a_{n-1}^{1}}{a_{n}^{1}}\sim
\frac{1}{z_0}\left[1-\frac{1}{n}\left(B_2+\frac{B_1}{z_0}-2
\right)\right].
\end{array}
\end{eqnarray*}
Thence, when $n\to\infty$
\begin{eqnarray*}
\begin{array}{l}\displaystyle
\frac{a_{n+1}^{1}
z^{-n-1}}{a_{n}^{1}z^{-n}}\sim
\frac{z_0}{z}\left[1+\frac{1}{n}\left(B_2+\frac{B_1}{z_0}-2
\right)\right]\Rightarrow\end{array} 
\begin{array}{l}\displaystyle
 \vline\frac{a_{n+1}^{1}
z^{-n-1}}
{a_{n}^{1}z^{-n}}\vline\sim
\frac{|z_0|}{|z|}\left[1+\frac{1}{n}\text{Re}
\left(B_2+\frac{B_1}{z_0}-2\right)\right].\end{array}
\end{eqnarray*}}
So, by the D'Alambert test the series converges absolutely for
$|z|>|z_0|$. However, by 
the Raabe test, the series converges even for $|z|=|z_0|$
provided that $\text{Re}\left[B_2+({B_1}/{z_0})\right]<1$.

%

\section*{Appendice C. Kernels for the reduced confluent Heun equation (RCHE)}
\protect\label{C}
\setcounter{equation}{0}
\renewcommand{\theequation}{C.\arabic{equation}}
%

In this appendix  C:
  \begin{itemize}
  \itemsep-3pt
  \item 
  {initially} we get the substitutions of variables which preserve the form of the
  equation for the kernels of the RCHE;
 \item 
  {in  C.1} we find a group of kernels containing products of two Bessel functions
  having an arbitrary {constant of separation  $\lambda$};  
  these kernels cannot be derived from {known}  kernels of the CHE by using the 
  Whittaker-Ince limit (\ref{ince});
  \item   %
  {in  C.2} we construct a group of kernels containing  products of Bessel and hypergeometric functions
 with an arbitrary constant of separation $\lambda$; these kernels may be derived as limits
   of kernels of the CHE; 
 \item
 {in  C.3,  C.4 and C.5,  by taking suitable values for $\lambda$ in the above cases, we get 
  kernels given by products of elementary  and  special functions;
  thus, in C.3 and C.4 the kernels are given by products of elementary and Bessel functions,
  while in C.5 the kernels are given by products of elementary and
  Gauss hypergeometric functions;
  \item 
 {in C.6 we find} an integral relation between a solution of a RCHE  
 in power series and a solution given by series of Bessel functions of first kind.}
\item 
{the above results are also valid  for the Mathieu equations
because these are particular instances of the RCHE.}
\end{itemize}

In Whittaker-Ince limit (\ref{ince}),  Eqs. (\ref{integral})
and (\ref{concomitant}) remain formally unchanged
but the operator (\ref{Lz1}) now reads
\begin{eqnarray}
L_{z}=z(z-z_{0})\frac{\partial^{2}}{\partial z^{2}}+
\left[B_{1}+B_{2}z\right]
\frac{\partial}{\partial z}
+qz,
\end{eqnarray}
and so the RCHE (\ref{incegswe}) and equation ({\ref{nucleus-gswe}})
for the corresponding kernels
$G(z,t)$ take the forms
\begin{eqnarray}\label{kernel-WI}
\left[ L_z+B_3-qz_0\right] U(z)=0,
\qquad\left[L_z-L_t \right] G(z,t)=0.
\end{eqnarray}
On the other side, if $U(z)=U(B_{1},B_{2},B_{3};z_{0},q;z)$ is
solution of the RCHE, new solutions
may be generated by the transformations 
$\mathscr{T}_1$, $\mathscr{T}_2$ and $\mathscr{T}_3$
given by \cite{lea-2}
\begin{eqnarray}\label{Transformacao2}
\begin{array}{l}
\mathscr{T}_{1}
U(z)=z^{1+\frac{B_1}{z_0}}\
U(C_{1},C_{2},C_{3};z_{0},q;z),\vspace{2mm}\\
%
\mathscr{T}_{2}
U(z)=(z-z_{0})^{1-B_{2}-\frac{B_1}{z_0}}\ U(B_{1},D_{2},D_{3};
z_{0},q;z),\vspace{2mm}\\
%
\mathscr{T}_{3}U(z)=
U(-B_{1}-B_{2}z_{0},B_{2},
B_{3}-q z_{0};z_{0},-q;z_{0}-z),
\end{array}
\end{eqnarray}
where $C_i$ and $D_i$ are defined in (\ref{constantes-C-D}). 
Similarly, we can check that, if $G(z,t)=G(B_{1},B_{2}; z_{0},q;z,t)$
is a kernel, 
new kernels
may generated by the transformations
\begin{eqnarray}\label{transformacao-K2}
\begin{array}{l}
\mathscr{R}_{1}G(z,t)=\left({z}{t}\right)^{1+\frac{B_1}{z_0}}
G(C_{1},C_{2};z_{0},q;z,t),\vspace{2mm}\\
\mathscr{R}_{2}(z)G(z,t)=
\left[ ({z-z_0})({t-z_0})\right] ^{1-B_{2}-\frac{B_1}{z_0}}G(B_{1},D_{2};
z_{0},q;z,t),\vspace{2mm}\\
\mathscr{R}_{3}G(z,t)=
G(-B_{1}-B_{2}z_{0},B_{2};z_{0},-q;z_{0}-z,z_0-t).
\end{array}
\end{eqnarray}

We will see that the kernels for the
RCHE reproduce all the 
known kernels \cite{arscott,McLachlan} for the Mathieu equation.
To this end, we write the last equation as
\begin{eqnarray}\label{mathieu}
\frac{d^2w}{du^2}+\sigma^2\big[\mathrm{a}-2k^2\cos(2\sigma u)\big]w=0,
\end{eqnarray}
where $\sigma=1$ or $\sigma=i$ for the Mathieu or
modified Mathieu equations, respectively. 
Then, by setting $z=\cos^{2}(\sigma u)$ and
$w(u)=U(z)$, Eq. (\ref{mathieu})
is converted into RCHE (\ref{incegswe}) with the
following parameters:
\begin{equation}
z_{0}=1,\qquad  B_{1}=-\frac{1}{2}, \qquad B_{2}=1, \qquad
B_{3}=\frac{k^2}{2}-\frac{\mathrm{a}}{4}, \qquad q=k^2.
\end{equation}
Besides this, putting $t=\cos^2({\sigma v})$ the integral
(\ref{integral}) reads
\begin{eqnarray}\label{mathieu-integral}
\mathcal{U}(z)=\int_{v_{1}}^{v_{2}}G[z(u),t(v)]\ U[t(v)]\ dv.
\end{eqnarray}
where $z(u)=\cos^{2}(\sigma u)$ and 
$t(v)=\cos^2({\sigma v})$.

\subsection*{C.1. First group of kernels: products of Bessel functions}
We find the set of kernels $G_{(\pm,\pm)}^{(i,j)}$
given by products of Bessel functions, namely,  
\begin{eqnarray}\label{ince-che}
 G_{(\pm,\pm)}^{(i,j)}(z,t)&=&
 \begin{array}{l}
 \left[\lambda\sqrt{\frac{(z-z_0)(t-z_0)}{z_0}} \right]^{1-B_2-\frac{B_1}{z_0}}
\left[ \sqrt{\frac{zt}{z_0}}\right] ^{1+\frac{B_1}{z_0}}\end{array}
\nonumber\\
&\times&\begin{array}{l}
Z_{\pm\left(1-B_2-\frac{B_1}{z_0}\right) }^{(i)}
\left[\lambda\sqrt{\frac{(z-z_0)(t-z_0)}{z_0}}\
\right]
%
Z_{\pm\left( 1+\frac{B_1}{z_0}\right) }^{(j)}
\left[\sqrt{\frac{1}{z_0}(\lambda^2+4q)zt}\ \right],
\end{array}
\end{eqnarray}
where $Z_{\alpha}^{(i)}$ are the four 
Bessel functions given in Eq. (\ref{Z2}) or linear
combinations of them.  
In addition, we find that the transformations $\mathscr{R}_1$ and
$\mathscr{R}_2$ and  $\mathscr{R}_3$ do not produce new kernels. 

In fact, the substitutions
\begin{equation}\label{substituicoes-reduced}
\begin{array}{l}\xi=\sqrt{\frac{(z-z_0)(t-z_0)}{z_0}}, \qquad
\zeta=\sqrt{\frac{zt}{z_0}},\qquad G(z,t)=H(\xi,\zeta),
\end{array}
\end{equation}
transform Eq. (\ref{kernel-WI}) into
\begin{eqnarray}\label{AGA}
\frac{\partial^2H}{\partial\xi^2}+\frac{2}{\xi}
\left( B_2+\frac{B_1}{z_0}-\frac{1}{2}\right) 
\frac{\partial H}{\partial\xi}-4qH-
\left[\frac{\partial^2H}{\partial\zeta^2}
 -\frac{2}{\zeta}\left(\frac{B_1}{z_0}+\frac{1}{2}\right)\frac{\partial H}{\partial\zeta}\right]=0.
\end{eqnarray}
Thence, by the separation of variable
\begin{eqnarray}\label{into}
H(\xi,\zeta)=\xi^{1-B_2-\frac{B_1}{z_0}}
\zeta^{1+\frac{B_1}{z_0}}\ X(\xi)\ Y(\zeta),
\end{eqnarray}
we find
\begin{eqnarray*}
\begin{array}{l}
\frac{1}{X}\left[\frac{d^2X}{d\xi^2}+\frac{1}{\xi} \frac{dX}{d\xi}
-\frac{1}{\xi^2}\left( B_2+\frac{B_1}{z_0}-1\right)^2X\right]
%
-\frac{1}{Y}\left[\frac{d^2Y}{d\zeta^2}
 +\frac{1}{\zeta}\frac{dY}{d\zeta}-
 \frac{1}{\zeta^2}\left(1+\frac{B_1}{z_0}
 \right)^2Y \right]-4q
=0,
\end{array}
\end{eqnarray*}
which leads to the Bessel equations 
\begin{eqnarray}
\begin{array}{l}
\xi^2\frac{d^2X}{d\xi^2}+
\xi\frac{dX}{d\xi}+\left[\lambda^2\xi^2-
 \left(1- B_2-\frac{B_1}{z_0}\right)^2\right]X=0,\qquad
%
\zeta^2\frac{d^2Y}{d\zeta^2}
 +\zeta \frac{dY}{d\zeta}+
 \left[\left(\lambda^2+4q\right)\zeta^2-
 \left(1+\frac{B_1}{z_0}\right)^2\right]Y=0,
\end{array}
\end{eqnarray}
where $\lambda^2$ is a constant of separation. Thus,
by taking
\begin{eqnarray*}\begin{array}{l}
y=\lambda\xi,\quad \alpha= \pm\left(1-B_2-\frac{B_1}{z_0}\right)
\quad\text{and}\quad
%
y=\sqrt{(\lambda^2+4q)}\zeta,\quad \alpha=\pm\left(
 1+\frac{B_1}{z_0}\right)
\end{array}\end{eqnarray*}
in the first and second equations, respectively, we obtain
\begin{eqnarray}\label{lambda-neq0}
X(\xi)=Z_{\pm\left(1-B_2-\frac{B_1}{z_0}\right) }^{(i)}
\left[\lambda\xi\right],\qquad
Y(\zeta)=Z_{\pm\left( 1+\frac{B_1}{z_0}\right) }^{(j)}
\left[\sqrt{(\lambda^2+4q)}\zeta\right].
\end{eqnarray}
Inserting these solutions into (\ref{into}) we
get the kernels (\ref{ince-che}).

Now we let that the transformations $\mathscr{R}_i$
transform the parameter $\lambda$ into $\lambda_i$. 
Since $\lambda$ and $\lambda_i$ are arbitrary, we 
conclude that $\mathscr{R}_i$  do not change the kernels. 
For instance, 
\begin{eqnarray*}\begin{array}{l}
\mathscr{R}_3G_{(\pm,\pm)}^{(i,j)}(z,t)=
\left[{\frac{(z-z_0)(t-z_0)}{z_0}} \right]^{\frac{1}{2}-\frac{B_2}{2}-\frac{B_1}{2z_0}}
\left[ \sqrt{\frac{zt}{z_0}}\right] ^{1+\frac{B_1}{z_0}}\end{array}
%
\begin{array}{l}Z_{\pm\left( 1+\frac{B_1}{z_0}\right) }^{(i)}
\left[\lambda_3\sqrt{\frac{zt}{z_0}}
\right] \ 
Z_{\pm\left( 1-B_2-\frac{B_1}{z_0}\right) }^{(j)}
\left[\sqrt{\frac{[\lambda^2_3-4q][z-z_0][t-z_0]}{z_0}}\ \right]\end{array}
\end{eqnarray*}
By setting $\lambda_3^2=\lambda^ 2+4q$, we see that
the right-hand side is $G_{(\pm,\pm)}^{(i,j)}(z,t)$.

For Mathieu equation, the kernels (\ref{ince-che}) become
\begin{eqnarray}
\begin{array}{l}
G_{(\pm,\pm)}^{(i,j)}(z,t)=\sqrt{\sin(2\sigma u)\sin(2\sigma v)} 
Z_{\pm\frac{1}{2} }^{(i)}
\left[\lambda\sin(\sigma u)\sin(\sigma v)
\right]
%
Z_{\pm\frac{1}{2}}^{(j)}
\left[\sqrt{\lambda^2+4q}\ \cos(\sigma u)\cos(\sigma v) \right],
\end{array} 
\end{eqnarray}
where the Bessel functions can be expressed in terms of 
elementary functions since \cite{gradshteyn}
\begin{eqnarray}\label{bessel-1/2}
\begin{array}{l}
\left(
\begin{array}{l}
J_{{1}/{2}}(x)=Y_{-1/2}(x)\vspace{2mm}\\
J_{-{1}/{2}}(x)=-Y_{1/2}(x)
\end{array}\right)
=\sqrt{\frac{2}{\pi x}}
\left(
\begin{array}{r}
 \sin{x}\vspace{2mm}\\
\cos{x}
\end{array}
\right),
\end{array}\qquad
%
\begin{array}{l}
\left(
\begin{array}{l}
H_{{1}/{2}}^{(1)}(x)=-iH_{-{1}/{2}}^{(1)}(x)\vspace{2mm}\\
H_{{1}/{2}}^{(2)}(x)=iH_{-{1}/{2}}^{(2)}(x)
\end{array}\right)
=\sqrt{\frac{2}{\pi x}}
\left(
\begin{array}{r}
 -ie^{ix}\vspace{2mm}\\
ie^{-ix}
\end{array}
\right).
\end{array}
\end{eqnarray}


\subsection*{C.2. Second group: products of Bessel
and hypergeometric functions}
The kernels given by products of 
Bessel and hypergeometric functions are written as
\begin{eqnarray}\label{CHE-ince-5}
G_{\ \pm}^{(i,j)}(z,t)=
\left[2\sqrt{q(z+t-z_0)}\right]^{1-{B_2}}
 Z_{\pm(2\lambda+1-B_2)}^{(i)}\left[ 2\sqrt{q(z+t-z_0)}\right] F
^{j}(\zeta),\qquad 
\left[
i=1,\cdots,4;\
j=1,\cdots,6\right] 
\end{eqnarray}
where $F^{j}$ denote the hypergeometric functions 
written in Eqs. (\ref{1-hipergeometricas}-\ref{6-hipergeometricas}). 
We can show that the transformations
$\mathscr{R}_i$ simply rearrange the previous kernels.

By using properties (\ref{J}) of the confluent 
hypergeometric functions, 
the above kernels may be obtained by applying the
Whittaker-Ince limit to the kernels (\ref{quinto}) 
of the CHE. 
To derive the kernels directly, we note that the substitutions
\begin{eqnarray}\label{usar1}
\xi=2\sqrt{q(z+t-z_0)},\qquad 
\zeta=\frac{zt}{z_0(z+t-z_0)},\qquad
G(z,t)=\xi^{1-B_2}\ H(\xi,\zeta),
\end{eqnarray}
transform the second Eq. (\ref{kernel-WI}) into
\begin{eqnarray} \label{ince-segundo}
\xi^2\frac{\partial^2H}{\partial\xi^2}+{\xi}
\frac{\partial H}{\partial\xi}+\left[\xi^2-(1-B_2)^2\right]H+
4\zeta(1-\zeta)\frac{\partial^2H}{\partial\zeta^2}+
4\left[-\frac{B_1}{z_0}-B_2\zeta\right]\frac{\partial H}{\partial\zeta}=0.
\end{eqnarray}
The separation of variables
\begin{eqnarray}\label{usar2}
H(\xi,\zeta)= X(\xi)\ Y(\zeta)\ \Rightarrow\ G(z,t)=\xi^{1-B_2}\ X(\xi)\ Y(\zeta),
\end{eqnarray}
leads to the following Bessel and hypergeometric equations, respectively,  
\begin{eqnarray}\label{usar3}
\begin{array}{l}
 \xi^2\frac{d^2X}{d\xi^2}+
\xi\frac{dX}{d\xi}+\left[\xi^2-
 \left(2\lambda+1- B_2\right)^2\right]X=0,\qquad
\zeta(1-\zeta)\frac{d^2Y}{d\zeta^2}
 +\left[-\frac{B_1}{z_0}-B_2\zeta\right] \frac{dY}{d\zeta}-
 \lambda(B_2-\lambda-1)Y=0,
\end{array}
\end{eqnarray}
where we have denoted the constant of separation by 
$4\lambda(\lambda+1-B_2)$.
Using Eqs. (\ref{usar1}-\ref{usar3}) we obtain the kernels (\ref{CHE-ince-5}).

To show that the transformations
$\mathscr{R}_i$ do not produce new kernels we use the fact
that the constants of separation are arbitrary. For example, since
\begin{eqnarray*}
\begin{array}{l}\mathscr{R}_1G_{ \ \pm}^{(i,1)}(z,t)\end{array}&\propto& \begin{array}{l}\left[\frac{zt}{z+t-z_0}
\right]^{1+\frac{B_1}{z_0}}
\left[2\sqrt{q(z+t-z_0)}\right]^{1-{B_2}}\times\end{array}\\
%
%
%
&\times& \begin{array}{l}
Z_{\pm\left(2\lambda_1-1-_2-\frac{2B_1}{z_0}\right)}^{(i)}\left[ 2\sqrt{q(z+t-z_0)}\right]
F\left[\lambda_1,1+B_2+\frac{2B_1}{z_0}-\lambda_1;2+\frac{B_1}{z_0};
\frac{zt}{z_0(z+t-z_0)}\right],
\end{array}
\end{eqnarray*}
by taking $\lambda_1=\lambda+1+(B_1/z_0)$ we find 
that the right-hand side
is $G_{\ \pm}^{(i,2)}(z,t)$. Analogously,
\begin{eqnarray*}
\mathscr{R}_2G_{\ \pm}^{(i,1)}(z,t)&\propto&\begin{array}{l}\left[(z-z_0)(t-z_0) 
\right]^{1-B_2-\frac{B_1}{z_0}}
\left[2\sqrt{q(z+t-z_0)}\right]^{{B_2}+\frac{2B_1}{z_0}-1}\end{array}
\\
&\times& \begin{array}{l}Z_{\pm\left(2\lambda_2-1+B_2+\frac{2B_1}{z_0}\right)}^{(i)}\left[ 2\sqrt{q(z+t-z_0)}\right]
%
F\left[\lambda_2,1-B_2-\frac{2B_1}{z_0}-\lambda_2;-\frac{B_1}{z_0};
\frac{zt}{z_0(z+t-z_0)}\right].
\end{array}
\end{eqnarray*}
Putting $\lambda_2=\lambda+1-B_2-({B_1}/{z_0})$ and using
Eq. (\ref{euler-1}), we find that the right-hand side is proportional
to $G_{\ \pm}^{(i,1)}$.

For the Mathieu equation, whenever appropriate we use
the relations \cite{gradshteyn} 
\begin{eqnarray*}\begin{array}{l}
F\left(\frac{\nu}{2},-\frac{\nu}{2};\frac{1}{2};y^2 \right)=
\cos(\nu\arcsin{y}),\qquad
F\left(\frac{1+\nu}{2},\frac{1-\nu}{2};\frac{3}{2};y^2 \right)=
\frac{\sin(\nu\arcsin{y})}{\nu y}\end{array}
\end{eqnarray*}
with $\nu=2\lambda$. Then, the kernels (\ref{CHE-ince-5})  
are rewritten as 
%
\begin{equation}\label{Mc1}
\begin{array}{l}
G_{\ \pm}^{(i,1)}(u,v)=\cos\left[2\lambda\arcsin \frac{\sqrt{2}
\cos(\sigma u)\cos(\sigma v)}
{\sqrt{\cos(2\sigma u)+\cos(2\sigma v)}}\right]
Z_{\pm2\lambda}^{(j)}\left[ k\sqrt{2\cos(2\sigma u)+2\cos(2\sigma v)}\right],\end{array}
\end{equation}
\begin{equation}\label{Mc2-0}\begin{array}{l}
G_{\ \pm}^{(i,2)}(u,v)=\sin\left[ 2\lambda\arcsin\frac{\sqrt{2}\cos(\sigma u)\cos(\sigma v)}
{\sqrt{\cos(2\sigma u)+\cos(2\sigma v)}}\right]
Z_{\pm 2\lambda}^{(i)}\left[ k\sqrt{2\cos(2\sigma u)+2\cos(2\sigma v)}\right],\end{array}
\end{equation}
\begin{eqnarray}\label{Mc3}\begin{array}{l}
G_{\ \pm}^{(i,3)}(u,v)=\cos\left[2\lambda\arcsin \frac{i\sqrt{2}
\sin(\sigma u)\sin(\sigma v)}
{\sqrt{\cos(2\sigma u)+\cos(2\sigma v)}}\right]
Z_{\pm 2\lambda}^{(i)}\left[ k\sqrt{2\cos(2\sigma u)+2\cos(2\sigma v)}\right],\end{array}
\end{eqnarray}
\begin{eqnarray}\label{Mc4}\begin{array}{l}
G_{\ \pm}^{(i,4)}(u,v)=\sin\left[ 2\lambda\arcsin\frac{i\sqrt{2}\sin(\sigma u)\sin(\sigma v)}
{\sqrt{\cos(2\sigma u)+\cos(2\sigma v)}}\right]
Z_{\pm 2\lambda}^{(i)}\left[ k\sqrt{2\cos(2\sigma u)+2\cos(2\sigma v)}\right],\end{array}
\end{eqnarray}
\begin{eqnarray}\label{Mc5}\begin{array}{l}
G_{\ \pm}^{(i,5)}(u,v)=\left[
\frac{\cos^2(\sigma u)\cos^2(\sigma v)}
{\cos(2\sigma u)+\cos(2\sigma v)}\right]^{-\lambda}
F\left[\lambda,\frac{1}{2}+\lambda;1+2\lambda;
\frac{\cos(2\sigma u)+\cos(2\sigma v)}{2\cos^2(\sigma u) \cos^2(\sigma v)}
 \right] \end{array}
%
Z_{\pm 2\lambda}^{(i)}
\left[ k\sqrt{2\cos(2\sigma u)+2\cos(2\sigma v)}\right] ,
\end{eqnarray}
\begin{eqnarray}\label{Mc6}\begin{array}{l}
G_{\ \pm}^{(i,6)}(u,v)=\left[
\frac{\cos^2(\sigma u)\cos^2(\sigma v)}
{\cos(2\sigma u)+\cos(2\sigma v)}\right]^{\lambda}
F\left[-\lambda,\frac{1}{2}-\lambda;1-2\lambda;
\frac{\cos(2\sigma u)+\cos(2\sigma v)}{2\cos^2(\sigma u) \cos^2(\sigma v)} \right] \end{array}
%
Z_{\pm 2\lambda}^{(i)}
\left[ k\sqrt{2\cos(2\sigma u)+2\cos(2\sigma v)}\right].
\end{eqnarray}
The kernels (\ref{Mc1}-\ref{Mc4}) are equivalent
to the ones given on pp. 190 and 191 of McLachlan
\cite{McLachlan}, but we have not found 
$G_{\ \pm}^{(i,5)}$ and $G_{\ \pm}^{(i,6)}$ in the literature. 

%
%
\subsection*{C.3. Third group: Bessel functions }
Up to a multiplicative constant, a initial set of kernels  
given by Bessel functions  is 
\begin{eqnarray}\label{G1-ince}
\begin{array}{l}
G_{1,\pm}^{(i)}(z,t)=\left[\sqrt{zt} \right]^{1+\frac{B_1}{z_0}}\
Z_{ \pm\left(1+\frac{B_1}{z_0}\right) }^{(i)}\left[ 2\sqrt{\frac{qzt}{z_0}}\right].
\end{array}
\end{eqnarray}
These kernels are obtained by supposing that $H(\xi,\zeta)$
depends only on $\zeta$ in Eq. (\ref{AGA}). Then, the substitution
\begin{eqnarray*}
G(z,t)=H(\xi,\zeta)=
\zeta^{1+\frac{B_1}{z_0}}\ Y(\zeta),\qquad \zeta=\sqrt{{zt}/{z_0}}
\end{eqnarray*}
leads to 
\begin{eqnarray*}
\begin{array}{l}
\zeta^2\frac{d^2Y}{d\zeta^2}
 +\zeta \frac{dY}{d\zeta}+
 \left[4q\zeta^2-
 \left(1+\frac{B_1}{z_0}\right)^2\right]Y=0,
\end{array}
\end{eqnarray*}
which is the Bessel equation (\ref{bessel}) with argument 
$y=2\sqrt{q}\zeta=2\sqrt{qzt/z_0}$ and 
order $\alpha=\pm[1+(B_1/z_0)]$. { In this manner, we find (\ref {G1-ince}).  
The remaining sets are obtained
by using the transformations (\ref{transformacao-K2}) as}
\begin{eqnarray*}\label{sequence}
G_{2,\pm}^{(i)}(z,t)=\mathscr{R}_2G_{1,\pm}^{(i)}(z,t),\qquad
G_{3,\pm}^{(i)}(z,t)=\mathscr{R}_3G_{2,\pm}^{(i)}(z,t),\qquad
G_{4,\pm}^{(i)}(z,t)=\mathscr{R}_1G_{3,\pm}^{(i)}(z,t).
\end{eqnarray*}
%
%
Thence,
\begin{eqnarray}
&&\begin{array}{l}
G_{2,\pm}^{(i)}(z,t)=\left[\sqrt{{zt}} \right] ^{ 1+\frac{B_1}{z_0} }
\left[(z-z_0)(t-z_0) \right]^{1-B_2-\frac{B_1}{z_0}}
Z_{\pm\left( 1+\frac{B_1}{z_0}\right) }^{(i)}\left[ 2\sqrt{\frac{qzt}{z_0}}\right],%
\end{array}\\
%
&&\begin{array}{l}
G_{3,\pm}^{(i)}(z,t)=[zt]^{1+\frac{B_1}{z_0}}
\left[\sqrt{(z-z_0)(t-z_0)} \right] ^{1-B_2-\frac{B_1}{z_0}}
Z_{\pm\left(1-B_2-\frac{B_1}{z_0}\right) }^{(i)}
\left[ 2\sqrt{-\frac{q}{z_0}(z-z_0)(t-z_0)}\right],
\end{array}\\
%
&&\begin{array}{l}
G_{4,\pm}^{(i)}(z,t)=
\left[\sqrt{(z-z_0)(t-z_0)} \right]^{1-B_2-\frac{B_1}{z_0}}
Z_{\pm\left(1-B_2-\frac{B_1}{z_0}\right) }^{(i)}
\left[2\sqrt{-\frac{q}{z_0}(z-z_0)(t-z_0)}\right].
\end{array}\quad
\end{eqnarray}
This group of kernels can as well be generated 
by applying the Whittaker-Ince 
limit to kernels of the CHE given by hypergeometric functions
in section 2.4. 

For the Mathieu equation, up to constant factors, 
we find 
\begin{eqnarray}
&&\begin{array}{l}
G_{1,\pm}^{(i)}(u,v)=\sqrt{\cos(\sigma u) \cos(\sigma v)}\ 
Z_{\pm{1}/{2}}^{(i)}
\left[ 2k \cos(\sigma u) \cos(\sigma v)\right],\end{array}\\ 
%
%
&&\begin{array}{l}
G_{2,\pm}^{(i)}(u,v)=\sin(\sigma u)\sin(\sigma v)
\sqrt{\cos(\sigma v) \cos(\sigma v)}
Z_{\pm{1}/{2}}^{(i)}
\left[ 2k \cos(\sigma u) \cos(\sigma v)\right],\end{array}\\
%
%
&&\begin{array}{l}
G_{3,\pm}^{(i)}(u,v)=\cos(\sigma u)\cos(\sigma v)
\sqrt{\sin(\sigma u) \sin(\sigma v)}
Z_{\pm{1}/{2}}^{(i)}
\left[ 2ik \sin(\sigma u) \sin(\sigma v)\right],\end{array}\\
%
&&\begin{array}{l}
G_{4,\pm}^{(i)}(u,v)=
\sqrt{\sin(\sigma u) \sin(\sigma v)}\
Z_{\pm{1}/{2}}^{(i)}
\left[ 2ik \sin(\sigma u) \sin(\sigma v)\right]. \end{array}
\end{eqnarray}
Kernels of this type 
have been used to generate solutions
in series of Bessel functions out of Fourier-like expansions
\cite{arscott,McLachlan} (that is, from
solutions in series of trigonometric or hyperbolic functions).

\subsection*{C.4. Fourth group: Bessel functions again }
Another group, {given by products of elementary and Bessel functions, is 
generated} from the initial set of the kernels
\begin{eqnarray}\label{wi-che1}
{G}_{1,\pm}^{(i)}(z,t)=\left[2 \sqrt{q(z+t-z_0)}\right]^{1-B_2}
%
Z_{\pm(B_2-1)}^{(i)}\left[ 2\sqrt{q(z+t-z_0)}\right],
\end{eqnarray}
which result when $H(\xi,\zeta)$ depends only on $\xi$ in Eq. (\ref{ince-segundo}),
that is, when $H(\xi,\zeta)=X(\xi)$. In effect, in this case  
we find 
\begin{eqnarray*}
G(z,t)\stackrel{(\ref{usar1})}{=}\xi^{1-B_2}\ X(\xi),\qquad \xi=2\sqrt{q(z+t-z_0)}
\end{eqnarray*}
where  $X(\xi)$ satisfies the Bessel equation
\begin{eqnarray*}
\begin{array}{l}
\displaystyle \xi^2\frac{d^2X}{d\xi^2}+
\xi\frac{dX}{d\xi}+\left[\xi^2-
 \left(B_2-1\right)^2\right]X=0
\end{array}
\end{eqnarray*}
Taking $X(\xi)=Z_{\pm(B_2-1)}^{(i)}(\xi)$ we obtain 
the kernels (\ref{wi-che1}). Then, by using the 
transformations $\mathscr{R}_1$ and $\mathscr{R}_2$ 
as ($\mathscr{R}_3$ is ineffective)
\begin{eqnarray*}
{G}_{2,\pm}^{(i)}(z,t)=\mathscr{R}_1{G}_{1,\pm}^{(i)}(z,t),\qquad
{G}_{3,\pm}^{(i)}(z,t)=\mathscr{R}_2{G}_{2,\pm}^{(i)}(z,t),
\qquad{G}_{4,\pm}^{(i)}(z,t)=\mathscr{R}_1{G}_{3,\pm}^{(i)}(z,t).
\end{eqnarray*}
{we find that} the other sets are
\begin{equation}\label{wi-che2}
{G}_{2,\pm}^{(i)}(z,t)=\left[zt\right]^{1+\frac{B_1}{z_0}}
\left[ 2\sqrt{q(z+t-z_0)}\right]^{-1-B_2-\frac{2B_1}{z_0}}
%
Z_{\pm\left(1+B_2+\frac{2B_1}{z_0}\right)}^{(i)}\left[ 2\sqrt{q(z+t-z_0)}\right],
\end{equation}
\begin{eqnarray}\label{wi-che3}\begin{array}{l}
{G}_{3,\pm}^{(i)}(z,t)=\left[zt\right]^{1+\frac{B_1}{z_0}}
\left[(z-z_0)(t-z_0)\right]^{1-B_2-
\frac{B_1}{z_0}} 
%
\left[ 2\sqrt{q(z+t-z_0)}\right] ^{B_2-3}
%
Z_{\pm(3-B_2)}^{(i)}\left[2\sqrt{q(z+t-z_0)}\right],\end{array}
\end{eqnarray}

\begin{eqnarray}\label{wi-che4}\begin{array}{l}
{G}_{4,\pm}^{(i)}(z,t)=
\left[(z-z_0)(t-z_0)\right]^{1-B_2-\frac{B_1}{z_0}}
%
\left[ 2\sqrt{q(z+t-z_0)}\right]^{B_2+\frac{2B_1}{z_0}-1}
%
Z_{\pm\left(1-B_2-\frac{2B_1}{z_0}\right)}^{(i)}\left[ 2\sqrt{q(z+t-z_0)}\right].\end{array}
\end{eqnarray}

The above kernels can as well be generated by applying the Whittaker-Ince
limit to kernels of the CHE given by confluent hypergeometric functions
in section 2.5. 
On the other  hand, these kernels are instances of the 
kernels $G_{\ \pm}^{(i,j)}(z,t)$ given in (\ref{CHE-ince-5}), 
corresponding to four choices of $\lambda$ 
which permit to write the hypergeometric functions
$F^{(j)}(\zeta)$ as elementary functions.  Indeed, up to
constant factors we find that: (i) the kernels $G_{1,\pm}^{(i)}$
correspond to $\lambda=0$ in $G_{\ \pm}^{(i,1)}$, 
$G_{\ \pm}^{(i,3)}$ and $G_{\ \pm}^{(i,5)}$;  $G_{2,\pm}^{(i)}$
correspond to $\lambda=-1-(B_1/z_0)$ in $G_{\ \pm}^{(i,2)}$;
 $G_{3,\pm}^{(i)}$
correspond to $\lambda=B_2-2$ in $G_{\ \pm}^{(i,5)}$;  
$G_{4,\pm}^{(i)}$
correspond to $\lambda=B_2+(B_1/z_0)-1$ in $G_{\ \pm}^{(i,5)}$.

Notice that $Z_{-\ell}^{(i)}(x)=(-1)^{\ell}Z_{\ell}^{(i)}(x)$
if $\ell$ is integer. So, up to multiplicative constants, for the Mathieu 
equation the previous kernels read 
\begin{eqnarray}\label{GG-mathieu}
&&\begin{array}{l}
{G}_{\ 1}^{(i)}(u,v)=Z_{\ 0}^{(i)}\left[ k\sqrt{2\cos(2\sigma u)+
2\cos(2\sigma v)}\right],\end{array}\\
%
&&\begin{array}{l}
{G}_{\ 2}^{(i)}(u,v)=\frac {\cos(\sigma u) \cos(\sigma v)}
{\sqrt{\cos(2\sigma u)+
\cos(2\sigma v)}}
%
Z_{\ 1}^{(i)}\left[k\sqrt{2\cos(2\sigma u)+
2\cos(2\sigma v)}\right],
\end{array}\\
&&\begin{array}{l}
{G}_{\ 3}^{(i)}(u,v)=\frac {\sin(2\sigma u) \sin(2\sigma v)}
{\cos(2\sigma u)+
\cos(2\sigma v)}
%
Z_{\ 2}^{(i)}\left[ k\sqrt{2\cos(2\sigma u)+
2\cos(2\sigma v)}\right],\end{array}\\
&&\begin{array}{l}
{G}_{\ 4}^{(i)}(u,v)=\displaystyle
\frac {\sin(\sigma u) \sin(\sigma v)}
{\sqrt{\cos(2\sigma u)+
\cos(2\sigma v)}}
%
Z_{\ 1}^{(i)}\left[k \sqrt{2\cos(2\sigma u)+
2\cos(2\sigma v)}\right].\end{array}
\end{eqnarray}
These kernels for the Mathieu equation 
are connected with 
particular values of $\lambda$ in the kernels 
(\ref{Mc1}-\ref{Mc6}). In fact: (i)  for ${G}_{\ 1}^{(i)}$
we take $\lambda=0$ in (\ref{Mc1}),
(\ref{Mc3}), (\ref{Mc5}) or (\ref{Mc6});
(ii) for ${G}_{\ 2}^{(i)}$, $\lambda=1/2$ in (\ref{Mc2-0});
(iii) for ${G}_{\ 3}^{(i)}$, $\lambda=-1$ in (\ref{Mc5})  
or $\lambda=1$ in (\ref{Mc6});  (iv)  
for ${G}_{\ 4}^{(i)}$, $\lambda=1/2$ in (\ref{Mc4}).

\subsection*{C.5. Fifth group: hypergeometric functions}
By taking $2\lambda+1-B_2=1/2$ in (\ref{CHE-ince-5}), we find
\begin{eqnarray}
G_{\ \pm}^{(i,j)}(z,t)=
\left[2\sqrt{q(z+t-z_0)}\right]^{1-{B_2}}
 Z_{\pm\frac{1}{2}}^{(i)}\left( 2\sqrt{q(z+t-z_0)}\right) 
 \Big[F^{(j)}(\zeta)\Big]_{\lambda=\frac{B_2}{2}-\frac{1}{4}},\qquad 
\left[
i=1,\cdots,4;\
j=1,\cdots,6\right],
\end{eqnarray}
where $F^{(j)}$ denote the hypergeometric functions 
written in Eqs. (\ref{1-hipergeometricas}-\ref{6-hipergeometricas}),
whereas $Z_{\pm1/2}^{(i)}$ are given by the elementary functions
(\ref{bessel-1/2}). For the Mathieu equation, the explicit form
of the kernels is obtained
by putting $\lambda=1/4$ in Eqs. (\ref{Mc1}-\ref{Mc6}).

\subsection*{C.6. Power series and series of Bessel functions for the RCHE}
In the Whittaker-Ince limit the power series solution (\ref{barber}) becomes
\begin{eqnarray}\label{barber-WI}
U_1^{\text{baber}}(z)=\displaystyle \sum_{n=0}^{\infty}a_{n}^{1}
(z-z_{0})^{n},\qquad (|z|=\text{finite})
\end{eqnarray}
where the series coefficients now satisfy ($a_{-1}^{1}=0$)
\begin{eqnarray}\label{WI-recorrencia}\begin{array}{l}
 z_{0}\left(n+B_{2}+\frac{B_{1}}{z_{0}}\right)
\left(n+1\right)a_{n+1}^{1}+
\big[ n\big(n+B_{2}-1\big)
+B_{3}\big]a_{n}^{1}
+qa_{n-1}^{1}=0.\end{array}
\end{eqnarray}
In the following we find that, if
\begin{equation}\label{if-then}
\text{if } \text{Re}\left({B_1}/{z_0} \right) <0 \ \ \text{ and }\ \ 
\text{Re}\left[n+B_2+({B_1}/{z_0}) \right]>0
\end{equation}
then, by means of an integral transformation, $U^{\text{baber}}_1(z)$
generates a known expansion $U_1(z)$ in series of 
Bessel functions given by \cite{lea-2}
\begin{equation}\label{besselabove}
U_{{1}}(z)=\sum_{n=0}^{\infty}
\begin{array}{l}
(-1)^nc_{n}^{1}
\big(\sqrt{qz}\big)^{-(n+B_2-1)} J_{n+B_2-1} \big(2\sqrt{qz}\big),
\end{array}
\end{equation}
where the recurrence relations for $c_{n}^{1}$
are obtained by writing
\begin{eqnarray}
\begin{array}{l}
c_{n}^{1}=C
(z_0)^ n\Gamma\left( n+B_2+\frac{B_1}{z_0}\right)
a_{n}^{1},
\end{array}\nonumber
\end{eqnarray}
$C$ being a constant independent of $n$. Thus, 
\begin{eqnarray}\begin{array}{l}
\left(n+1\right)c_{n+1}^{1}+
\big[ n\big(n+B_{2}-1\big)+B_{3}\big]c_{n}^{1}
+qz_0\left(n +B_2+\frac{B_1}{z_0}-1\right)c_{n-1}^{1}=0.\end{array}
\end{eqnarray}
%

In fact, by inserting $U_1(t)=U^{\text{baber}}_1(t)$  and the kernel $G(z,t)=G_{1,-}^{(1)}(z,t)$ 
given in (\ref{G1-ince})
into (\ref{integral}), and taking $t_1=0$ and $t_2=z_0$,
we find
\begin{eqnarray*}
\mathcal{U}(z)=
 z^{\frac{1}{2}+\frac{B_1}{2z_0}}\displaystyle\sum_{n=0}^{\infty}
a_{n}^{(1)}\int_{0}^{z_0}dt\bigg[t^{-\frac{1}{2}-\frac{B_1}{2z_0}}
%
 (t-z_0)^{n+B_2+\frac{B_1}{z_0}-1}J_{-1-\frac{B_1}{z_0}}
\left(2\sqrt{qz{t}/{z_0}}\  \right)
\bigg].
\end{eqnarray*}
By using the integral  \cite{erdelyi}
\begin{eqnarray*}\label{int-2'}
\int_{0}^{y}x^{\frac{1}{2}\nu}(y-x)^{\mu-1}
J_{\nu}(a\sqrt{x}\ )dx
%
=2^\mu y^{\frac{1}{2}(\mu+\nu)}\Gamma(\mu)a^{-\mu}
J_{\mu+\nu}(\sqrt{y}a),\qquad 
\left[  \text{Re}(\mu)>1,\quad  \text{Re}(\nu)>-1\right] ,
\nonumber
\end{eqnarray*}
we get $\mathcal{U}=U_1$, where $U_1$ is given in (\ref{besselabove}). 
On the other side, 
since
\[\frac{d}{dy}\left[y^{-\nu}J_{\nu}(y)\right]=-
y^{-\nu}J_{\nu+1}(y),\]
the bilinear concomitant (\ref{concomitant})
takes the form
\begin{eqnarray*}
P_{1}(z,t)&=&-z^{\frac{1}{2}+\frac{B_1}{2z_0}}\ t^{\frac{1}{2}-
\frac{B_1}{2z_0}}\ 
(t-z_0)^{B_2+\frac{B_1}{z_0}}
\\
&\times&\bigg[ \sqrt{\frac{qz}{tz_0}}J_{-\frac{B_1}{z_0}}
\left(2\sqrt{{qzt}/{z_0}}\right) \sum_{n=0}^{\infty}a_{n}^{1}
(t-z_{0})^{n}+ J_{-1-\frac{B_1}{z_0}}
\left(2\sqrt{{qzt}/{z_0}}\right)\sum_{n=1}^{\infty}na_{n}^{1}
(t-z_{0})^{n-1}\bigg].
\end{eqnarray*}
Thence, the conditions (\ref{if-then})
assure that $P_1(z,0)=P_1(z,z_0)=0$.
%

%
\section*{Appendice D. Kernels for the double-confluent Heun equation (DHE)}
\protect\label{D}
\setcounter{equation}{0}
\renewcommand{\theequation}{D.\arabic{equation}}
As $z_0\rightarrow 0$ the CHE (\ref{gswe}) 
reduces to the double-confluent Heun equation (DHE)
\begin{equation}
\label{dche}
\left[L_z+B_3\right]U=z^2\frac{d^{2}U}{dz^{2}}+[B_{1}+B_{2}z]\frac{dU}{dz}
+\left[\omega^{2}z^2-2\eta\omega z+B_3\right]U=0,
\end{equation}
where $z=0$ and $z=\infty$ are both irregular points. In this appendix D:
  \begin{itemize}
  \itemsep-3pt
  \item
 {initially}  we get the substitutions of variables which preserve the form of the
  equation for the kernels of the DHE;
 \item 
{in  D.1 we find kernels containing products of two confluent hypergeometric functions and
 presenting an arbitrary constant of separation $\lambda$; they may be derived 
 by applying the Leaver limit ($z_0\to 0$) to kernels of the CHE 
 given by } products of hypergeometric and confluent hypergeometric
 functions (section 2.3);
 \item
{in  D.2}, by  taking appropriate values for $\lambda$,  we get 
  kernels given products of elementary and confluent hypergeometric functions;
  \item 
 {in  D.3} we  obtain kernels given by elementary functions; {these kernels 
 cannot be derived as limits of known kernels of the CHE.}
  \end{itemize}

Since $\lim_{x\rightarrow 0}(1+x)^{1/x}=e$, {when $z_0\to 0$ }
the integral (\ref{integral}) assumes the form
\begin{eqnarray}
\label{integral-dche}
\mathcal{U}(z)=\int_{t_{1}}^{t_{2}}K(z,t)U(t)dt= 
\int_{t_{1}}^{t_{2}}t^{B_2-2}
e^{-\frac{B_1}{t}}G(z,t)U(t)dt,\qquad K(z,t)=w(z,t)G(z,t)=t^{B_2-2}
e^{-\frac{B_1}{t}}G(z,t), 
%
\end{eqnarray}
where $G(z,t)$ is determined from the equation
\begin{equation}\label{Lz1-dche}
 %
%
\begin{array}{l}
\left[z^2\frac{\partial^{2}}{\partial z^{2}}+
\left(B_{1}+B_{2}z\right)
\frac{\partial}{\partial z}
+\left(\omega^{2}z^2-
2\eta\omega z\right)\right]G
%
%
-\left[t^2\frac{\partial^{2}}{\partial t^{2}}+
\left(B_{1}+B_{2}t\right)
\frac{\partial}{\partial t}
+\left(\omega^{2}t^2-
2\eta\omega t\right)\right]G=0.
\end{array}
\end{equation}
Similarly, the expression (\ref{concomitant}) for the bilinear concomitant now reads
\begin{eqnarray}\label{M-dche}
\begin{array}{l}
P(z,t)=
t^2\left[U(t)\frac{\partial K(z,t)}{\partial t}
-K(z,t)\frac{d U(t)}{d t}\right]+
\left[\left(2-B_{2}\right)t -B_{1}\right]U(t)K(z,t)=t^{B_2}\ e^{-\frac{B_1}{t}}
\left[U(t)\frac{\partial G(z,t)}{\partial t}-G(z,t)\frac{d U(t)}{d t}\right],
\end{array}
\end{eqnarray}
{
In general the solutions $U(t)$ of the DHE and RDHE converge in a domain 
including only one of the singular points, $0$ or $\infty$.  
For this reason we must avoid using intervals of integration extending from 
$t_1=0$ to $t_2=\infty$. In reference \cite{eu3} this requirement was
satisfied  by using endpoints $t_i$ which depend
on the variable $z$, that is, 
\begin{eqnarray} 
\mathcal{U}(z) =\int_{t_{1}(z)}^{t_{2}(z)}K(z,t)U(t)dt
=\int_{t_{1}(z)}^{t_{2}(z)}t^{B_2-2}
e^{-\frac{B_1}{t}}G(z,t)U(t)dt.
\end{eqnarray}
Then, {the formula
\begin{eqnarray}
\label{derivada}
\frac{d}{dz}\int_{t_{1}(z)}^{t_{2}(z)}F(z,t)dt=
\int_{t_{1}(z)}^{t_{2}(z)}\frac{\partial F(z,t)}{\partial z}dt+
F(z,t_{2})\frac{dt_{2}}{dz}
-F(z,t_{1})\frac{dt_{1}}{dz}
\end{eqnarray}
implies that Eq. (\ref{condition1}) must be replaced by}
\begin{eqnarray}
\label{condition2}
[L_{z}+B_{3}]\mathcal{U}(z)=
P(z,t_{2})+Q(z,t_1)-[P(z,t_{1})+Q(z,t_2)],
\end{eqnarray}
%
%
%
where ($i=1,2$)
\begin{eqnarray}
\label{Q}
Q(z,t_{i})&=&\begin{array}{l}
\left[z^2\frac{d^2t_{i}}{dz^2}+\left(B_{1}+
B_{2}z\right)\frac{d t_{i}}{d z}\right]U(t_{i})K(z,t_{i})\end{array}
\nonumber\\
&+&\begin{array}{l}
z^2U(t_{i})\left[\frac{\partial K(z,t_{i})}{\partial
t_{i}}\left(\frac{dt_{i}}{dz}\right)^2+2\frac{\partial K(z,t_{i})}
{\partial z}\frac{d t_{i}}{d z}
\right]+z^2\left(\frac{dt_{i}}{dz}\right)^2\frac{d U(t_{i})}{d t_{i}}K(z,t_{i}).\end{array}
\end{eqnarray}
%
%
%
%
%
%
%
From Eq. (\ref{condition2}) we see that the condition $P(z,t_2)=P(z,t_1)$ must be replaced by 
$P(z,t_2)+Q(z,t_2)=P(z,t_1)+Q(z,t_1)$.}

Since the differential operators in
Eqs. (\ref{dche}) and (\ref{Lz1-dche}) 
have the same functional form, from the transformations
of the DHE (\ref{dche}) we get the transformations for its kernels. In fact,
if $U(z)=U(B_1,B_2,B_3;\omega,\eta;z)$ denotes a solution of the DHE,
the substitutions which preserve the form of the equation 
are represented by  
the transformations $t_{1}$, $t_{2}$ and $t_{3}$ \cite{decarreau1,eu-eu} 
\begin{eqnarray}\label{d5}
\begin{array}{l}
t_{1}U(z)=e^{\frac{B_{1}}{z}}z^{2-B_{2}}U(-{B}_{1},4-{B}_{2},
{B}_{3}+2-B_2; \omega,\eta;z),\vspace{3mm}\\
t_{2}U(z)=e^{i\omega z+\frac{B_1}{2z}}z^{-i\eta-\frac{B_2}{2}}
U\left(B_{1}^{'},B_{2}^{'},B_{3}^{'};\omega^{'},\eta^{'};\vartheta=\frac{iB_1}{2z}\right),\vspace{3mm}\\
t_{3}U(z)=U({B}_{1},{B}_{2},
{B}_{3}; -\omega,-\eta;z),
\end{array}
\end{eqnarray}
where
\begin{eqnarray*}\begin{array}{l}
B_{1}^{'}=\omega B_{1},  \quad
B_{2}^{'}=2+2i\eta,  \quad
B_{3}^{'}=B_{3}-\left(\frac{B_{2}}{2}+i\eta\right)
\left(\frac{B_{2}}{2}-i\eta-1\right),
\quad
\omega^{'}=1, \quad
i\eta^{'}=\frac{B_{2}}{2}-1, 
\end{array}
\end{eqnarray*}
%
From (\ref{d5}) we obtain the transformations for the kernels of 
the DHE, namely,
\begin{eqnarray}\label{transfomacao-r}
\begin{array}{l}
r_{1}G(z,t)=e^{\frac{B_{1}}{z}+\frac{B_{1}}{t}}(zt)^{2-B_{2}}U(-{B}_{1},4-{B}_{2},; \omega,\eta;z,t),\vspace{3mm}\\
r_{2}G(z,t)=e^{i\omega (z+t)+\frac{B_1}{2z}+\frac{B_1}{2t}}(zt)^{-i\eta-\frac{B_2}{2}}
U\left(B_{1}^{'},B_{2}^{'};\omega^{'},\eta^{'};\frac{iB_1}{2z},\frac{iB_1}{2t}\right),\vspace{3mm}\\
r_{3}G(z,t)=U({B}_{1},{B}_{2}; -\omega,-\eta;z,t).
\end{array}
\end{eqnarray}
%


\subsection*{D.1. Kernels with products of two confluent hypergeometric functions} 
We write the kernels before explaining how they are obtained. 
A group of solutions for (\ref{Lz1-dche}) is given by the 16 kernels
\begin{eqnarray}\label{heunconfluente}
G^{(i,j)}(z,t)=
e^{-i\omega(z+t)}(zt)^{-\lambda}
\varphi^{i}(\xi)\;\bar{\varphi}^{j}(\zeta),\qquad [i,j=1,2,3,4]
\end{eqnarray}
where $\lambda$ is a constant of separation,
and $\varphi^{i}(\xi)$ and $\bar{\varphi}^{j}(\zeta)$ are
the confluent hypergeometric functions
(\ref{confluent1}) with the following arguments and parameters :
\begin{eqnarray}\label{parametros1}
\begin{array}{llll} 
\varphi^{i}(\xi):&\quad \xi=2i\omega(z+t),&\qquad
{a}=\frac{B_2}{2}-i\eta-\lambda,&\qquad
{c}=B_2-2\lambda;\vspace{2mm}\\
%
%
\bar{\varphi}^{j}(\zeta):&\quad \zeta=\frac{B_1(z+t)}{zt}, &\qquad
a= \lambda,&\qquad
c=2\lambda+2-B_2.
    \end{array}
\end{eqnarray}
The kernels given by regular confluent
hypergeometric functions are
\begin{eqnarray}
\begin{array}{l}
G^{(1,1)}(z,t)
=
e^{-i\omega(z+t)}\left[zt\right]^{-\lambda}
\Phi\left[\frac{B_2}{2}-i\eta-\lambda,B_2-2\lambda;
2i\omega (z+t)\right] 
\Phi\left[\lambda,2\lambda+2-B_2;
\frac{B_1 (z+t)}{zt}\right],
\end{array}
\end{eqnarray}
\begin{eqnarray}
G^{(1,2)}(z,t)
&=&\begin{array}{l}
e^{-i\omega(z+t)+\frac{B_1}{z}+\frac{B_1}{t}}\left[zt\right]^{\lambda+1-B_2}[z+t]^{B_2-1-2\lambda}
\Phi\left[\frac{B_2}{2}-i\eta-\lambda,B_2-2\lambda;
2i\omega (z+t)\right] \end{array}\nonumber\vspace{2mm}\\
&\times&\begin{array}{l}
\Phi\left[1-\lambda,B_2-2\lambda;-
\frac{B_1 (z+t)}{zt}\right],
\end{array}
\end{eqnarray}
\begin{eqnarray}
G^{(2,1)}(z,t)
& = &\begin{array}{l}
e^{i\omega(z+t)}\left[zt\right]^{-\lambda}[z+t]^{1+2\lambda-B_2}
\Phi\left[1+\lambda+i\eta-\frac{B_2}{2}-,2+2\lambda-B_2;
-2i\omega (z+t)\right] \end{array} \nonumber\vspace{2mm}\\
&\times& \begin{array}{l}\Phi\left[\lambda,2\lambda+2-B_2;
\frac{B_1 (z+t)}{zt}\right],
\end{array}
\end{eqnarray}
\begin{eqnarray}\label{g22}
G^{(2,2)}(z,t)
&=&\begin{array}{l}
e^{i\omega(z+t)+\frac{B_1}{z}+\frac{B_1}{t}}\left[zt\right]^{\lambda+1-B_2}
\Phi\left[1+i\eta+\lambda-\frac{B_2}{2},2+2\lambda-B_2;
-2i\omega (z+t)\right] \end{array}\nonumber\vspace{2mm}\\
&\times&\begin{array}{l}
\Phi\left[1-\lambda,B_2-2\lambda;-
\frac{B_1 (z+t)}{zt}\right].
\end{array}
\end{eqnarray}
The full group is obtained by replacing
one or both functions $\Phi$ by $\Psi$. By using these explicit forms of the kernels, 
we can show that the transformations $r_1$, $r_2$ and $r_3$
simply rearrange the kernels. For example, we find
\begin{eqnarray}
r_1G^{(i,j)}(z,t)=
e^{-i\omega(z+t)+\frac{B_1}{z_0}+\frac{B_1}{z_0}}\;[zt]^{2-B_2-\lambda_1}\;
\tilde{\varphi}^{i}(\xi)\;\tilde{\bar{\varphi}}^{j}(\zeta),\qquad [i,j=1,2,3,4]
\end{eqnarray}
where we have transformed $\lambda$ into $\lambda_1$,
and now $\tilde{\varphi}^{i}(\xi)$ and $\tilde{\bar{\varphi}}^{j}(\zeta)$ are
the confluent hypergeometric functions
(\ref{confluent1}) with the following arguments and parameters :
\begin{eqnarray*}
\begin{array}{llll} 
\tilde{\varphi}^{i}(\xi):&\quad \xi=2i\omega(z+t),&\qquad
{a}=2-i\eta-\lambda_1-\frac{B_2}{2},&\qquad
{c}=4-B_2-2\lambda_1;\vspace{2mm}\\
\tilde{\bar{\varphi}}^{j}(\zeta):&\quad \zeta=-\frac{B_1(z+t)}{zt}, &\qquad
a= \lambda_1,&\qquad
c=2\lambda_1-2+B_2.
    \end{array}
\end{eqnarray*}
In particular,
\begin{eqnarray*}
r_1G^{(1,1)}(z,t)&=&\begin{array}{l}
e^{-i\omega(z+t)+\frac{B_1}{z}+\frac{B_1}{t}}\left[zt\right]^{2-B_2-\lambda_1}
\Phi\left[1-i\eta-\lambda_1-\frac{B_2}{2},4-B_2-2\lambda_1;
2i\omega (z+t)\right] \end{array}\nonumber\vspace{2mm}\\
&\times&\begin{array}{l}
\Phi\left[\lambda_1,2\lambda_1-2+B_2;-
\frac{B_1 (z+t)}{zt}\right].
\end{array}
\end{eqnarray*}
By taking $\lambda_1=1-\lambda$ and using (\ref{kummer}), we see that the right-hand side
of the above equation is $G^{(2,2)}(z,t)$ given in (\ref{g22}). {Thus, in fact $r_1$
simply rearranges the kernels (\ref{heunconfluente}). The same is true of $r_2$ and $r_3$.}
%
%
%
%
%
%

The kernels (\ref{heunconfluente}) can be found by solving Eq. 
(\ref{Lz1-dche}), or by applying the limit when $z_0\to 0$
to the kernels (\ref{quinto}) of the CHE. The latter
procedure transforms the Gauss hypergeometric functions
given in Eqs. (\ref{1-hipergeometricas}-\ref{6-hipergeometricas})
into confluent hypergeometric functions due to the relations 
\cite{erdelii}
\begin{eqnarray}\label{limites-0}
\begin{array}{l}
\displaystyle\lim_{\mathrm{c}\rightarrow \infty}
F\left(\mathrm{a},\mathrm{b};\mathrm{c};1-\frac{\mathrm{c}}{u}\right)=
\lim_{\mathrm{c}\rightarrow \infty}
F\left(\mathrm{a},\mathrm{b};\mathrm{c};-\frac{\mathrm{c}}{u}\right)
=u^{\mathrm{a}}\Psi(\mathrm{a},\mathrm{a}+1-\mathrm{b};u),\qquad
\vspace{3mm}\\
\displaystyle\lim_{\mathrm{b}\rightarrow \infty}
F\left(\mathrm{a,b;c};\frac{u}{\mathrm{b}}\right)=\Phi(\mathrm
{a,c};u),\qquad \lim_{y\to\infty}\left(1+\frac{x}{y}\right)^y=e^x.
\end{array}
\end{eqnarray}
Thus, up to a multiplicative constant, we find
\begin{eqnarray}\label{limites}
\begin{array}{l} \displaystyle
\lim_{z_0\to 0}F^1=\lim_{z_0\to 0}F^3=
\begin{array}{l} [zt]^{-\lambda}[z+t]^{\lambda} \Psi\left(\lambda,2+2\lambda-B_2;
\frac{B_1(z+t)}{zt}\right)\end{array},\vspace{3mm}\\
\displaystyle
\lim_{z_0\to 0}F^2=\lim_{z_0\to 0} F^4= \begin{array}{l} e^{\frac{B_1}{z}+\frac{B_1}{t}}[zt]^{\lambda-1-B_2}[z+t]^{B_2+1-\lambda} 
\Psi\left(1-\lambda,B_2-2\lambda;
-\frac{B_1(z+t)}{zt}\right),\end{array}\vspace{3mm}\\
\displaystyle
\lim_{z_0\to 0}F^5=\begin{array}{l}  [zt]^{-\lambda}[z+t]^{\lambda} \Phi\left(\lambda,2+2\lambda-B_2;
\frac{B_1(z+t)}{zt}\right),\end{array}\vspace{3mm}\\
\displaystyle
\lim_{z_0\to 0}F^6=\begin{array}{l} e^{\frac{B_1}{z}+\frac{B_1}{t}}[zt]^{\lambda+1-B_2}[z+t]^{B_2-1-\lambda} 
\Phi\left(1-\lambda,B_2-2\lambda;-
\frac{B_1(z+t)}{zt}\right).\end{array}
\end{array}
\end{eqnarray}
{To get these limits, in some cases} we have to rewrite the functions (\ref{1-hipergeometricas}-\ref{6-hipergeometricas})
in a convenient form. For example, using (\ref{euler-1}) we find
\begin{eqnarray*}
\begin{array}{l} 
F^2=\frac{zt}{z+t-z_0}\left[\frac{(z-z_0)(t-z_0)}{z+t-z_0}\right]^{1-B_2}
\left[1-\frac{z_0}{zt}(z+t-z_0)\right]^{-\frac{B_1}{z_0}}F\left(1-\lambda,2+\lambda-B_2;2+\frac{B_1}{z_0};
\frac{zt}{z_0(z+t-z_0)}\right)
\end{array}
\end{eqnarray*}
by suppressing a multiplicative constant depending on $z_0$. After this, we use 
the limits (\ref{limites-0}). 

\subsection*{D.2. Kernels with one confluent hypergeometric function} 
For particular values of 
$\lambda$, the kernels (\ref{heunconfluente}) present only
one of the confluent hypergeometric functions (\ref{confluent1}). As an initial set we take
\begin{eqnarray}\label{kernel-DCHE-2}
{G}_{\  1}^{(i)}(z,t)=G^{(i,1)}(z,t)\big|_{\lambda=0}=
e^{-i\omega(z+t)}
\varphi^{i}(\xi),\qquad \xi=2i\omega(z+t),\qquad a=\frac{B_2}{2}-i\eta,\qquad c=B_2.
\end{eqnarray}
This set may also obtained 
by setting $z_0=0$ in the kernels (\ref{GG1}) for the CHE  and 
in their partners in terms of $\Psi(a,c;\xi)$. The kernels in terms of the regular
functions $\Phi(a,c;\xi)$ read
%
\begin{eqnarray}
\begin{array}{l}
\begin{array}{l}
{G}_{\ 1}^{(1)}(z,t)
=e^{-i\omega (z+t)}
\Phi\left[ \frac{B_{2}}{2}-i\eta, B_2;
2i\omega(z+t) \right],\end{array}\vspace{2mm}\\
\begin{array}{l}
G_{\ 1}^{(2)}(z,t) =
e^{i\omega (z+t)}[z+t]^{1-B_2}
\Phi\left[1+i\eta -\frac{B_{2}}{2}, 2-B_2;
-2i\omega(z+t) \right],
\end{array}\end{array}
\end{eqnarray}
while two other kernels result by replacing $\Phi(a,c;\xi)$  by $\Psi(a,c;\xi)$.
From the transformations (\ref{transfomacao-r}) we obtain three additional sets generate as
\begin{eqnarray}\begin{array}{l}\label{tres}
{G}_{\  2}^{(i)}(z,t)=r_1{G}_{\  1}^{(i)}(z,t),\qquad 
{G}_{\  3}^{(i)}(z,t)=r_2{G}_{\  1}^{(i)}(z,t),\qquad
{G}_{\  4}^{(i)}(z,t)=r_2{G}_{\  2}^{(i)}(z,t).
\end{array}\end{eqnarray}
The transformation $r_3$ does not produce new kernels. Thus, the kernels with 
$\Phi(a,c;\xi)$ are
%
%
\begin{eqnarray}
\begin{array}{l}
\begin{array}{l}
G_{\ 2}^{(1)}(z,t) =r_1G_{\ 1}^{(1)}(z,t)=
e^{-i\omega (z+t)+\frac{B_1}{z}+\frac{B_1}{t}} [zt]^{2-B_2}
\Phi\left[2-i\eta- \frac{B_{2}}{2},
4- B_2;2i\omega(z+t) \right],\end{array}\vspace{3mm}\\
\begin{array}{l}G_{\ 2}^{(2)}(z,t) =r_1G_{\ 1}^{(2)}(z,t)=
e^{i\omega (z+t)+\frac{B_1}{z}+\frac{B_1}{t}}[zt]^{2-B_2}
[z+t]^{B_2-3} 
\Phi\left[i\eta -1+\frac{B_{2}}{2},B_2-2 ;
-2i\omega(z+t) \right];
\end{array}\end{array}
\end{eqnarray}
%
%
%
%
\begin{eqnarray} 
\begin{array}{l}
G_{\ 3}^{(1)}=r_2G_{\ 1}^{(1)}=G^{(2,2)}(z,t)=
e^{i\omega(z+t)+\frac{B_1}{z}+\frac{B_1}{t}}\left[zt\right]^{-i\eta-\frac{B_2}{2}}
\Phi\left[2+i\eta-\frac{B_2}{2},2+2i\eta;-
\frac{B_1 (z+t)}{zt}\right],\vspace{2mm}\\
%
G_{\ 3}^{(2)}=r_2G_{\ 1}^{(2)}=G^{(2,1)}(z,t)= 
e^{i\omega(z+t)}\left[zt\right]^{1+i\eta-\frac{B_2}{2}}[z+t]^{-1-2i\eta}
\Phi\left[\frac{B_2}{2}-i\eta-1,-2i\eta;
\frac{B_1 (z+t)}{zt}\right];
\end{array}
\end{eqnarray}
%
%
%
\begin{eqnarray}
\begin{array}{l}
G_{ \ 4}^{(1)}=r_2G_{\ 2}^{(1)}=
e^{-i\omega(z+t)+\frac{B_1}{z}+\frac{B_1}{t}}\left[zt\right]^{i\eta-\frac{B_2}{2}}
\Phi\left[2-i\eta-\frac{B_2}{2},2-2i\eta;-
\frac{B_1 (z+t)}{zt}\right],\vspace{2mm}\\
%
%
G_{ \ 4}^{(2)}=r_2G_{\ 2}^{(2)}=
e^{-i\omega(z+t)}\left[zt\right]^{1-i\eta-\frac{B_2}{2}}[z+t]^{2i\eta-1}
\Phi\left[i\eta-1+\frac{B_2}{2},2i\eta;
\frac{B_1 (z+t)}{zt}\right].
\end{array}
\end{eqnarray}

\subsection*{D.3. Kernels given by elementary functions} 
The kernel 
\begin{eqnarray}\label{kernel-DCHE-1}
\begin{array}{l}
G_{1}(z,t)=
e^{-i\omega (z+t)}
\left[1+\frac{2i\omega zt}{B_1}\right]^{i\eta-\frac{B_{2}}{2}}
\end{array}
\end{eqnarray}
has the same form as a kernel found by Schmidt and Wolf \cite{schmidt} who have considered
a DCHE with only four parameters. To obtain (\ref{kernel-DCHE-1}), we insert
\begin{eqnarray*}
G(z,t)=e^{-i\omega(z+t)}f(z,t),
\end{eqnarray*}
into Eq. (\ref{Lz1-dche}). This leads to
\begin{eqnarray*}
\begin{array}{l}
z^2\frac{\partial^{2}f}{\partial z^{2}}+\Big[B_{1}+
B_{2}z-
2i\omega z^2\Big]\frac{\partial f}{\partial z}
-t^2\frac{\partial^{2}f}{\partial t^{2}}-\Big[B_{1}+
B_{2}t-
2i\omega t^2\Big]\frac{\partial f}{\partial t}
-2i\omega \left(\frac{B_{2}}{2}-i\eta\right)(z-t)f=0.
\end{array}
\end{eqnarray*}
By supposing that $f(z,t)$ depends on $z$ and $t$ through  the product 
$2i\omega zt/B_1=y$, the previous equation gives
\begin{eqnarray*}
\begin{array}{l}
\left[1+y\right]\frac{d f}{dy}-\left[i\eta-\frac{B_2}{2}\right]f=0
\qquad \Rightarrow\qquad  f(z,t)=\left[1+\frac{2i\omega zt}{B_1}\right]^{i\eta-\frac{B_2}{2}}.
\end{array}
\end{eqnarray*}
Hence we obtain the kernel (\ref{kernel-DCHE-1}). The transformations (\ref{transfomacao-r})
generate three additional kernels given  by
\begin{eqnarray}\begin{array}{l}
{G}_{  2}(z,t)=r_1{G}_{  1}(z,t),\qquad 
{G}_{  3}^{(i)}(z,t)=r_3{G}_{ 1}(z,t),\qquad
{G}_{  4}^{(i)}(z,t)=r_3{G}_{ 2}(z,t).
\end{array}\end{eqnarray}
Thus, we have
\begin{eqnarray} \label{kernel-DCHE-29}
\begin{array}{l}
G_{2}(z,t)=
e^{-i\omega (z+t)+\frac{B_1}{z}+\frac{B_1}{t}}\;[zt]^{2-B_2}\;
\left[1-\frac{2i\omega zt}{B_1}\right]^{i\eta-2+\frac{B_{2}}{2}},
\end{array}
\end{eqnarray}
while $G_{3}(z,t)$ and $G_{4}(z,t)$ are obtained by substituting $(\eta,\omega)$ 
for $(-\eta,-\omega)$ in $G_{1}(z,t)$ and $G_{2}(z,t)$.  
%

%
\section*{Appendice E. Kernels for the reduced double-confluent Heun equation (RDHE)}
\protect\label{E}
\setcounter{equation}{0}
\renewcommand{\theequation}{E.\arabic{equation}}

For the reduced double-confluent Heun equation (RDHE), 
\begin{equation}
\label{incedche}
\left[L_z+B_3\right]U=z^2\frac{d^{2}U}{dz^{2}}+[B_{1}+B_{2}z]\frac{dU}{dz}
+\left[q z+B_3\right]U=0,
\end{equation}
in this appendix E:
  \begin{itemize}
  \itemsep-3pt
  \item
  {initially we get} a substitution of variables which preserve the form of the
  equation for the kernels;
 \item {in E.1 we find} kernels given by products of Bessel and confluent hypergeometric functions; they 
 present an arbitrary constant of separation $\lambda$ and may be derived as limits
 of kernels of the DHE ($z_0\to 0$) or of the RCHE (Whitaker-Ince limit);
 \item {in E.2, by putting  $\lambda=0$, we obtain
  kernels given  by products of elementary and Bessel functions;
 \item
  in E.3, by choosing appropriate values for $\lambda$, we obtain
  kernels given  by products of elementary and confluent hypergeometric functions;
  \item 
 in E.4 we find two kernels given by products of elementary functions; these can be derived by 
 applying the Whittaker-Ince limit to kernels of the DHE.}
  \end{itemize}

For the RDHE, the integral (\ref{integral-dche}) remains formally 
unaltered, that is, 
\begin{eqnarray}
\label{integral-dche-2}
\mathcal{U}(z)
=\int_{t_{1}}^{t_{2}}t^{B_2-2}
e^{-{B_1}/{t}}G(z,t)U(t)dt,
\end{eqnarray}
while the equation (\ref{Lz1-dche}) for the kernels becomes
\begin{equation} \label{Lz1-rdche}
\begin{array}{l}
\left[z^2\frac{\partial^{2}}{\partial z^{2}}+
\left(B_{1}+B_{2}z\right)
\frac{\partial}{\partial z}
+q z\right]G
-\left[t^2\frac{\partial^{2}}{\partial t^{2}}+
\left(B_{1}+B_{2}t\right)
\frac{\partial}{\partial t}
+q t\right]G=0.
\end{array}
\end{equation}
For fixed endpoints of integrations the bilinear concomitant (\ref{M-dche}) is again
\begin{eqnarray} 
\displaystyle
P(z,t)=t^{B_2}\ e^{-B_1/t}
\left[U(t)\frac{\partial G(z,t)}{\partial t}-G(z,t)\frac{d U(t)}{d t}\right].
\end{eqnarray}
If the endpoints depend on $z$, we proceed as in appendix D.

On the other side, if $U(z)=U(B_{1},B_{2},B_{3}; q;z)$ denotes a known
solutions of RDHE, other solution is generated by the transformation
$T$ defined by
\begin{eqnarray}
TU(z)=e^{\frac{B_{1}}{z}}z^{2-B_{2}}\;U(-B_{1},4-B_{2},
B_{3}+2-B_{2}; q;z),
\end{eqnarray}
as we can show by substitutions of variables. Similarly, 
if $G(z,t)=G(B_{1},B_{2}; q;z,t)$ denotes a 
solution of Eq. (\ref{Lz1-rdche}), the corresponding 
transformation ${R}$ for this kernel is
\begin{eqnarray}\label{d3}
R G(z,t)=e^{\frac{B_{1}}{z}+\frac{B_{1}}{t}}(zt)^{2-B_{2}}\;G(-B_{1},4-B_{2}; q;z,t),
\end{eqnarray}
\subsection*{E.1. Kernels with products of Bessel and confluent hypergeometric functions}
We obtain the kernels given by products of 
Bessel and confluent hypergeometric functions 
by taking the limits when $z_0\to 0$ of the kernels 
(\ref{CHE-ince-5}) for the RCHE. Thus, up to a multiplicative constant
\begin{eqnarray} 
G_{\ \pm}^{(i,j)}(z,t)=\left[z+t\right]^{\frac{1}{2}-\frac{B_2}{2}}
 Z_{\pm(2\lambda+1-B_2)}^{(i)}\left[ 2\sqrt{q(z+t)}\right]\;\lim_{z_0\to 0} F
^{j}(\zeta)
\end{eqnarray}
where the $\displaystyle\lim_{z_0\to 0}F^{j}$ are given in Eqs. (\ref{limites}). 
Explicitly, from 
$F^{1}$, $F^{2}$,  $F^{5}$ and $F^{6}$ we get, respectively, 
\begin{eqnarray}\label{explicitly}
\begin{array}{l} 
G_{\ \pm}^{(i,1)}(z,t)=
\begin{array}{l} 
[zt]^{-\lambda}\;[z+t]^{\lambda+\frac{1}{2}-\frac{B_2}{2}} 
\; Z_{\pm(2\lambda+1-B_2)}^{(i)}\left[ 2\sqrt{q(z+t)}\right]\;
\Psi\left[\lambda,2+2\lambda-B_2;
\frac{B_1(z+t)}{zt}\right]\end{array},\vspace{3mm}\\
G_{\ \pm}^{(i,2)}(z,t)= \begin{array}{l} 
e^{\frac{B_1}{z}+\frac{B_1}{t}}\;[zt]^{\lambda+1-B_2}\;[z+t]^{-\lambda-\frac{1}{2}+\frac{B_2}{2}} 
\; Z_{\pm(2\lambda+1-B_2)}^{(i)}\left[ 2\sqrt{q(z+t)}\right]\;
\Psi\left[1-\lambda,B_2-2\lambda;
-\frac{B_1(z+t)}{zt}\right],\end{array}\vspace{3mm}\\
G_{\ \pm}^{(i,3)}(z,t)=\begin{array}{l} 
 [zt]^{-\lambda}\;[z+t]^{\lambda+\frac{1}{2}-\frac{B_2}{2}} 
 \; Z_{\pm(2\lambda+1-B_2)}^{(i)}\left[ 2\sqrt{q(z+t)}\right]\;
 \Phi\left[\lambda,2+2\lambda-B_2;
\frac{B_1(z+t)}{zt}\right],\end{array}\vspace{3mm}\\
G_{\ \pm}^{(i,4)}(z,t)=\begin{array}{l}
 e^{\frac{B_1}{z}+\frac{B_1}{t}}[zt]^{\lambda+1-B_2}\;[z+t]^{-\lambda-\frac{1}{2}+\frac{B_2}{2}} 
 \; Z_{\pm(2\lambda+1-B_2)}^{(i)}\left[ 2\sqrt{q(z+t)}\right]\;
\Phi\left[1-\lambda,B_2-2\lambda;-
\frac{B_1(z+t)}{zt}\right].\end{array}
\end{array}
\end{eqnarray}
The transformation $R$ given in Eq. (\ref{d3}) simply rearranges the previous
kernel provided that we transform the arbitrary constant $\lambda$ into
another arbitrary constant $\bar{\lambda}$. For example, we find
\begin{eqnarray*}
\begin{array}{l} 
RG_{\ \pm}^{(i,1)}(z,t)=e^{\frac{B_1}{z}+\frac{B_1}{t}}\; 
[zt]^{2-B_2-\bar{\lambda}}\;[z+t]^{\bar{\lambda}-\frac{3}{2}+ \frac{B_2}{2}} 
\; Z_{\pm(2\bar{\lambda}-3-B_2)}^{(i)}\left[ 2\sqrt{q(z+t)}\right]\;
\Psi\left[\bar{\lambda},2\bar{\lambda}-2+B_2;
-\frac{B_1(z+t)}{zt}\right].
\end{array}
\end{eqnarray*}
Putting $\bar{\lambda}=\lambda+2-B_2$ and using the second relation in (\ref{kummer}),
we find that the right-hand side of the above equation is a constant multiple of 
$G_{\ \pm}^{(i,2)}(z,t)$.

\subsection*{E.2. Kernels given by products of elementary and Bessel functions}
If $\lambda=0$ in $G_{\ \pm}^{(i,1)}$ or $G_{\ \pm}^{(i,3)}$, we have the kernels
\begin{eqnarray}
\begin{array}{l} 
G_{ \pm}^{(i)}(z,t)=[z+t]^{\frac{1}{2}-\frac{B_2}{2}} 
\; Z_{\pm(1-B_2)}^{(i)}\left[ 2\sqrt{q(z+t)}\right].
\end{array}
\end{eqnarray}
If $\lambda=1$ in $G_{\ \pm}^{(i,2)}$ or $G_{\ \pm}^{(i,4)}$, we have the kernels
\begin{eqnarray}
\begin{array}{l} 
\tilde{G}_{ \pm}^{(i)}(z,t)= 
e^{\frac{B_1}{z}+\frac{B_1}{t}}\;[zt]^{2-B_2}\;[z+t]^{-\frac{3}{2}+\frac{B_2}{2}} 
\; Z_{\pm(3-B_2)}^{(i)}\left[ 2\sqrt{q(z+t)}\right].
\end{array}
\end{eqnarray}
$G_{ \pm}^{(i)}$ and $\tilde{G}_{ \pm}^{(i)}$ are connected by the transformation $R$
given in (\ref{d3}).
\subsection*{E.3. Kernels given by products of elementary and confluent hypergeometric functions}
If $\lambda=({B_2}/{2})-({1}/{4})$, the order of the Bessel functions is $\pm 1/2$ in
the kernels (\ref{explicitly}) and, according to (\ref{bessel-1/2}), these functions reduce
to elementary functions. Then we have the following kernels given by products of elementary and confluent
hypergeometric functions:
\begin{eqnarray}
\begin{array}{l} 
G_{\ \pm}^{(i,1)}(z,t)=
\begin{array}{l} 
[zt]^{\frac{1}{4}-\frac{B_2}{2}}\;[z+t]^{\frac{1}{4}} 
\; Z_{\pm\frac{1}{2}}^{(i)}\left[ 2\sqrt{q(z+t)}\right]\;
\Psi\left[\frac{B_2}{2}-\frac{1}{4},\frac{3}{2};
\frac{B_1(z+t)}{zt}\right]\end{array},\vspace{3mm}\\
G_{\ \pm}^{(i,2)}(z,t)= \begin{array}{l} 
e^{\frac{B_1}{z}+\frac{B_1}{t}}\;[zt]^{\frac{3}{4}-\frac{B_2}{2}}\;[z+t]^{-\frac{1}{4}} 
\; Z_{\pm\frac{1}{2}}^{(i)}\left[ 2\sqrt{q(z+t)}\right]\;
\Psi\left[\frac{5}{4}-\frac{B_2}{2},\frac{1}{2};
-\frac{B_1(z+t)}{zt}\right],\end{array}\vspace{3mm}\\
G_{\ \pm}^{(i,3)}(z,t)=\begin{array}{l} 
 [zt]^{\frac{1}{4}-\frac{B_2}{2}}\;[z+t]^{\frac{1}{4}} 
\; Z_{\pm\frac{1}{2}}^{(i)}\left[ 2\sqrt{q(z+t)}\right]\;
\Phi\left[\frac{B_2}{2}-\frac{1}{4},\frac{3}{2};
\frac{B_1(z+t)}{zt}\right],\end{array}\vspace{3mm}\\
G_{\ \pm}^{(i,4)}(z,t)=\begin{array}{l}
e^{\frac{B_1}{z}+\frac{B_1}{t}}\;[zt]^{\frac{3}{4}-\frac{B_2}{2}}\;[z+t]^{-\frac{1}{4}} 
\; Z_{\pm\frac{1}{2}}^{(i)}\left[ 2\sqrt{q(z+t)}\right]\;
\Phi\left[\frac{5}{4}-\frac{B_2}{2},\frac{1}{2};
-\frac{B_1(z+t)}{zt}\right].\end{array}
\end{array}
\end{eqnarray}
The transformation (\ref{d3}) simply rearranges the above kernels.
\subsection*{E.4. Kernels given by elementary functions}
We find two kernels given by elementary functions. Up to multiplicative constants, we have
\begin{eqnarray}
\begin{array}{l} 
{G}_{1}(z,t)= 
\exp{\left[\frac{qzt}{B_1}\right]},\qquad {G}_{2}(z,t)= r{G}_{1}(z,t)=[zt]^{2-B_2}
\exp{\left[-\frac{qzt}{B_1}+\frac{B_1}{z}+\frac{B_1}{t}\right]}.
\end{array}
\end{eqnarray}
These kernels can be obtained by applying the Whittaker-Ince limit (\ref{ince})
to the kernels (\ref{kernel-DCHE-1}) and (\ref{kernel-DCHE-29}) of the RCHE. 
Alternatively, we can compute ${G}_{1}(z,t)$ by  supposing that $G(z,t)$ depends on $z$ and $t$ 
through  the product $q zt/B_1=y$, in which case Eq.(\ref{Lz1-rdche}) becomes
$dG/dy=G$, whose solution is the kernel ${G}_{1}$.

{At last we mention that there is an equation called doubly reduced 
double-confluent Heun equation \cite{lay}. However, as we have found no relation
of such equation with the equations discussed here, we do not 
consider its kernels.}

%

\end{document}